\documentclass[a4paper, 12pt]{article}

\pdfoutput=1 % if your are submitting a pdflatex (i.e. if you have images in pdf, png or jpg format)

\usepackage{jheppub} % for details on the use of the package, please see the JHEP-author-manual

\usepackage{
bbm,			% for doucle struck symbols like 1, ...
hyperxmp,		% xmp metadata
mathtools,		% improved maths
mleftright,		% better spacing for \left and \right
orcidlink, 		% link to orcid
}

\hypersetup{
    unicode=true,
	bookmarksdepth=3,
	bookmarksnumbered=true,
	pdfauthor={Matijn Fran\c{c}ois, Alba Grassi, Tommaso Pedroni},				% added by hyperxmp and arXiv
	pdfkeywords={supersymmetric gauge theory, spectral theory, quantum mechanics, quantum geometry, difference equations, eigenfunctions},
	pdflang={en-GB},
	pdfcontactaddress={Section de Mathématiques, Université de Genève, Rue du Général Dufour 24},
	pdfcontactpostcode={1211},
	pdfcontactcity={Genève 4},
	pdfcontactregion={GE},
	pdfcontactcountry={CH},
	pdfcontactemail={matijn.francois@unige.ch, alba.grassi@cern.ch, tpedroni@sissa.it},
    pdfcontacturl={https://orcid.org/0009-0002-8099-3374, https://orcid.org/0000-0003-0654-1759, https://orcid.org/0000-0003-0752-0247},
	pdfversionid={2},															% any human readable data, prefer incremented numbers as for arXiv versions
	pdfdoi={10.48550/arXiv.2511.10636},											% include only the doi name without any url prefix
	pdfurl={https://arxiv.org/abs/2511.10636},
	pdfpubtype={other},															% one of the prism aggregation types (book, journal, magazine, manual, report, whitepaper, other), use ``other'' for arXiv
	pdfpubstatus={AO},															% AO, SMUR, AM, P, VoR, CVoR, EVoR (Author’s Original, Submitted Manuscript Under Review, Accepted Manuscript, Proof, Version of Record, Corrected Version of Record, Enhanced Version of Record), use ``AO'' for arXiv
	pdfapart=1,
	pdfaconformance=B,
}

\usepackage[type={CC}, modifier={by}, version={4.0}, imageposition=left]{doclicense}

\usepackage{array, booktabs, multirow, ragged2e, tabularx}

\newcolumntype{Y}{>{\RaggedRight\arraybackslash}X}

\newcommand{\ResultCell}[1]{%
  \begin{minipage}[t]{\linewidth}
    \vspace{0pt}
    \RaggedRight
    \setlength{\abovedisplayskip}{0.4em}
    \setlength{\belowdisplayskip}{0.5em}
    #1
  \end{minipage}%
}

\NewDocumentCommand{\NewMathFunction}{mo}{
	\expandafter\NewDocumentCommand\csname #1\endcsname{o}
	{\IfValueTF{#2}{#2}{\operatorname{#1}}\IfValueT{##1}{\mleft( ##1 \mright)}}
}

\NewDocumentCommand{\NewMathFunctionLim}{mo}{
	\expandafter\NewDocumentCommand\csname #1\endcsname{o}
	{\IfValueTF{#2}{#2}{\operatorname*{#1}}\IfValueT{##1}{\mleft( ##1 \mright)}}
}

\NewDocumentCommand{\RenewMathFunction}{mo}{
	\expandafter\RenewDocumentCommand\csname #1\endcsname{o}
	{\IfValueTF{#2}{#2}{\operatorname{#1}}\IfValueT{##1}{\mleft( ##1 \mright)}}
}

\NewDocumentCommand{\RenewMathFunctionLim}{mo}{
	\expandafter\RenewDocumentCommand\csname #1\endcsname{o}
	{\IfValueTF{#2}{#2}{\operatorname*{#1}}\IfValueT{##1}{\mleft( ##1 \mright)}}
}

\mleftright % renews the commands \left and \right to \mleft and \mright respectively, for better spacing

\RenewMathFunction{Re}
\RenewMathFunction{Im}

\RenewMathFunctionLim{min}
\RenewMathFunctionLim{max}
\RenewMathFunctionLim{inf}
\RenewMathFunctionLim{sup}
\RenewMathFunctionLim{lim}
\RenewMathFunctionLim{liminf}
\RenewMathFunctionLim{limsup}
\RenewMathFunctionLim{gcd}		% \limits seems to be the standard behaviour of \gcd
\RenewMathFunctionLim{det}		% \limits seems to be the standard behaviour of \det

\RenewMathFunction{arg}
\RenewMathFunction{deg}
\RenewMathFunction{dim}
\RenewMathFunction{ker}

\RenewMathFunction{exp}
\RenewMathFunction{ln}
\RenewMathFunction{log}
\RenewMathFunction{lg}

\RenewMathFunction{cos}
\RenewMathFunction{sin}
\RenewMathFunction{tan}
\RenewMathFunction{sec}
\RenewMathFunction{csc}
\RenewMathFunction{cot}

\RenewMathFunction{arccos}
\RenewMathFunction{arcsin}
\RenewMathFunction{arctan}

\RenewMathFunction{cosh}
\RenewMathFunction{sinh}
\RenewMathFunction{tanh}
\RenewMathFunction{coth}

\NewDocumentCommand{\rH}{}{\mathrm{H}}
\NewDocumentCommand{\rJ}{}{\mathrm{J}}
\NewDocumentCommand{\rT}{}{\mathrm{T}}

\NewDocumentCommand{\rd}{}{\mathrm{d}}
\NewDocumentCommand{\rp}{}{\mathrm{p}}

\NewDocumentCommand{\x}{}{\mathrm{x}}
\NewDocumentCommand{\p}{}{\mathrm{p}}

\NewDocumentCommand{\bs}{o}{\IfValueTF{#1}{\boldsymbol{#1}}{\boldsymbol}}

\NewDocumentCommand{\bsa}{}{\bs{a}}
\NewDocumentCommand{\bse}{}{\bs{e}}
\NewDocumentCommand{\bsf}{}{\bs{f}}
\NewDocumentCommand{\bh}{}{\bs{h}}
\NewDocumentCommand{\bn}{}{\bs{n}}

\NewDocumentCommand{\balpha}{}{\bs{\alpha}}

\NewDocumentCommand{\blam}{}{\bs{\lambda}}

\NewDocumentCommand{\CD}{}{\mathcal{D}}
\NewDocumentCommand{\CE}{}{\mathcal{E}}
\NewDocumentCommand{\CN}{}{\mathcal{N}}
\NewDocumentCommand{\CP}{}{\mathcal{P}}
\NewDocumentCommand{\CQ}{}{\mathcal{Q}}
\NewDocumentCommand{\CS}{}{\mathcal{S}}
\NewDocumentCommand{\CV}{}{\mathcal{V}}
\NewDocumentCommand{\CW}{}{\mathcal{W}}

\NewDocumentCommand{\til}{o}{\IfValueTF{#1}{\widetilde{#1}}{\widetilde}}
\NewDocumentCommand{\tilE}{}{\widetilde{E}}
\NewDocumentCommand{\tilpsi}{}{\widetilde{\psi}}

\DeclarePairedDelimiter{\pr}{(}{)}
\DeclarePairedDelimiter{\ps}{[}{]}
\DeclarePairedDelimiter{\pc}{\lbrace}{\rbrace}

\DeclarePairedDelimiter{\abs}{\lvert}{\rvert}

\NewDocumentCommand{\br}{m}{\pr*{#1}}
\NewDocumentCommand{\sbr}{m}{\ps*{#1}}
\NewDocumentCommand{\cbr}{m}{\pc*{#1}}

\NewDocumentCommand{\deq}{}{\coloneqq}

\NewDocumentCommand{\pints}{}{\mathbb{N}_{>0}}
\NewDocumentCommand{\nnints}{}{\mathbb{N}}
\NewDocumentCommand{\ints}{}{\mathbb{Z}}

\NewDocumentCommand{\prats}{}{\mathbb{Q}_{>0}}

\NewDocumentCommand{\preals}{}{\mathbb{R}_{>0}}

\NewDocumentCommand{\reals}{}{\mathbb{R}}

\NewDocumentCommand{\comps}{}{\mathbb{C}}

\NewDocumentCommand{\pnaturals}{}{\pints}
\NewDocumentCommand{\prationals}{}{\prats}

\NewDocumentCommand{\IZ}{}{\mathbb{Z}}
\NewDocumentCommand{\IR}{}{\mathbb{R}}
\NewDocumentCommand{\IC}{}{\mathbb{C}}
\NewDocumentCommand{\IF}{}{\mathbb{F}}

\NewDocumentCommand{\re}{}{\mathrm{e}}
\NewDocumentCommand{\ri}{}{\mathrm{i}}

\NewMathFunction{bigo}[\mathcal{O}]
\NewMathFunction{sgn}

\NewMathFunction{sech}
\NewMathFunction{csch}

\NewMathFunction{arcsec}
\NewMathFunction{arccsc}
\NewMathFunction{arccot}

\NewMathFunction{arccosh}
\NewMathFunction{arcsinh}
\NewMathFunction{arctanh}
\NewMathFunction{arcsech}
\NewMathFunction{arccsch}
\NewMathFunction{arccoth}

\NewMathFunction{perm}[\sigma]
\NewMathFunction{gammaf}[\Gamma]
\NewMathFunction{loggamma}[\operatorname{log\Gamma}]

\NewMathFunction{fns}[F_{\mathrm{NS}}]
\NewMathFunction{qdilog}[\Phi_b]

\NewDocumentCommand{\indicator}{mm}{\mathbbm{1}_{#2}{\br{#1}}}

\NewDocumentCommand{\real}{m}{\Re[#1]}
\NewDocumentCommand{\imaginary}{m}{\Im[#1]}
\NewDocumentCommand{\bigO}{m}{\bigo[#1]}

\NewDocumentCommand{\be}{}{\begin{equation}}
\NewDocumentCommand{\ee}{}{\end{equation}}
\NewDocumentCommand{\ba}{}{\begin{aligned}}
\NewDocumentCommand{\ea}{}{\end{aligned}}

\preprint{CERN-TH-2025-230}

\title{Eigenfunctions of deformed Schr\"odinger equations}

\date{\today}

\author[a, b]{Matijn Fran\c{c}ois\,\orcidlink{0009-0002-8099-3374},}
\author[a, b]{Alba Grassi\,\orcidlink{0000-0003-0654-1759}}
\author[c, d, e]{and Tommaso Pedroni\,\orcidlink{0000-0003-0752-0247}}

\affiliation[a]{Section de Math\'ematiques, Universit\'e de Gen\`eve, 1211 Gen\`eve 4, Switzerland}
\affiliation[b]{Theoretical Physics Department, CERN, 1211 Geneva 23, Switzerland}
\affiliation[c]{SISSA, Via Bonomea 265, 34136 Trieste, Italy}
\affiliation[d]{INFN, Sezione di Trieste, Trieste, Italy}
\affiliation[e]{Institute for Geometry and Physics, IGAP, via Beirut 2, 34136 Trieste, Italy}

\emailAdd{matijn.francois@unige.ch}
\emailAdd{alba.grassi@cern.ch}
\emailAdd{tpedroni@sissa.it}

\abstract{We study the spectral problems associated with the finite-difference operators $\rH_N = 2 \cosh(p) + V_N(x)$, where $V_N(x)$ is an arbitrary polynomial potential of degree $N$.
These systems can be regarded as a solvable deformation of the standard Schr\"odinger operators $p^2 + V_N(x)$, and they arise naturally from the quantization of the Seiberg-Witten curve of four-dimensional, $\CN = 2$, SU($N$) supersymmetric Yang–Mills theory.
Using the open topological string/spectral theory correspondence, we construct exact, generalized eigenfunctions of $\rH_N$, valid for arbitrary polynomial potentials and describing both bound and resonant states. We also comment on the case with a $\sinh(p)$ kinetic term. Our solutions are entire in $x$ for all generalized eigenvalues, and become square-integrable for a discrete subset of those.
An interesting feature is the existence of special loci in the parameter space of the potential, where the eigenfunctions exhibit enhanced decay, leading to spectral degeneracies for confining potentials and to a real energy spectrum for unbounded ones.
Our results provide a rare example of a quantum-mechanical spectral problem that is exactly solvable, admitting explicit, analytic eigenfunctions for both bound and resonant states.}

\begin{document}

% To create a "Abstract" entry in the contents and/or the pdf bookmarks.
\phantomsection % Creates an anchor for the hyperlink
%\addcontentsline{toc}{section}{Abstract} % Adds an entry to both the TOC & pdf bookmarks
\pdfbookmark[1]{Abstract}{Abstract} % Adds an entry to the pdf bookmarks only

\maketitle

\flushbottom

\clearpage

%%%%%%%%%%%%%%%%%%%%%%%%%%%%%%%%%%%%%%%%%%%%%%%%%%%%%%%%%%%%%%%%%%%%%%%%%%%%%%%%%%%%
%%%%%%%%%%%%%%%%%%%%%%%%%%%%%%%%%%%%%%%%%%%%%%%%%%%%%%%%%%%%%%%%%%%%%%%%%%%%%%%%%%%%

\section{Introduction}

Exactly solvable models have been playing a central role in quantum mechanics, providing
insights into its mathematical structure and phenomenology. Over the past decade, advances in topological string theory and supersymmetric gauge theory have uncovered many new examples of solvable quantum spectral problems, often arising from the quantization of mirror curves; see \cite{Marino:2015nla, Marino:2024tbx} for a review and a list of references. Regarding quantities that encode the energy spectrum (such as exact quantization conditions, Fredholm determinants, and spectral traces), the situation is now well understood for operators associated with mirror curves of arbitrary genus. In contrast,
less is known about the corresponding eigenfunctions, particularly for mirror curves of genus greater than one, which are related to gauge theories with higher-rank gauge groups. In this work, we fill this gap for a specific class of geometries. More precisely, we analyse the spectral problem associated with the difference operator
\begin{equation}
\label{eq:hamil}
    \rH_{N} = 2 \Lambda^N \cosh[\p] + V_N(\x) \, ,
    \qquad \qquad
    \sbr{\x, \p} = \ri \hbar \, ,
\end{equation}
where $V_N(x) = \sum_{k=0}^{N-1} (-1)^k h_k \, x^{N-k}${, $h_0 = 1$,} is a generic polynomial potential of degree $N$.
This system can be regarded as a deformation of the standard quantum-mechanical anharmonic oscillator
\begin{equation}
\label{eq:anh}
    \p^2 + V_N(\x) \, ,
    \qquad \qquad
    \sbr{\x, \p} = \ri \hbar \, .
\end{equation}
As we will show, the deformation \eqref{eq:hamil} has a remarkable property: it defines a spectral problem that is exactly solvable in terms of explicit analytic functions. This stands in sharp contrast to the quantum anharmonic oscillator \eqref{eq:anh}, for which no explicit, closed-form analytic expressions for the eigenfunctions are known.\footnote{One notable exception occurs when $N=6$ and the potential takes the form $V_6(x) = x^6 + 2b\,x^4 + \bigl(b^2 - (4m+2p+3)\bigr)x^2$, with $m \in \pnaturals$ and $p \in \cbr{0, 1}$. In this case, certain special eigenfunctions can be constructed explicitly \cite{TURBINER1987181}.}

The exact solvability of \eqref{eq:hamil} originates from a beautiful connection with supersymmetric gauge theory.
In particular, the operator \eqref{eq:hamil} can be identified with a quantization\footnote{When quantizing such a curve \cite{adkmv,acdkv,mirmor}, one must decide which variable plays the role of position and which that of its conjugate momentum. Our convention differs from the standard one by a Fourier transform; as a result, the quantized curve is a difference equation rather than a higher order differential equation \cite{Ito:2021boh,Yan:2020kkb}.} of the Seiberg--Witten (SW) curve associated with four-dimensional \( \mathcal{N}=2 \), SU(\( N \))  supersymmetric Yang--Mills (SYM) theory \cite{adkmv,acdkv,mirmor,ns}. This correspondence provides access to powerful tools from supersymmetric localization \cite{ns} and topological string theory \cite{ghm},
enabling the explicit computation of spectral quantities via Nekrasov--Shatashvili (NS) functions. By contrast, from the perspective of supersymmetric QFT, the differential operator \eqref{eq:anh} is related to Argyres--Douglas theories \cite{gmn,Gaiotto:2014bza,Grassi:2018spf,Ito:2017ypt}, which are non-Lagrangian.
Consequently, at present there is no known analogue of NS functions for these theories, and explicit closed-form expressions for the eigenfunctions and other spectral quantities remain out of reach (for progress in this direction, see e.g.~\cite{Ito:2019twh,Hollands:2021itj,Ito:2018eon,Fucito:2023txg,Ito:2024wxw,Bonelli:2025owb}).

The deformed Hamiltonian \eqref{eq:hamil} was first analysed in \cite{Grassi:2018bci} and later in \cite{ggm}, where its Fredholm determinant and an exact quantization condition for the energy spectrum were derived.
Interestingly, the same deformed Hamiltonian also appears in the context of the $T \bar{T}$ deformation \cite{Chakrabarti:2023czz}.
In this paper, we provide explicit analytic formulas for the {generalized} eigenfunctions of the operator \eqref{eq:hamil}, valid for arbitrary polynomial potentials $V_N(x)$ and capturing both bound and resonant states. {We will also give some comments on the case where the kinetic term is $\sinh(\rp)$ rather than $\cosh(\rp)$.}

The eigenvalue equation associated with the Hamiltonian \eqref{eq:hamil} is the finite-difference equation
\begin{equation}
\label{eq:diffin}
    \Lambda^N \br{\psi{\br{x + \ri \hbar, \bs{h}}} + \psi{\br{x - \ri \hbar, \bs{h}}}} + V_N{\br{x}} \psi{\br{x, \bs{h}}}
    = E \psi{\br{x, \bs{h}}} \, .
\end{equation}
Difference equations of this type typically admit many formal solutions. The solutions we construct stand out by the following properties:
\begin{enumerate}

    \item For \emph{generic} values of the parameters $h_k$ and the energy $E$, our solutions solve \eqref{eq:diffin} and are entire in $x$. It is worth emphasizing that {analyticity is non-trivial} for a finite-difference equation. In our construction, the entireness ultimately follows from background independence in topological string theory: the special eigenfunctions we construct arise as specific limits of non-perturbative open topological string partition functions, obtained via the open topological string/spectral theory (TS/ST) correspondence \cite{Marino:2016rsq,Marino:2017gyg,Francois:2025wwd}. We refer to these as \emph{generalized} or \emph{off-shell} eigenfunctions. Their explicit expressions are given in  \eqref{eq:eigenfeven} for $N$ even and in \eqref{eq:eigenfodd} for $N$ odd.

    \item  For generic values of \(E\), the off-shell solutions are not square-integrable. 
    They become $L^2$-normalizable (physical) eigenfunctions only at the discrete set of energies $\cbr{E_k}_{k \geqslant 0}$ determined by the quantization condition of \cite{Grassi:2018bci}. We refer to the eigenfunctions evaluated at these energies as \emph{on-shell} eigenfunctions.

    \item Starting from the solutions to \eqref{eq:diffin}, one can also construct solutions to some closely related finite-difference equations. In particular, one obtains solutions for the problem with a $\sinh[\rp]$ kinetic term and/or with an inverted potential $-V_N(x)$. While not the main focus of this paper, we will look at these related problems in \autoref{sec:relatedproblems}. See \autoref{tab:summary} for an overview of the solutions we find.

    \item  As noted in \cite{Grassi:2018bci}, the deformed Hamiltonian \eqref{eq:hamil} exhibits several novel features with no counterpart in the standard Schr\"odinger Hamiltonian \eqref{eq:anh}.

    For instance, when $N$ is even, \eqref{eq:hamil} admits degenerate ground states at special values of the moduli $h_k$. Moreover, the oscillation theorem \cite[p.~66, thm.~3.5]{BS91} that applies to the standard Hamiltonian \eqref{eq:anh} no longer holds for the deformed Hamiltonian \eqref{eq:hamil}.

    Similarly, when $N$ is odd, there exist special values of the moduli $h_k$ for which \eqref{eq:hamil} admits $L^2$-normalizable eigenfunctions with real energies, even though the potential is unbounded from below.

\end{enumerate}

Let us also note that the difference equation \eqref{eq:diffin} arises in the integrability context as the Baxter equation of the $N$-particle closed quantum Toda lattice \cite{Gutzwiller1980, Gutzwiller1981, Sklyanin1985, PasquierGaudin1992, Kozlowski:2010tv}. In that setting, one imposes {specific} boundary conditions on the solutions, beyond just square-integrability. As a consequence, the solutions considered in \cite{Gutzwiller1980, Gutzwiller1981, Sklyanin1985, PasquierGaudin1992, Kozlowski:2010tv} exist only at special values of the parameters $h_k$ with $k = 2, \cdots, N$. {These special values coincide precisely with those at which the Hamiltonian \eqref{eq:hamil} exhibits {the features described above under point 4}}. We will return to this in \autoref{sec:toda}.

This paper is organized as follows. In \autoref{sec:sec2} we define the spectral problem and present explicit eigenfunctions; see \eqref{eq:eigenfeven} and \eqref{eq:eigenfodd}. In \autoref{sec:relatedproblems} we present solutions for the inverted potential and the $\sinh[p]$ kinetic term. The relation to the quantum Toda lattice is discussed in \autoref{sec:toda}. In \autoref{sec:tsst} we derive the results of \autoref{sec:sec2} from the TS/ST correspondence.
In addition, four appendices complement the main text. In \autoref{app:GT} we review conventions for the SU($N$) root system. In \autoref{sec:faddev} we summarize the definition and relevant properties of the Faddeev quantum dilogarithm. In \autoref{sec:sf} we review several gauge-theory special functions, with particular emphasis on Nekrasov-Shatashvili functions. Finally, in \autoref{sec:figs} we present additional figures that provide further numerical evidence for our results.

%%%%%%%%%%%%%%%%%%%%%%%%%%%%%%%%%%%%%%%%%%%%%%%%%%%%%%%%%%%%%%%%%%%%%%%%%%%%%%%%%%%%
%%%%%%%%%%%%%%%%%%%%%%%%%%%%%%%%%%%%%%%%%%%%%%%%%%%%%%%%%%%%%%%%%%%%%%%%%%%%%%%%%%%%

% To create a "Acknowledgments" entry in the contents and/or the pdf bookmarks.
\phantomsection % Creates an anchor for the hyperlink
%\addcontentsline{toc}{section}{Acknowledgments} % Adds an entry to both the TOC & pdf bookmarks
\pdfbookmark[1]{Acknowledgments}{Acknowledgments} % Adds an entry to the pdf bookmarks only

\acknowledgments

We would like to thank Jonah Baermann, Giulio Bonelli, Matteo Gallone, Saebyeok Jeong, Marcos Mari{\~n}o, Giovanni Ravazzini, Lukas Schimmer, Alessandro Tanzini and Joerg Teschner for interesting and valuable discussions.
The work of AG and MF is partially supported by the Swiss National Science Foundation Grants No.~185723, 218510 and the NCCR SwissMAP. The work of TP is partly supported by the INFN Iniziativa Specifica GAST, Indam GNFM, the MIUR PRIN Grant 2020KR4KN2 “String Theory as a bridge between Gauge Theories and Quantum Gravity''. Furthermore, TP acknowledges funding from the EU project Caligola (HORIZON-MSCA-2021-SE-01), Project ID: 101086123, and CA21109 -- COST Action CaLISTA. Finally, TP gratefully acknowledges the hospitality of CERN and the University of Geneva during the completion of this work.

\vspace{5mm}

\doclicenseThis

\clearpage

\begin{table}[p]
    \centering
    \footnotesize
    \setlength{\tabcolsep}{6pt}
    \renewcommand{\arraystretch}{1.1}

    \begin{tabularx}{\linewidth}{
        @{}
        >{\centering\arraybackslash}p{1.5cm}
        Y
        Y
        @{}
    }
        \toprule
        \(N\) even &  \(+V_N\) & \(-V_N\) \\
        \midrule

        \(\cosh[\rp]\)
        &
        \ResultCell{%
            \textbf{Quantization condition: \eqref{eq:QCeven}}\\[3pt]
            \emph{Eigenvalues:} $E_n\in\reals$, $E_{n} \leqslant E_{n+1}$.\\
            \emph{Solutions:} $\psi_n \in L^{2}\br{\mathbb{R} + \ri \alpha}$, $\alpha \in \mathbb{R}$.\\
            \emph{Decay:} exponential at $\pm \infty$.
        }
        &
        \ResultCell{%
            \textbf{No quantization condition}\\[3pt]
            \emph{Eigenvalues:} any $E \in\comps $, see \eqref{eq:continuous}.\\
            \emph{Solutions:} $\psi_E \in L^{2}\br{\mathbb{R}+\ri \alpha}$, $|\alpha|<\alpha_N$.\\
            \emph{Decay:} power-law at $\pm \infty$.\\

            \medskip
            \textbf{Quantization condition: \eqref{eq:QCeven}}\\[3pt]
            \emph{Eigenvalues:} $E_n \in \reals$, $E_{n} \geqslant E_{n+1}$.\\
            \emph{Solutions:} $\psi^{\pm}_n \in L^{2}\br{\mathbb{R}+\ri \alpha}$, $|\alpha| < \alpha_N$.\\
            \emph{Decay:} exponential at $\mp \infty$, power-law at $\pm \infty$.
        }
        \\

        \midrule

        \(\sinh[\rp]\)
        &
        \ResultCell{%
            \textbf{Quantization condition: \eqref{eq:QCeven}}\\[3pt]
            \emph{Eigenvalues:} $E_n\in\comps$.\\
            \emph{Solutions:} $\psi_n\in L^{2}\br{\mathbb{R}+\ri \alpha}$, $\alpha>-\alpha_N$.\\
            \emph{Decay:} power-law at $-\infty$, exponential at $+ \infty$.
        }
        &
        \ResultCell{%
            \textbf{Quantization condition: \eqref{eq:QCeven}}\\[3pt]
            \emph{Eigenvalues:} $E_n\in\comps$.\\
            \emph{Solutions:} $\psi_n \in L^{2}\br{\mathbb{R}+\ri \alpha}$, $\alpha < \alpha_N$.\\
            \emph{Decay:} exponential at $- \infty$, power-law at $+\infty$.

            \medskip
            \textbf{Quantization condition: \eqref{eq:QCeven_alt}}\\[3pt]
            \emph{Eigenvalues:} $E_n\in\comps$.\\
            \emph{Solutions:} $\psi_n \in L^{2}\br{\mathbb{R}+\ri \alpha}$, $\alpha <\alpha_N$.\\
            \emph{Decay:} power-law at $\pm \infty$.
        }
        \\

        % Clear separation between even and odd N
        \specialrule{1.2pt}{7pt}{3pt}

        $N$ odd & $+V_N$ & $-V_N$ \\
        \midrule

        \(\cosh[\rp]\)
        &
        \ResultCell{%
            \textbf{Quantization condition: \eqref{eq:QCodd}}\\[3pt]
            \emph{Eigenvalues:} $E_n\in\comps$.\\
            \emph{Solutions:} $\psi \in L^{2}\br{\mathbb{R}+\ri \alpha}$, $\alpha >-\alpha_N $.\\
            \emph{Decay:} power-law at $-\infty$, exponential at $+ \infty$.
            
            \medskip
            \textbf{Quantization condition: \eqref{eq:QCodd_alt}}\\[3pt]
            \emph{Eigenvalues:} $E_n\in\comps$.\\
            \emph{Solutions:} $\psi \in L^{2}\br{\mathbb{R}+\ri \alpha}$, $|\alpha| < \alpha_N$.\\
            \emph{Decay:} power-law at $-\infty$, exponential at $+ \infty$.
        }
        &
        \ResultCell{%
            \textbf{Quantization condition: \eqref{eq:QCodd}}\\[3pt]
            \emph{Eigenvalues:} $E_n \in\comps $.\\
            \emph{Solutions:} $\psi \in L^{2}\br{\mathbb{R}+\ri \alpha}$, $|\alpha|<\alpha_N$.\\
            \emph{Decay:} exponential at $-\infty$, power-law at $+\infty$.\\

            \medskip
            \textbf{Quantization condition: \eqref{eq:QCodd_alt}}\\[3pt]
            \emph{Eigenvalues:} $E_n\in\comps$.\\
            \emph{Solutions:} $\psi \in L^{2}\br{\mathbb{R}+\ri \alpha}$, $\alpha < \alpha_N$.\\
            \emph{Decay:} exponential at $- \infty$, power-law at $+\infty$.
        }
        \\

        \midrule

        \(\sinh[\rp]\)
        &
        \ResultCell{%
            \textbf{No solutions in $L^2(\reals)$ generically}

            \medskip
            \textbf{Quantization conditions: \eqref{eq:QCodd} \& \eqref{eq:QCodd_alt}}\\[3pt]
            \emph{Note}: only possible for special $V_N$.\\
            \emph{Eigenvalues:} $E_n \in \mathbb{C}$.\\
            \emph{Solutions:} $\psi\in L^{2}\br{\mathbb{R}+\ri \alpha}$, $\alpha \in \mathbb{R}$.\\ 
            \emph{Decay:} exponential at $\pm\infty$.
        }
        &
        \ResultCell{%
            \textbf{Quantization condition: \eqref{eq:QCodd}}\\[3pt]
            \emph{Eigenvalues:} $E_n\in\comps$.\\
            \emph{Solutions:} $\psi \in L^{2}\br{\mathbb{R}+\ri \alpha}$, $|\alpha| < \alpha_N$.\\
            \emph{Decay:} power-law at $\pm\infty$.

            \medskip
            \textbf{Quantization condition: \eqref{eq:QCodd_alt}}\\[3pt]
            \emph{Eigenvalues:} $E_n\in\comps$.\\
            \emph{Solutions:} $\psi \in L^{2}\br{\mathbb{R}+\ri \alpha}$, $|\alpha| <\alpha_N$.\\
            \emph{Decay:} power-law at $\pm \infty$.
        }
        \\

        \bottomrule
    \end{tabularx}

    \caption{Quantization conditions and square-integrable solutions
    for even and odd \(N\). Throughout this table we assume that the
    parameters \(h_k\) entering the potentials \(\pm V_N\) are real
    and generic, so that the solutions exhibit no accidental
    enhancement of their decay. The parameter \(\alpha_N\) is defined
    by \(\alpha_N=(\hbar/2)\br{1-1/N}\).}
    \label{tab:summary}
\end{table}

\clearpage

%%%%%%%%%%%%%%%%%%%%%%%%%%%%%%%%%%%%%%%%%%%%%%%%%%%%%%%%%%%%%%%%%%%%%%%%%%%%%%%%%%%%
%%%%%%%%%%%%%%%%%%%%%%%%%%%%%%%%%%%%%%%%%%%%%%%%%%%%%%%%%%%%%%%%%%%%%%%%%%%%%%%%%%%%

\section{The spectral problem and its solution}
\label{sec:sec2}

%%%%%%%%%%%%%%%%%%%%%%%%%%%%%%%%%%%%%%%%%%%%%%%%%%%%%%%%%%%%%%%%%%%%%%%%%%%%%%%%%%%%

\subsection{The spectral problem}
\label{sec:spectp}

We consider the quantized Seiberg-Witten curve
\begin{equation}
\label{eq:hn}
    \rH_{N} = \Lambda^N \br{\re^{\p} + \re^{-\p}} + V_N{\br{\x}} \, ,
    \qquad \qquad
    \sbr{\x, \p} = \ri \hbar \, ,
    \qquad \qquad
    \Lambda, \hbar \in \preals \, ,
\end{equation}
\begin{equation}
\label{eq:vn}
    V_N{\br{x}} = \sum_{k=0}^{N-1} \br{-1}^k h_k \, x^{N-k} \, ,
    \qquad \qquad
    h_k \in \reals \, ,
\end{equation}
where we set $h_0 = 1$ and $h_1 = 0$, without loss of generality.\footnote{The restriction $\Lambda, \hbar \in \preals$ and $h_k \in \reals$ is not essential: the solutions we construct remain well-defined even for other choices of the parameters. However, this assumption simplifies the discussion of the operator’s spectral properties, and we will adopt this convention.}
A natural domain $\CD$ on which the operator $\rH_N: \CD \to  L^2(\reals)$ is symmetric is given by
\begin{equation}
\label{eq:domain}
   \CD = (\CD_1 \cap \CD_2) \subset L^2(\reals) \, ,
   \qquad
   \CD_1 = \cbr{\psi(x) \in L^2(\reals) \; \middle |\; V_N(x)\psi\br{x} \in L^2(\IR)} \, ,
\end{equation}
while $\CD_2$ consists of all $\psi(x + \ri y)$ that admit an analytic continuation to the strip $- \hbar < y < \hbar$ such that $\psi{\br{x + \ri y}} \in L^2{\br{\reals}}$ for all fixed $- \hbar < y < \hbar$, and for which the limits
\begin{equation}
    \lim_{y \to \hbar^-} \psi{\br{x \pm \ri y}}
\end{equation}
exist in the sense of convergence in $L^2(\reals)$. For $N$ even, one can in fact consider a canonical self-adjoint extension of this operator, see below.

The eigenvalue equation associated with \eqref{eq:hn} is a difference equation and reads\footnote{To simplify notation, we suppress the explicit dependence of the eigenfunctions on $\Lambda$ and $\hbar$.}
\begin{equation}
\label{eq:diffm}
    \Lambda^N\left(\psi(x+\ri \hbar,{\bs h})+\psi(x-\ri \hbar,{\bs h})\right)+V_N(x) \psi(x, {\bs h})
    = E \psi(x, {\bs h})\,,
\end{equation}
where $\bs{h}$ contains $h_2, \cdots, h_N$ with
\begin{equation}
    h_N = \br{-1}^{N-1} E \, ,
\end{equation}
and $E$ plays the role of energy.
Note that the spectral properties of $\rH_N$ depend on whether $N$ is even or odd.

\paragraph{N even: confining potential.}

When $N$ is even, the potential $V_N(x)$ is confining. One can consider a canonical self-adjoint extension of the operator $\rH_N$, the so-called Friedrichs extension.\footnote{We will not explicitly construct this extension, and all our on-shell eigenfunctions will in fact turn out to be in the domain $\CD$ in \eqref{eq:domain}.} It was shown in \cite{LST19}, based on \cite{LST16}, that this extension has a purely discrete spectrum corresponding to bound states, and has a trace-class inverse. Numerically, the spectrum and eigenfunctions can be computed via Hamiltonian truncation in the harmonic oscillator basis.

\paragraph{N odd: unbounded potential.}

When $N$ is odd, the potential $V_N(x)$ is confining as $x \to +\infty$ but unbounded from below as $x \to -\infty$. {One therefore expects the existence of resonant states with complex energies.}\footnote{\label{footnote:extensions}{Note that while the existence of a self-adjoint extension of $\rH_N$ is guaranteed by von Neumann's theorem \cite[thm.~X.3]{RS75} \cite[prop.~13.25]{S12}, there is no clear \emph{canonical} choice for $N$ odd. In fact, it is known that for the Schrödinger case $p^2 + V_N(x)$ with $N$ odd there is an infinite family of self-adjoint extensions, all of which have a discrete spectrum \cite[lemma.~2.2]{Caliceti}. }}

\begin{figure}[t]
    \centering
    \includegraphics[width=1\textwidth]{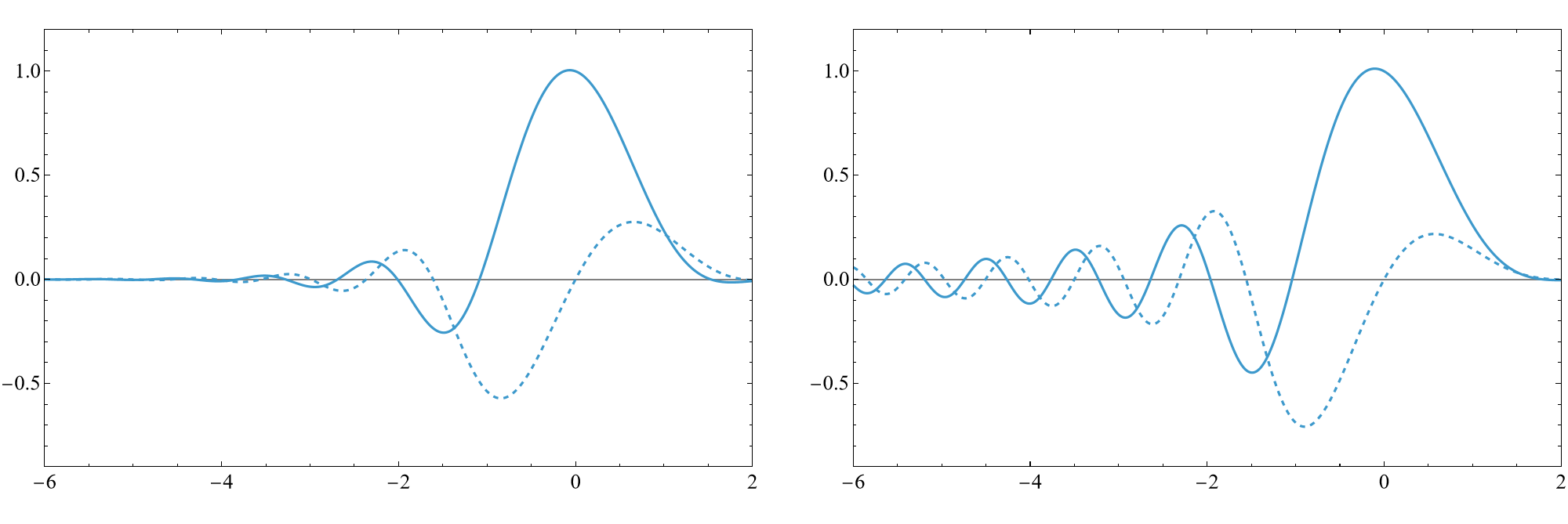}
    \caption{Numerical ground state eigenfunction ($E_0\approx 0.54151 + \ri \, 0.25905$) of \eqref{eq:hn} with $V_3(x) = x^3$, obtained via complex dilation. Left: rotated eigenfunction $\psi_0^{\theta}(x,\bs{h})$. Right: true eigenfunction $\psi_0(x,\bs{h}) = \psi_0^{\theta}(\re^{-\ri \theta}x,\bs{h})$. Here we set $\Lambda=1/2$, $\hbar=1$ and $\theta=-1/10$. The eigenfunction is normalized so that $\psi_0(0,\boldsymbol{h})=1$. Solid lines denote the real part; dashed lines denote the imaginary part.}
    \label{fig:complexdil}
\end{figure}

Such potentials have been studied numerically for the standard Schr\"odinger equation \eqref{eq:anh} using the method of complex dilation \cite{complexdila}, and rigorous analyses of resonances are available in \cite{Caliceti, Caliceti1983, Maioli}.
As emphasized in \cite{Grassi:2018bci}, complex dilation can be applied directly to the deformed Hamiltonian \eqref{eq:hn}; this is the numerical method we adopt here.
Let us briefly recall the main idea.
For $N$ odd, the eigenfunctions of \eqref{eq:hn} decay only as a power-law for $x \to -\infty$
and numerical diagonalization in the harmonic oscillator basis does not seem to be convergent. 
To circumvent this problem, we perform a small complex rotation of the position and momentum operators by an angle $\theta$. This yields the rotated Hamiltonian
\begin{equation}
\label{eq:hamiltonianhnth}
    \rH_{N, \theta} = 2\Lambda^N \cosh[\re^{-\ri\theta}\p] + V_{N}\pr*{\re^{\ri \theta}\x} \, 
\end{equation}
which admits exponentially decaying, square-integrable eigenfunctions \be\label{eq:eigrot} \psi^{(\theta)}_n(x,{\bs h})\ee
with complex eigenvalues. Therefore, $\psi^{(\theta)}_n(x,{\bs h})$ can be computed efficiently by diagonalizing \eqref{eq:hamiltonianhnth} in the harmonic oscillator basis. The eigenfunctions of the original Hamiltonian \eqref{eq:hn} are then obtained by analytic continuation:
\be  \label{eq:eigtoeig}
\psi_n(x,{\bs h})=\psi^{(\theta)}_n(\re^{-\ri \theta}x,{\bs h}) \, .
\ee
An explicit example is shown in \autoref{fig:complexdil} for the cubic potential $V_3(x)=x^3$. One observes the exponential decay of the rotated eigenfunctions \eqref{eq:eigrot} (left), in contrast with the slower power-law decay of the true eigenfunctions \eqref{eq:eigtoeig} (right).

\hfill

\noindent
In the next section, we construct explicit solutions to \eqref{eq:diffm} for arbitrary $N$. These solutions are entire in \(x\) for {\emph{all}} values of $E$ and \(h_k\), \(k = 2, \ldots, N-1 \), and we refer to them as \emph{off-shell} eigenfunctions. As we will see, such solutions become square-integrable wavefunctions only when \(E\) satisfies an appropriate quantization condition. The corresponding eigenfunctions, evaluated at these discrete energy values, will be denoted as \emph{on-shell} eigenfunctions.

%%%%%%%%%%%%%%%%%%%%%%%%%%%%%%%%%%%%%%%%%%%%%%%%%%%%%%%%%%%%%%%%%%%%%%%%%%%%%%%%%%%%

\subsection{Gauge theoretic building blocks}
\label{sec:Matone}

The explicit construction of our eigenfunctions is based on the open TS/ST correspondence \cite{Marino:2016rsq,Marino:2017gyg,Francois:2025wwd}.
A key lesson from studying spectral problems through the lens of supersymmetric gauge theory and topological strings is that eigenfunctions — and spectral quantities more generally — take a simpler form when written in terms of the Coulomb branch (or K\"ahler) parameters $\{a_I\}_{I=1}^{N}$, rather than the complex moduli $\{h_k\}_{k=2}^{N}$ that appear in the difference equation \eqref{eq:diffm}. The map relating the two sets of parameters has a precise geometrical meaning: it is known as the quantum mirror map \cite{mirmor,acdkv,Mironov:2009dv}. In the framework of five-dimensional, $\mathcal{N}=1$ supersymmetric gauge theory, the inverses of these maps are interpreted as Wilson loop expectation values \cite{Gaiotto:2014ina,Bullimore:2014awa}, and in the four-dimensional limit they reduce to the generalized Matone relations \cite{matone,Flume:2004rp, Grassi:2018bci}. At leading order in $\Lambda$, we have
\begin{equation}
\label{eq:htoam}
    \boxed{
    h_k = s_{(1^k)}(\bsa) + \bigO{\Lambda^{2 N}}
    = \sum_{1 \leqslant I_1 < I_2 < \cdots < I_k \leqslant N} a_{I_1} a_{I_2} \cdots a_{I_k} + \bigO{\Lambda^{2 N}}
    \, ,
    }
\end{equation}
where $s_R$ denotes the Schur polynomial associated to the Young tableau $R$, and $(1^k)$ refers to the tableau consisting of one column with $k$ boxes. {Thus, $s_{(1^k)}$ is an elementary symmetric polynomial, and the roots of $V_{N}(x)-E$, with the potential \eqref{eq:vn}, coincide with the Coulomb branch parameters $a_I$ to leading order in $\Lambda^N$.}
We refer to \eqref{eq:htoam} as the generalized Matone relation. Explicit examples are
\begin{equation}
    \begin{gathered}
        h_2 = -\frac{1}{2}\sum_{I=1}^N a_I^{2} +\bigO{\Lambda^{2 N}} \, ,
        \qquad \qquad
        h_3 = \frac{1}{3}\sum_{I=1}^N a_I^{3} + \bigO{\Lambda^{2 N}} \, ,
        \\
        h_4 = \frac{1}{8}\left(\left(\sum_{I=1}^N a_I^{2}\right)^{2} -2 \sum_{I=1}^N a_I^{4} \right) + \bigO{\Lambda^{2 N}} \, ,
    \end{gathered}
\end{equation}
recalling that the Coulomb branch parameters $a_I$ sum to zero in the $\mathrm{SU}(N)$ case,
\begin{equation}
    \sum_{I=1}^{N}a_I=0 \, .
\end{equation}
The complete expression for the map, including all orders in $\Lambda$ and $\hbar$ dependence, is reviewed in \autoref{sec:wl}; see in particular equation \eqref{eq:genmat}.  One can invert \eqref{eq:htoam} and \eqref{eq:genmat} to obtain an expression for $a_{I}$ as a function of $h_k$, $\Lambda$ and $\hbar$. Geometrically, the quantities $a_{I}(\boldsymbol{h}, \Lambda, \hbar)$, for $I = 1, \dots, N-1$, represent the quantum A-periods associated with the SW curve \cite{acdkv,mirmor,Mironov:2009dv}
\begin{equation}
\label{eq:swcurve}
  2 \Lambda^N \cosh(p) + V_N(x) = E, \qquad \qquad x, p \in \mathbb{C} \, .
\end{equation}
As noted in the introduction, \eqref{eq:diffm} coincides with a quantization of the SW curve \eqref{eq:swcurve} describing the low-energy dynamics of four-dimensional, $\mathcal{N}=2$, SU($N$) SYM. It is therefore natural to expect that the eigenfunctions inherit an SU($N$) structure. We thus begin by recalling a few representation-theoretic notions related to SU($N$).

Let $\mathcal{W}_N$ denote the Weyl group of SU($N$). Consider a vector
\begin{equation}
    {\bs v} = \sum_{I=1}^N v_I {\bse}_I \, ,
\end{equation}
where $\bs{e}_I$ are the weights of the fundamental representation, see \autoref{app:GT} for our conventions. The Weyl orbit of ${\bs v}$ is defined by
\begin{equation}
    \mathcal{W}_N \cdot {\bs v}
    =  \pc*{w({\bs v}) \, \middle| \, w \in \mathcal{W}_N}
    = \pc*{\sum_{I=1}^N v_{\perm[I]} {\bse}_I \, \middle| \, \perm \in \CS_N}
    \, ,
\end{equation}
where $\CS_N$ is the permutation group of $N$ elements. We define
\begin{equation}
\label{eq:gammadef}
    \bs{\gamma}
    = \frac{1}{2} \sum_{I=1}^N (-1)^{I-1} \bs{e}_I
    = \sum_{k = 1}^{N-1}(-1)^{k-1} \bs{\lambda}_k \, ,
\end{equation}
where $\bs \lambda_k$ is a fundamental weight \eqref{eq:latticedef},
and we organize the Coulomb branch parameters $a_I$ of $\mathcal{N}=2$, SU($N$) SYM into the vector
\begin{equation}
    \bs{a} = \sum_{I=1}^N a_I \bs{e}_I \, ,
    \qquad \qquad
    \sum_{I=1}^N a_I = 0 \, .
\end{equation}
It is also useful to introduce the linear involution
\begin{equation}
\label{eq:fa}
    \bsf(\bsa) =- \sum_{I=1}^N a_{N-I+1} \, \bs{e}_I
    \,,
\end{equation}
which acts on the parameters $h_k$ according to $h_k(\bsf(\bsa)) = h_k(-\bsa) = (-1)^k h_k(\bsa)$.

The starting point of our construction is the following defect partition function:
\begin{equation}
\label{eq:ZDfullmain}
    \boxed{
    Z_D(x,{\bs a}, \Lambda, \hbar)
    =
    \br{\frac{\Lambda}{\hbar}}^{\ri N \frac{{x}}{\hbar}} \re^{\pi \br{1 + \frac{N}{2}} \frac{x}{\hbar}} \cbr{\prod_{I=1}^N
    \gammaf[\frac{\ri \, \bs{e}_I \cdot \bs{a} - \ri \, x }{\hbar}]} \, Z_D^{\rm inst}\br{x, \bs{a}, \Lambda, \hbar}
    \, ,
    }
\end{equation}
where the instanton part $Z_D^{\rm inst}$ is defined in \eqref{eq:Zdinst}.
It is known \cite{Kozlowski:2010tv, Alday:2010vg,Kanno:2011fw,Gaiotto:2014ina,Sciarappa:2017hds,Jeong:2021rll} that this function provides a formal solution to the difference equation \eqref{eq:diffm}. However, $Z_D$ has poles at
\begin{equation}
    x = a_I + \ri \hbar k \, , \qquad \qquad k \in \IZ \, , \qquad \qquad I \in \{1, \cdots, N\}\, .
\end{equation}
Hence, the function \eqref{eq:ZDfullmain} alone is insufficient to construct entire, square-integrable eigenfunctions.
In what follows, we will combine two copies of \eqref{eq:ZDfullmain} with an additional function in such a way that the resulting expression becomes entire in $x$ for \emph{any} values of $a_I$, and hence \emph{any} value of $h_k$. The function we shall use is
\begin{equation}
\label{eq:Hndef}
    \boxed{
    \CP_{\bs{n}}(x,{\bs a}, \Lambda, \hbar) =
    \frac{\exp[\left(\frac{\ri}{\hbar}\partial_{\bs a} F_{\rm NS} - \frac{\pi \bs{a}}{\hbar} \indicator{N}{2 \nnints + 1} \right)\cdot\bs{n}]}
    {\prod_{\bs{\alpha}\in \Delta_{+}} \br{2\sinh\br{\frac{\pi \bs{\alpha} \cdot \bs{a}}{\hbar}}}^{({\bs \alpha} \cdot \bs{n})^2}}
    \prod_{I=1}^{N} \bigl(1 - \re^{\frac{2\pi}{\hbar}\left({\bs e}_I \cdot {\bs a}-x\right)} \bigr)^{\frac{1}{2} - n_I} \, ,
    }
\end{equation}
\begin{equation}\label{eq:nadmiss}
    \bs{n} = \sum_{I = 1}^N n_I \bs{e}_I \, ,
    \qquad \qquad
     n_I  \in \cbr{- \frac{1}{2}, + \frac{1}{2}} \, ,
\end{equation}
where $\indicator{N}{2 \nnints + 1}$ is the indicator function vanishing for $N$ even, $\Delta_+$ is the set of positive roots  \eqref{eq:positroot}, the $\bs e_I$ are the weights of the fundamental representation in \eqref{eq:latticedef}, and $\bs n$ is a particular element in the weight lattice, see later.
The dual quantum period is given by\footnote{The derivatives with respect to $a_I$ are taken before imposing the constraint $\sum_{I=1}^N a_I=0$.}
\begin{multline}
    \frac{\ri}{\hbar} \partial_{\bs{a}} F_{\mathrm{NS}}
    =
    \sum_{I=1}^N \frac{\ri}{\hbar} \frac{\partial F^{\rm inst}_{\rm NS}}{\partial a_I} \bs{e}_I
    \\
    - \sum_{\balpha \in \Delta_+}
    \sbr{\ri \, \frac{\pi}{2}
    + \ri \, 2 \br{\balpha \cdot \frac{\bs{a}}{\hbar}} \log\!{\br{\frac{\Lambda}{\hbar}}}
    + \log\!{\br{\frac{
    \Gamma\!{\br{1 - \ri \br{\balpha \cdot \frac{\bs{a}}{\hbar}}}}}
    {\Gamma\!{\br{1 + \ri \br{\balpha \cdot \frac{\bs{a}}{\hbar}}}}}
    }}} \balpha \, ,
\end{multline}
where $ F^{\rm inst}_{\rm NS}$ is defined in \eqref{fns4d}.
Note that the functions $\CP_{\bs{n}}$ are invariant under the shifts appearing in the difference operator in \eqref{eq:diffm}.

Finally, to analyse the asymptotic behaviour of the resulting solutions, it is convenient to introduce the shorthand notation
\begin{equation}
    \begin{split}
        \CQ_{\bs{n}}^0({\bs a}, \Lambda, \hbar)
        & =
        \frac{\exp[\left(\frac{\ri}{\hbar}\partial_{\bs a} F_{\rm NS} - \frac{\pi \bs{a}}{\hbar} \indicator{N}{2 \nnints + 1} \right)\cdot\bs{n}]}
        {\prod_{\bs{\alpha}\in \Delta_{+}} \br{2\sinh\br{\frac{\pi \bs{\alpha} \cdot \bs{a}}{\hbar}}}^{({\bs \alpha} \cdot \bs{n})^2}}
        \, ,
        \\
        \CQ^1_{\bs{n}}({\bs a}, \Lambda, \hbar)
        & =
        - \CQ_{\bs{n}}^0({\bs a}, \Lambda, \hbar)
        \sum_{I = 1}^N \pr*{\frac{1}{2} - n_I} \exp[2 \pi \bse_I \cdot \frac{\bsa}{   {\hbar}}]
        \, ,
        \\
        \CQ^M_{\bs{n}}({\bs a}, \Lambda, \hbar)
         & =
        (-1)^M \CQ_{\bs{n}}^0({\bs a}, \Lambda, \hbar)
        \exp[\sum_{I = 1}^N \pr*{\frac{1}{2} - n_I} 2 \pi \bse_I \cdot \frac{\bsa}{   {\hbar}}]
        \, ,
    \end{split}
\end{equation}
where $M = \sum_{I = 1}^N (1/2-n_I) \in \nnints$.
These are the coefficients of the  leading, subleading and most suppressed contributions to the large $x$ expansion of $\CP_{\bn}$ \eqref{eq:Hndef}.
Using the building blocks \eqref{eq:ZDfullmain} and \eqref{eq:Hndef}, we then obtain the following entire, off-shell solutions to the difference equation \eqref{eq:diffm}.

\subsection{The eigenfunctions}
\label{sec:Eigenfunctions4D}

\paragraph{N even: confining potential.}

For $N$ even, we have:
 \begin{align}\label{eq:eigenfeven}
\boxed{%
\begin{aligned}
\psi_N^{\rm even}(x,{\bs h}) & =
   Z_D(x, \bs{a}, \Lambda, \hbar)
   \sum_{\bs{n}\in \mathcal{W}_{N}\cdot \bs{\gamma}}
   \CP_{\bs n}(x, \bs{a}, \Lambda,\hbar) \\
 & \qquad + \ri \,
   \re^{\frac{2\pi x}{\hbar}}Z_D(-x, {\bsf}{\br{\bs{a}}}, \Lambda, \hbar)
   \sum_{\bs{n}\in \mathcal{W}_{N}\cdot \left(\bs{\gamma}+\bs{e}_N\right)}
   \CP_{\bs n}(-x, {\bsf}(\bs{a}),\Lambda, \hbar) \, ,
\end{aligned}
}
\end{align}
where we recall that ${\bsf}(\bs{a})$ is defined in \eqref{eq:fa}, and $\bs{a}$ is related to $\bs{h}$ via the generalized Matone relations; see \eqref{eq:htoam} and \eqref{eq:genmat}.\footnote{Note that $\CW_N \cdot \bs{\gamma} = \CW_N \cdot \bs{\lambda}_{N/2}$ and $\CW_N \cdot \br{\bs{\gamma} + \bs{e}_N} = \CW_N \cdot \bs{\lambda}_{(N/2)+1}$ for $N$ even, with $\bs{\lambda}_{N/2}, \bs{\lambda}_{(N/2)+1} \in \Lambda_w$ fundamental weights.} We will refer to the first and second terms in \eqref{eq:eigenfeven} as saddles, reflecting their origin in topological string theory \cite{Marino:2016rsq,Marino:2017gyg,Francois:2025wwd}; see \autoref{sec:tsst}.

Let us now analyse the asymptotic behaviour of \eqref{eq:eigenfeven}. For $N$ even, the Weyl orbit of $\bs{\gamma}$ contains $\binom{N}{N/2}$ elements, while the orbit of $\bs{\gamma}+\bs{e}_N$ contains $\binom{N}{N/2-1}$ elements. Moreover, in this case the number of non-trivial factors appearing in the product in \eqref{eq:Hndef} equals $N/2$ for elements in the orbit of $\bs{\gamma}$, and $N/2-1$ for those in the orbit of $\bs{\gamma}+\bs{e}_N$. This leads to the following asymptotic behaviour as $\Re[x] \to \pm \infty$, with $\Im[x]$ held fixed:
\begin{multline}
\label{eq:asen}
    \psi_N^{\rm even}(x,{\bs h}) \simeq
    \\
    \re^{\frac{\pi x}{\hbar}}\left(\frac{2 \pi \hbar}{\abs{{\rm Re}(x)}}\right)^{\frac{N}{2}} \br{\frac{\abs{{\rm Re}(x)}}{\Lambda}}^{N \frac{{\rm Im}(x)}{\hbar}} u_N(x) \,
    \sum_{\bs{n}\in \mathcal{W}_{N}\cdot \bs{\gamma}}
    \CQ_{\bn}^0(\bsa, \Lambda, \hbar)
    ,
    \qquad
    \Re[x] \to + \infty
    ,
\end{multline}
\begin{multline}
\label{eq:asymevenminus}
    \psi_N^{\rm even}(x,{\bs h} ) \simeq \re^{\frac{\pi x}{\hbar}} \left(\frac{2 \pi \hbar}{|{\rm Re }(x)|}\right)^{\frac{N}{2}}
    \left\{\br{\frac{|{\rm Re}(x)|}{\Lambda}}^{N \frac{{\rm Im}(x)}{\hbar}}u_N^{-1}(x)
    \sum_{\bs{n}\in \mathcal{W}_{N}\cdot \bs{\gamma}}
    \CQ_{\bn}^{\frac{N}{2}}(\bsa, \Lambda, \hbar) \right.
    \\
    \left. + \, \ri \br{\frac{|{\rm Re}(x)|}{\Lambda}}^{-N \frac{{\rm Im}(x)}{\hbar}} u_N(x)
    \sum_{\bs{n}\in \mathcal{W}_{N}\cdot \left(\bs{\gamma}+\bs{e}_N\right)}
    \CQ_{\bn}^{0}(\bsf(\bsa), \Lambda, \hbar) \right\}
    \, ,
    \qquad
    \Re[x] \to - \infty
    \, ,
\end{multline}
where $u_N(x)$ is an $x$-dependent factor of unit modulus,
\begin{equation}
\label{eq:cdef}
    u_N(x) \deq \re^{\ri \frac{\pi}{4}N}
   \br{\frac{\re \, \Lambda}{\abs{{\rm Re} (x)}}}^{\ri N \frac{\abs{{\rm Re} (x)}}{\hbar}}
    \, .
\end{equation}
Due to the exponential growth near positive infinity, \eqref{eq:eigenfeven} is not square-integrable unless we impose
\begin{equation}
\label{eq:QCeven}
    \sum_{\bs{n}\in \mathcal{W}_{N}\cdot \bs{\gamma}}
    \CQ_{\bn}^0(\bsa, \Lambda, \hbar)
    =
    \sum_{\bs{n}\in \mathcal{W}_{N}\cdot \bs{\gamma}}
    \frac{\exp[\frac{\ri}{\hbar}\partial_{\bs a} F_{\rm NS}\cdot\bs{n}]}
    {\prod_{\bs{\alpha}\in \Delta_{+}} \br{2\sinh\br{\frac{\pi \bs{\alpha} \cdot \bs{a}}{\hbar}}}^{({\bs \alpha} \cdot \bs{n})^2}}
    = 0 \, ,
\end{equation}
which is precisely the quantization condition for the energy $E$ obtained in \cite{Grassi:2018bci} by explicitly constructing the Fredholm determinant. Note that we are implicitly using the generalized Matone relations \eqref{eq:htoam} to write $\bs{a}=\bs{a}(\bs{h})$.
After imposing this condition, one has
\begin{multline}
\label{eq:asymevenplus}
    \psi_N^{\rm even}(x,{\bs h})\Big|_{\text{\eqref{eq:QCeven}}} \simeq
    \re^{-\frac{\pi x}{\hbar}} \left(\frac{2 \pi \hbar}{{\rm Re} (x)}\right)^{\frac{N}{2}}
    \left\{ \br{\frac{{\rm Re}(x)}{\Lambda}}^{N \frac{{\rm Im}(x)}{\hbar}} u_N(x)
    \sum_{\bs{n}\in \mathcal{W}_{N}\cdot \bs{\gamma}} \CQ_{\bn}^1(\bsa, \Lambda, \hbar)
    \right.
    \\
    \left.+ \, \ri \br{\frac{{\rm Re}(x)}{\Lambda}}^{-N \frac{{\rm Im}(x)}{\hbar}}u_N^{-1}(x)
    \sum_{\bs{n}\in \mathcal{W}_{N}\cdot \left(\bs{\gamma}+\bs{e}_N\right)}
    \CQ_{\bn}^{\frac{N}{2}-1}(\bsf(\bsa), \Lambda, \hbar)
    \right\}
    \, ,
    \qquad
    \Re[x] \to + \infty
    \, ,
\end{multline}
so the on-shell eigenfunctions are decaying exponentially also for ${\rm Re} (x) \to +\infty$.

In summary, the functions \eqref{eq:eigenfeven} are entire solutions of the difference equation \eqref{eq:diffm}, and they are square-integrable and belong to the domain $\CD$ in \eqref{eq:domain} if and only if the quantization condition \eqref{eq:QCeven} is satisfied. At these quantized values of $E$, they thus define genuine eigenfunctions of $\rH_N$ in \eqref{eq:hn}.

\paragraph{N odd: unbounded potential.}

For $N$ odd, the potential is unbounded from below as $x \to - \infty$, and we find the following off-shell solution of \eqref{eq:diffm}:
\begin{align}\label{eq:eigenfodd}
\boxed{%
\begin{aligned}
\psi_N^{\rm odd}(x,{\bs h} ) & =
    Z_D(x, {\bs a}, \Lambda, \hbar)
   \sum_{\bs{n}\in \mathcal{W}_{N}\cdot \bs{\gamma}}
   \CP_{\bs n}(x, \bs{a},\Lambda, \hbar) \\
   & \qquad + \re^{\frac{\pi x}{\hbar}}Z_D(-x, {\bsf}({\bs a}), \Lambda, \hbar)
   \sum_{\bs{n}\in \mathcal{W}_{N}\cdot \bs{\gamma}}
   \CP_{\bs n}(-x, {\bsf}(\bs{a}),\Lambda, \hbar)\, ,
\end{aligned}
}
\end{align}
where, again, we recall that ${\bsf}({\bs a})$ is defined in \eqref{eq:fa} and $\bs a$ is related to $\bs h$ via the generalized Matone relations; see \eqref{eq:htoam}.\footnote{Note that $\CW_N \cdot \bs{\gamma} = \CW_N \cdot \bs{\lambda}_{(N+1)/2}$ for $N$ odd, with $\bs{\lambda}_{(N+1)/2} \in \Lambda_w$ a fundamental weight.} Once more, we will refer to the first and second terms in \eqref{eq:eigenfodd} as saddles.
Let us now analyse the asymptotic behaviour of this solution. For $N$ odd, the Weyl orbit of $\bs \gamma$ contains $\binom{N}{(N-1)/2}$ elements. In this case, the number of non-trivial factors appearing in the product in equation \eqref{eq:Hndef} is $(N-1)/2$. Hence, for $\real{x}\to\pm\infty$ with $\Im[x]$ held fixed, we find:
\begin{multline}
\label{eq:Nodd_asymp+}
    \psi_N^{\rm odd}(x,{\bs h} ) \simeq
    \\
    \re^{\frac{\pi x}{\hbar}} \left(\frac{2 \pi \hbar}{{\rm Re}(x)}\right)^{\frac{N}{2}} \br{\frac{{\rm Re}(x)}{\Lambda}}^{N \frac{{\rm Im}(x)}{\hbar}} u_N(x)
    \sum_{\bs{n}\in \mathcal{W}_{N} \cdot \bs{\gamma}}
    \CQ_{\bn}^0(\bsa, \Lambda, \hbar)
    \, ,
    \qquad
    \Re[x] \to + \infty
    \, ,
\end{multline}
\begin{multline}
\label{eq:Nodd_asymp-}
    \psi_N^{\rm odd}(x,{\bs h}) \simeq
    \\
    \left(\frac{2 \pi \hbar}{|{\rm Re}(x)|}\right)^{\frac{N}{2}}\br{\frac{|{\rm Re}(x)|}{\Lambda}}^{-N \frac{{\rm Im}(x)}{\hbar}} u_N(x)
    \sum_{\bs{n}\in \mathcal{W}_{N}\cdot \bs{\gamma}}
    \CQ_{\bn}^0(\bsf(\bsa), \Lambda, \hbar)
    \, ,
    \qquad
    \Re[x] \to - \infty
    \, .
\end{multline}
Let us emphasize that, although at ${\rm Re} (x) \to -\infty$ there is only power-law decay, the solution remains square-integrable in this direction as long as
\begin{equation}\label{eq:upperplane}
    \Im[x] > -\frac{\hbar}{2}\br{1-\frac{1}{N}}.
\end{equation}
In contrast, the exponential growth at $\real{x} \to + \infty$ must be eliminated to achieve square-integrability, which requires imposing
\begin{equation}
\label{eq:QCodd}
    \sum_{\bs{n}\in \mathcal{W}_{N}\cdot \bs{\gamma}}
    \CQ_{\bn}^0(\bsa, \Lambda, \hbar)
    =
    \sum_{\bs{n}\in \mathcal{W}_{N}\cdot \bs{\gamma}}
    \frac{\exp[\pr*{\frac{\ri}{\hbar}\partial_{\bs a} F_{\rm NS} - \frac{\pi \bsa}{\hbar}} \cdot \bs{n}]}
    {\prod_{\bs{\alpha}\in \Delta_{+}} \br{2\sinh\br{\frac{\pi \bs{\alpha} \cdot \bs{a}}{\hbar}}}^{({\bs \alpha} \cdot \bs{n})^2}}
    = 0 \, .
\end{equation}
Once again, this is precisely the quantization condition found in \cite{Grassi:2018bci}. After imposing this condition, the solution decays exponentially as $\Re[x] \to + \infty$,
\begin{multline}
    \psi_N^{\rm odd}(x,{\bs h} ) \Big|_{\text{\eqref{eq:QCodd}}} \simeq \re^{-\frac{\pi x}{\hbar}} \br{\frac{2 \pi \hbar}{{\rm Re} (x)}}^{\frac{N}{2}} \left\{\br{\frac{{\rm Re}(x)}{\Lambda}}^{N \frac{{\rm Im}(x)}{\hbar}} u_N(x)
    \sum_{\bs{n}\in \mathcal{W}_{N} \cdot \bs{\gamma}}
    \CQ_{\bn}^1(\bsa, \Lambda, \hbar) \right.
    \\
    \left.+\br{\frac{{\rm Re}(x)}{\Lambda}}^{-N \frac{{\rm Im}(x)}{\hbar}}u^{-1}_N(x) 
    \sum_{\bs{n}\in \mathcal{W}_{N} \cdot \bs{\gamma}}
    \CQ_{\bn}^{\frac{N-1}{2}}(\bsf(\bsa), \Lambda, \hbar) \right\}\,,
    \qquad
    \Re[x] \to + \infty
    \, .
\end{multline}
and becomes a square-integrable solution to the difference equation \eqref{eq:diffm}. As we will discuss below, the same square-integrable solutions also appear to be captured by the complex dilation method.

However, the quantization condition \eqref{eq:QCodd} is not the only one that gives rise to square-integrable solutions of the difference equation \eqref{eq:diffm}. Indeed, introducing
\begin{equation} \label{eq:eigenfodd2}
    \re^{-\frac{2\pi x}{\hbar}} \psi_N^\text{odd}(x, \bh) \, ,
\end{equation}
one obtains a new family of square-integrable solutions upon imposing the alternative quantization condition
\begin{equation}
\label{eq:QCodd_alt}
    \sum_{\bs{n}\in \mathcal{W}_{N}\cdot \bs{\gamma}}
    \CQ_{\bn}^0(\bsf(\bsa), \Lambda, \hbar)
    =
    \sum_{\bs{n}\in \mathcal{W}_{N}\cdot \bs{\gamma}}
    \frac{\exp[\pr*{\frac{\ri}{\hbar}\partial_{\bsf(\bsa)} F_{\rm NS} - \frac{\pi \bsf(\bsa)}{\hbar}} \cdot \bs{n}]}
    {\prod_{\bs{\alpha}\in \Delta_{+}} \br{2\sinh\br{\frac{\pi \bs{\alpha} \cdot \bsf(\bsa)}{\hbar}}}^{({\bs \alpha} \cdot \bs{n})^2}}
    = 0
    \, .
\end{equation}
This second family of solutions once again decays exponentially as $\real{x} \to + \infty$, as is clear from \eqref{eq:Nodd_asymp+}. For $\real{x}\to-\infty$, on the other hand, the solutions exhibit power-law decay,
\begin{multline}
    \re^{-\frac{2\pi x}{\hbar}} \psi_N^{\rm odd}(x,{\bs h} ) \Big|_{\text{\eqref{eq:QCodd_alt}}} \simeq \br{\frac{2 \pi \hbar}{   {\abs{\Re[x]}}}}^{\frac{N}{2}} \left\{\br{\frac{   {\abs{\Re[x]}}}{\Lambda}}^{N \frac{{\rm Im}(x)}{\hbar}} u_N^{-1}(x)
    \sum_{\bs{n}\in \mathcal{W}_{N} \cdot \bs{\gamma}}
    \CQ_{\bn}^{\frac{N-1}{2}}(\bsa, \Lambda, \hbar) \right.
    \\
    \left.+\br{\frac{   {\abs{\Re[x]}}}{\Lambda}}^{-N \frac{{\rm Im}(x)}{\hbar}}u_N(x) 
    \sum_{\bs{n}\in \mathcal{W}_{N} \cdot \bs{\gamma}}
    \CQ_{\bn}^{1}(\bsf(\bsa), \Lambda, \hbar) \right\}
    \qquad
    \Re[x] \to - \infty
    \, ,
\end{multline}
and are therefore square-integrable along lines parallel to the real axis provided that
\begin{equation}
\label{eq:smallstrip}
    \abs{\imaginary{x}} < \frac{\hbar}{2}\br{1-\frac{1}{N}} \, .
\end{equation}

In summary, \eqref{eq:eigenfodd} are entire solutions of the difference equation \eqref{eq:diffm}, and are square-integrable in the shifted upper half plane \eqref{eq:upperplane} if and only if \eqref{eq:QCodd} holds. On the other hand, the \eqref{eq:eigenfodd2} are square-integrable in the strip \eqref{eq:smallstrip} if and only if \eqref{eq:QCodd_alt} holds.
However, neither of them is in the domain $\CD$ in \eqref{eq:domain}, and the eigenvalues solving the quantization conditions \eqref{eq:QCodd} and \eqref{eq:QCodd_alt} are generically complex.\footnote{As discussed in \autoref{sec:toda}, the eigenvalues are real if we take the parameters of the potential to be at the Toda points $\{h_k^T\}$. At these special points, the eigenfunctions are expected to belong to $\CD$, suggesting that the corresponding difference operator has qualitatively different spectral properties and may even be essentially self-adjoint.}
Thus, these square-integrable solutions do not belong to the domain of any symmetric or self-adjoint realization of the difference operator.\footnote{\label{footnote:adjointext}Since they are square-integrable solutions of the difference equation, one may speculate that they are instead eigenfunctions of the adjoint $\rH_N^*$. This is indeed what happens for the usual Schr\"odinger equation, but at present it is not clear whether the same holds true for difference equations. We thank Lukas Schimmer for an illuminating discussion on this point.} Moreover, the solutions to the quantization condition \eqref{eq:QCodd_alt} do not appear to be captured by the complex dilation method.

\paragraph{Discussion.}

We conclude this section with a few remarks on the expressions \eqref{eq:eigenfeven} and \eqref{eq:eigenfodd}.
\begin{enumerate}
    \item Whereas the Schr\"{o}dinger equation \eqref{eq:anh} admits only entire solutions, this is not the case for the difference equations associated with the deformed operator \eqref{eq:hamil}. For instance, each of the two terms in \eqref{eq:eigenfeven} and \eqref{eq:eigenfodd} is, by itself, a solution of \eqref{eq:diffm}, but each has poles at
    \be
        x = a_I + \ri \hbar n, \quad n \in \IZ.
    \ee
    Only for the special linear combinations \eqref{eq:eigenfeven} and \eqref{eq:eigenfodd} do these poles cancel, yielding a function that is entire in $x$ for any value of the parameters $h_k$, $k = 2, \cdots, N$.\footnote{Although we do not have a rigorous proof of this statement, we have checked the pole cancellation for $N=2,3,4,5,6$ up to three orders in the $\Lambda^{2 N}$ expansion.}

    \item Eigenfunctions are defined up to an overall normalization. In our conventions, we choose the normalization so that the asymptotic behaviour \eqref{eq:asen} and \eqref{eq:Nodd_asymp+} reproduces the Fredholm determinant of the corresponding spectral problem as computed in \cite{Grassi:2018bci,ggm}. This choice is natural from the perspective of the TS/ST correspondence, where eigenfunctions are required to be entire both in the open and closed string moduli. However, as we discuss in \autoref{sec:toda}, this normalization is not convenient for studying the Toda points.

    \item A noteworthy property of the deformed Hamiltonian \eqref{eq:diffm} is that its bound states violate the standard oscillation theorem \cite[p.~66, thm.~3.5]{BS91}. For the deformed Schr\"odinger equation \eqref{eq:diffm} it is no longer true that the $n^{\text{th}}$ eigenfunction has $n$ zeroes, even if the potential is confining. This behaviour occurs even for the simplest potential $V_2(x)= x^2$, see \cite{Francois:2025wwd}.\footnote{A similar phenomenon has been observed for difference equations arising in the context of the open TS/ST correspondence \cite{Marino:2016rsq,Marino:2017gyg,Francois:2025wwd}.} This violation is evident from the oscillatory asymptotics \eqref{eq:asymevenminus} and \eqref{eq:asymevenplus}, see also  \autoref{fig:su4_almostsymm} for a graphical representation.

    \item For $N$ odd, the asymptotic expression \eqref{eq:Nodd_asymp-} exhibits an outgoing oscillatory behaviour for ${\rm Re} (x) \to -\infty$, characteristic of resonant states. See also \autoref{fig:su3_3rd}.

     \item As noted in \cite{Grassi:2018bci}, there are special values of the moduli $h_k$ -- those corresponding to eigenvalues of the quantum Toda Hamiltonians -- at which the spectrum and eigenfunctions exhibit special features. For $N$ even, certain eigenstates, including the ground state, become degenerate, i.e.~tunnelling is suppressed at these special points.
     A similar phenomenon occurs for $N$ odd. For generic values of $h_k$, the energies associated with square-integrable solutions are complex. Nevertheless, at these special values of $h_k$, the eigenfunctions exhibit enhanced exponential decay as $\Re[x] \to -\infty$, and the corresponding energies are real, despite the potential being unbounded from below.\footnote{Recently, these phenomena were also studied from the point of view of WKB in \cite{Gu:2026xgp}.} We will return to these special points in \autoref{sec:toda}.

    \item We recover \cite[eq.~(4.16)]{Francois:2025wwd} by evaluating \eqref{eq:eigenfeven} for $N=2$.

\end{enumerate}

\subsection{Examples}
\label{sec:test}

\subsubsection{Cubic potentials}

\begin{figure}
    \centering
    \includegraphics[width=1\textwidth]{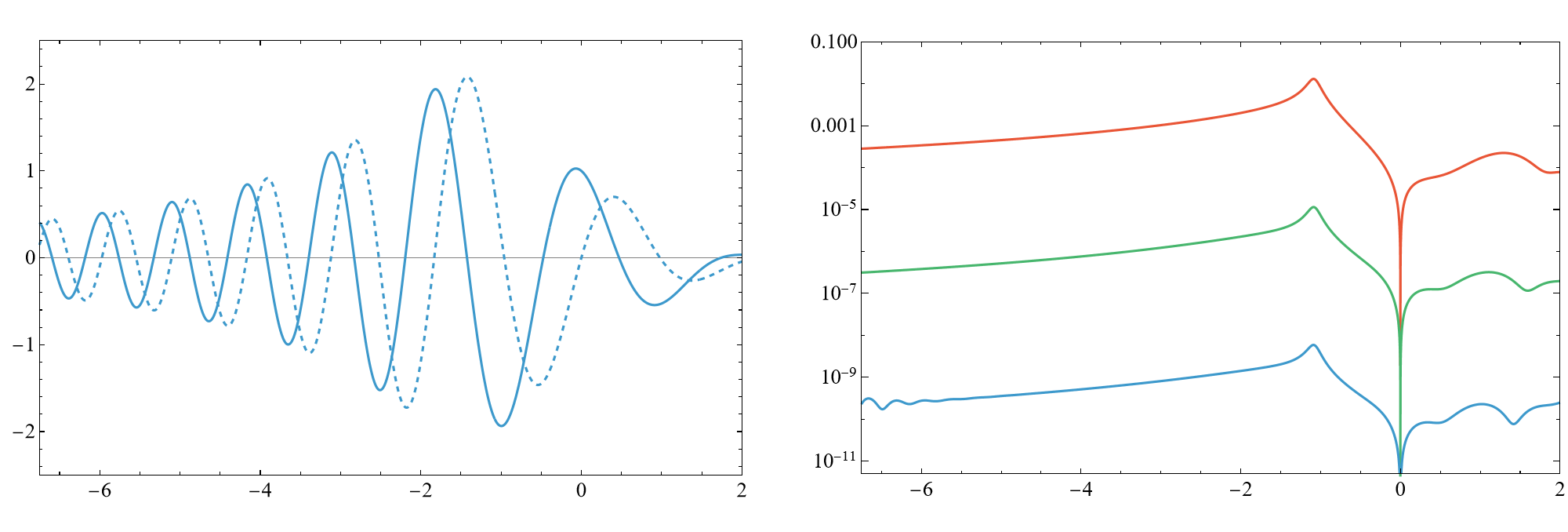}
    \caption{Left: third excited state of the $V_3(x)$ potential with $h_2=0$, $\hbar=1$, and $\Lambda=\tfrac{1}{2}$. Dashed lines denote the imaginary part of the eigenfunction, while solid lines denote the real part. Right: difference between the numerical eigenfunction and the analytic expression from \eqref{eq:eigenfodd}. The coloured curves show the effect of including an increasing number of terms in the $\Lambda$-expansion of the eigenfunction: red (0 terms), green (1 term), blue (2 terms).}
    \label{fig:su3_3rd}
\end{figure}

Let us consider the $N=3$ case:
\begin{equation}
     \rH_3 = 2 \, \Lambda^3 \cosh\!{\br{\p}} + \x^3 + h_2 \, \x
     \, .
\end{equation}
The potential is unbounded from below, and for generic values of $h_k$ the spectrum contains resonant states with complex energies. Such resonances always occur in complex-conjugate pairs, and throughout this paper we focus on those having a positive imaginary part of the energy. For $N=3$, the Weyl orbit of $\bs \gamma$ is given by:
\be \mathcal{W}_N\cdot {\bs \gamma}=\left\{\tfrac{1}{2} \big({\bs e}_1-{\bs e}_2+{\bs e}_3\big),\;
\tfrac{1}{2} \big({\bs e}_1+{\bs e}_2-{\bs e}_3\big),\;
\tfrac{1}{2} \big(-{\bs e}_1+{\bs e}_2+{\bs e}_3\big)\right\}\, .\ee
Examples of on-shell eigenfunctions are shown in \autoref{fig:su3_3rd}, \autoref{fig:su3_ground}, \autoref{fig:su3_1st} and \autoref{fig:su3_2nd}. In \autoref{fig:su3_3rd}, we consider the case $h_2 = 0$ and focus on the third excited state, with complex energy:
\be
E_3 \approx 2.511486 + \ri\, 2.512027 \, .
\ee
Through the generalized Matone relations, this corresponds to:
\be
a_1 \approx 1.473672 + \ri\, 0.395307, \quad a_2 \approx -0.394390 - \ri \, 1.473919. %\quad a_3 \approx -1.079282 + 1.078612 \ri.
\ee
The power-law decay of the on-shell wavefunctions as $x\to -\infty$ is particularly evident in this figure. Similarly, \autoref{fig:su3_ground}, \autoref{fig:su3_1st} and \autoref{fig:su3_2nd} illustrate the ground, first and second excited state for \( h_2 = -5 \), \( h_2 = -3 \) and \( h_2 = 2 \) respectively.

\begin{figure}
    \centering
    \includegraphics[width=1\textwidth]{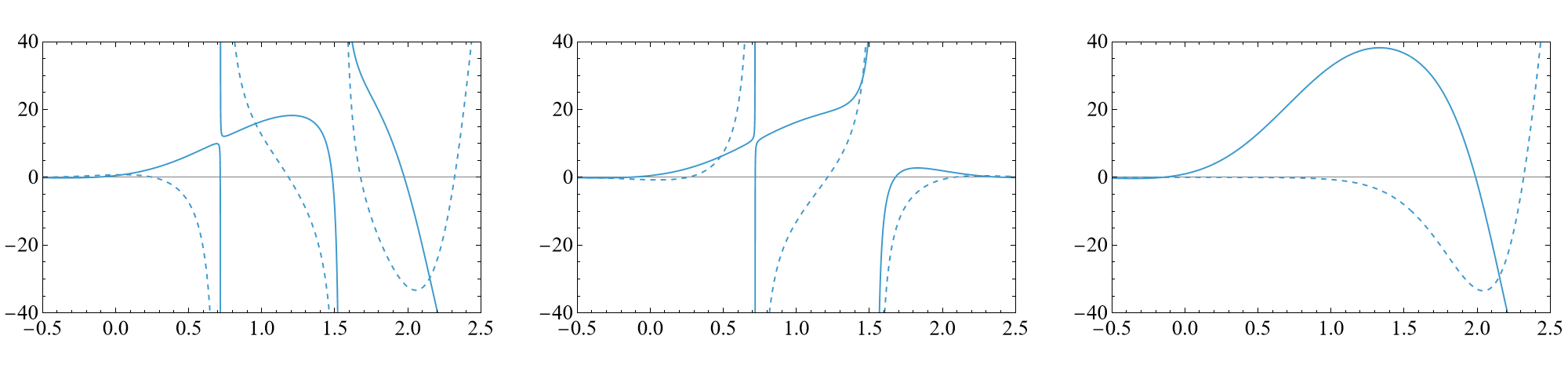}
    \caption{Off-shell eigenfunction of the potential $V_3(x)$ with
     $h_2=-4$, $h_3= -2.5$, $\hbar=1$ and $\Lambda=2/5$. From left to right: first saddle, second saddle and the full eigenfunction given by \eqref{eq:eigenfodd}. Dashed lines denote the imaginary part, while solid lines denote the real part.}
    \label{fig:su3_offshell}
\end{figure}

Each figure shows excellent agreement between the analytic expression \eqref{eq:eigenfodd} and the numerical results obtained via complex dilation. In addition, \autoref{fig:su3_1st} clearly shows the cancellation of poles between the two terms in \eqref{eq:eigenfodd}. Finally, an example of off-shell eigenfunction is shown in \autoref{fig:su3_offshell}. The figure also clearly shows that each individual saddle in \eqref{eq:eigenfodd} develops poles, whereas the complete sum is free of singularities. Since we are off-shell, the resulting eigenfunction does not belong to $L^2(\mathbb{R})$.


\subsubsection{Quartic potentials}

\begin{figure}
    \centering
    \includegraphics[width=1\textwidth]{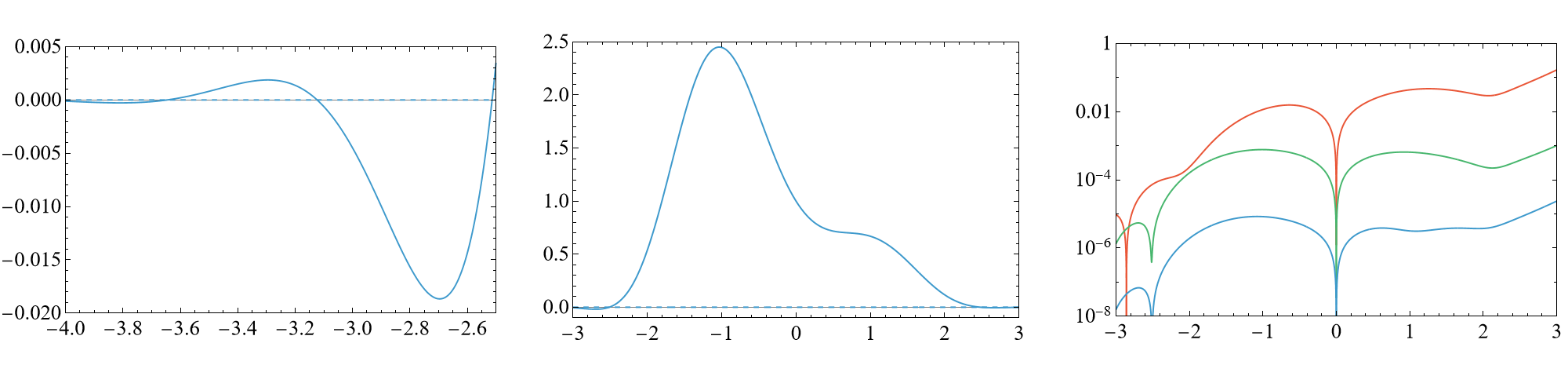}
    \caption{The ground state of the $V_4(x)$ potential with $h_2=-3$, $h_3=-0.1$, $\Lambda=(1/2)^{1/2}$ and $\hbar=1$; the corresponding energy is $E_0 \approx -0.527521$. Dashed lines represent the imaginary part of the eigenfunction, while solid lines represent the real part. Left: illustration of the failure of the oscillation theorem. Centre: eigenfunction obtained from \eqref{eq:eigenfeven}. Right: difference between the numerical eigenfunction and the analytic expression from \eqref{eq:eigenfeven}. The coloured curves show the effect of including an increasing number of terms in the $\Lambda$-expansion of the eigenfunction: red (0 terms), green (1 term), blue (2 terms).}
    \label{fig:su4_almostsymm}
\end{figure}
Let us consider the $N=4$ Hamiltonian:
\begin{equation}
\label{eq:exh4}
    \rH_4 = 2 \, \Lambda^4 \cosh\!{\br{\p}} + \x^4 + h_2 \, \x^2 - h_3 \, \x
    \, .
\end{equation}
In this case, the potential is confining and the operator has a purely real and discrete spectrum, corresponding to bound states. The Weyl orbits of $\bs \gamma$  and $\bs \gamma+{\bs e}_N$ are, respectively:
\be\ba \mathcal{W}_N\cdot {\bs\gamma}=&\left\{
\frac{1}{2} (\boldsymbol{e}_1 - \boldsymbol{e}_2 + \boldsymbol{e}_3 - \boldsymbol{e}_4),\,
\frac{1}{2} (\boldsymbol{e}_1 - \boldsymbol{e}_2 - \boldsymbol{e}_3 + \boldsymbol{e}_4),\right.\\
&\left.\frac{1}{2} (\boldsymbol{e}_1 + \boldsymbol{e}_2 - \boldsymbol{e}_3 - \boldsymbol{e}_4),\,\frac{1}{2} (-\boldsymbol{e}_1 + \boldsymbol{e}_2 + \boldsymbol{e}_3 - \boldsymbol{e}_4),\right.\\
&\left.\frac{1}{2} (-\boldsymbol{e}_1 + \boldsymbol{e}_2 - \boldsymbol{e}_3 + \boldsymbol{e}_4),\,
\frac{1}{2} (-\boldsymbol{e}_1 - \boldsymbol{e}_2 + \boldsymbol{e}_3 + \boldsymbol{e}_4)
\right\},
\ea\ee
\be\ba \mathcal{W}_N\cdot \left({\bs\gamma}+{\bs e}_N \right)=
\left\{
\frac{1}{2} (\boldsymbol{e}_1 - \boldsymbol{e}_2 + \boldsymbol{e}_3 + \boldsymbol{e}_4),\,
\frac{1}{2} (\boldsymbol{e}_1 + \boldsymbol{e}_2 - \boldsymbol{e}_3 + \boldsymbol{e}_4),\,\right.\\
\left.\frac{1}{2} (\boldsymbol{e}_1 + \boldsymbol{e}_2 + \boldsymbol{e}_3 - \boldsymbol{e}_4),\,
\frac{1}{2} (-\boldsymbol{e}_1 + \boldsymbol{e}_2 + \boldsymbol{e}_3 + \boldsymbol{e}_4)
\right\}.
\ea\ee
Examples of on-shell solutions are shown in \autoref{fig:su4_almostsymm}, \autoref{fig:su4_1st} and \autoref{fig:su4_2nd}. In \autoref{fig:su4_almostsymm}, we exhibit the ground state eigenfunction for $h_2 = -3$ and $h_3=-0.1$, corresponding to the eigenvalue:
\begin{equation}
    E_0\approx -0.527521.
\end{equation}
\begin{figure}
    \centering
    \includegraphics[width=1\textwidth]{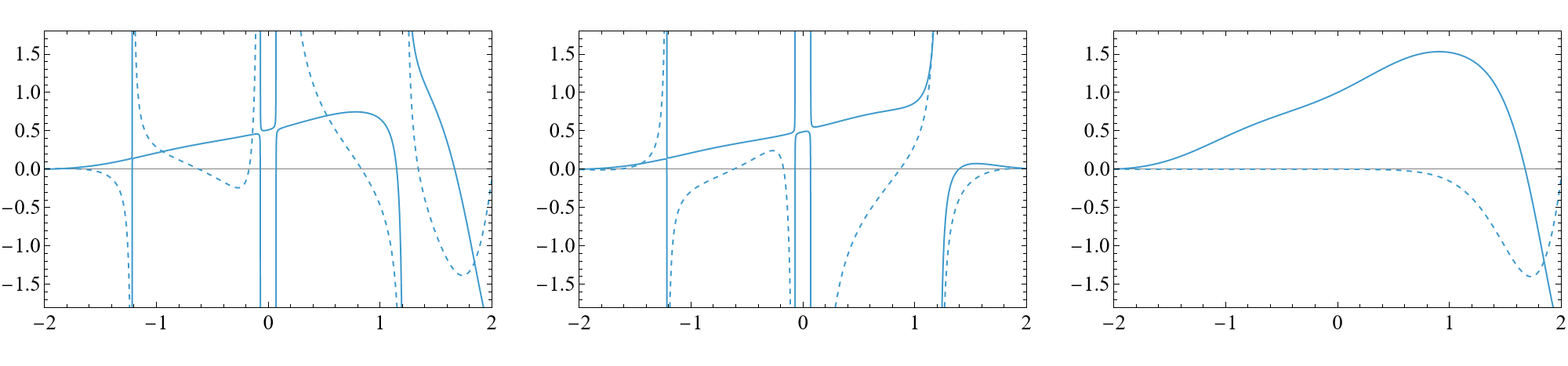}
    \caption{Off-shell eigenfunction of the potential $V_4(x)$ with $h_2 = -1.6$, $h_3 = 0.45$, $h_4 = 0.06$, $\hbar = 1$ and $\Lambda = 13/23$. From left to right: first saddle, second saddle and the full eigenfunction given by \eqref{eq:eigenfeven}. Dashed lines represent the imaginary part, while solid lines the real part.}
    \label{fig:su4_offshell}
\end{figure}
Through the generalized Matone relations, this corresponds to:
\begin{equation}
    a_1 \approx 0.446118, \qquad a_2 \approx 1.656437, \qquad a_3 \approx -1.694744.
\end{equation}
In this figure, the quartic potential is a slight deformation of the symmetric double well (the symmetric case has $h_2<0$ and $h_3=0$), and the corresponding ground state clearly reflects this small asymmetry. Similarly, \autoref{fig:su4_1st} and \autoref{fig:su4_2nd} show the first and second excited state for \( h_2 = 3.5,\, h_3=0 \) and \( h_2 = 1.2,\, h_3=-4.7 \), respectively. Again, each figure shows excellent agreement between the analytic expression \eqref{eq:eigenfeven} and the results obtained by numerically diagonalizing the Hamiltonian \eqref{eq:exh4}. The cancellation of poles is clearly visible for the on-shell eigenfunction in \autoref{fig:su4_1st} and in the off-shell eigenfunction in \autoref{fig:su4_offshell}. In the latter case, the resulting eigenfunction does not belong to $L^2(\mathbb{R})$, being off-shell.

%%%%%%%%%%%%%%%%%%%%%%%%%%%%%%%%%%%%%%%%%%%%%%%%%%%%%%%%%%%%%%%%%%%%%%%%%%%%%%%%%%%%
%%%%%%%%%%%%%%%%%%%%%%%%%%%%%%%%%%%%%%%%%%%%%%%%%%%%%%%%%%%%%%%%%%%%%%%%%%%%%%%%%%%%

\section{Related spectral problems}
\label{sec:relatedproblems}

One can obtain solutions to some variants of \eqref{eq:diffm} from \eqref{eq:eigenfeven} and \eqref{eq:eigenfodd} by multiplying them by appropriate exponential factors. In this section, we will discuss the solutions to the case where the variant of $\rH_N$ in \eqref{eq:hn} has an inverted potential $-V_N(x)$ and/or a $\sinh(p)$ kinetic term. We will not go into the details of these related spectral problems, but simply discuss the solutions we obtain and their properties.

\subsection{Inverted potential}
\label{sec:inverted}

Given a solution $\psi$ of \eqref{eq:diffm}, for either even or odd $N$, one also has a solution of the corresponding equation with an inverted potential,
\begin{equation}
\label{eq:diffeq:invert_pot}
    \Lambda^N \pr*{\tilpsi_\pm( x + \ri \hbar, \bh) + \tilpsi_\pm( x - \ri \hbar, \bh)} - V_N(x) \tilpsi_\pm(x, \bh)
    = \tilE \tilpsi_\pm(x, \bh)
    \, ,
\end{equation}
\begin{equation}
    \tilpsi_\pm(x) := \exp[\pm \frac{\pi x}{\hbar}] \psi(x) \, , \qquad \qquad \tilE := - E \, ,
\end{equation}
where the associated symmetric operator again has domain $\CD$, defined in \eqref{eq:domain}.\footnote{See \autoref{footnote:extensions} for a comment on the existence of self-adjoint extensions, which also holds for the case of the inverted potential for both even and odd $N$.}
Here, the subscript $\pm$ labels two distinct solutions corresponding to the same eigenvalue $\tilE$. The analytic properties of $\tilpsi_\pm$ are inherited directly from those of $\psi$, as is evident from their definition.

\paragraph{Inverted potential for N even.}

The two choices of sign for $\tilpsi_{\pm,N}^{\text{even}}$ lead to qualitatively different classes of solutions. For $\tilpsi_{+, N}^{\text{even}}$ one finds
\begin{multline}
    \tilpsi_{+, N}^{\text{even}}(x, \bh) \simeq
    \\
    \re^{\frac{2 \pi x}{\hbar}}\left(\frac{2 \pi \hbar}{\abs{{\rm Re}(x)}}\right)^{\frac{N}{2}} \br{\frac{\abs{{\rm Re}(x)}}{\Lambda}}^{N \frac{{\rm Im}(x)}{\hbar}} u_N(x) \,
    \sum_{\bs{n}\in \mathcal{W}_{N}\cdot \bs{\gamma}}
    \CQ_{\bn}^0(\bsa, \Lambda, \hbar)
    ,
    \qquad
    \Re[x] \to + \infty
    ,
\end{multline}
\begin{multline}
    \tilpsi_{+, N}^{\text{even}}(x, \bh) \simeq
    \re^{\frac{2 \pi x}{\hbar}} \left(\frac{2 \pi \hbar}{|{\rm Re }(x)|}\right)^{\frac{N}{2}}
    \left\{\br{\frac{|{\rm Re}(x)|}{\Lambda}}^{N \frac{{\rm Im}(x)}{\hbar}}u_N^{-1}(x)
    \sum_{\bs{n}\in \mathcal{W}_{N}\cdot \bs{\gamma}}
    \CQ_{\bn}^{\frac{N}{2}}(\bsa, \Lambda, \hbar) \right.
    \\
    \left. + \, \ri \br{\frac{|{\rm Re}(x)|}{\Lambda}}^{-N \frac{{\rm Im}(x)}{\hbar}} u_N(x)
    \sum_{\bs{n}\in \mathcal{W}_{N}\cdot \left(\bs{\gamma}+\bs{e}_N\right)}
    \CQ_{\bn}^{0}(\bsf(\bsa), \Lambda, \hbar) \right\}
    \, ,
    \qquad
    \Re[x] \to - \infty
    \, ,
\end{multline}
which follows from \eqref{eq:asen} and \eqref{eq:asymevenminus}.
Thus, $\tilpsi_{+, N}^{\text{even}}$ is square-integrable parallel to the real line if and only if the quantization condition \eqref{eq:QCeven} is satisfied and
\begin{equation}
\label{eq:toosmall}
    \abs{\imaginary{x}} < \frac{\hbar}{2}\br{1-\frac{1}{N}} \, .
\end{equation}
It then decays exponentially for $\real{x} \to - \infty$, while it exhibits a power-law decay for $\real{x} \to + \infty$, which can be read off from \eqref{eq:asymevenplus}. We therefore obtain square-integrable solutions of the difference equation \eqref{eq:diffeq:invert_pot} for a discrete set of \emph{real} eigenvalues bounded from \emph{above}. These eigenvalues are precisely the negatives of those of the corresponding operator with a non-inverted potential.

On the other hand, it is evident from \eqref{eq:asen} and \eqref{eq:asymevenminus} that $\tilpsi_{-, N}^{\text{even}}$ has a power-law decay for both $\real{x} \to \pm \infty$, and is square-integrable parallel to the real line inside the strip \eqref{eq:toosmall}.
This holds for \emph{any value} of
\begin{equation}\label{eq:continuous}
    E \in \comps~.
\end{equation} We therefore obtain a one-parameter family of square-integrable solutions of the difference equation. Note that this also means that we have two linearly independent, square-integrable solutions when the quantization condition \eqref{eq:QCeven} holds, which are simply related by the factor $\exp[2 \pi x/\hbar]$.

It is important to note that even when $\tilpsi_{\pm, N}^{\text{even}}$ are entire and square-integrable solutions to the difference equation \eqref{eq:diffeq:invert_pot}, neither of them is in the domain $\CD$ of the original quantized curve as defined around \eqref{eq:domain}.\footnote{\label{footnote:GenericVsSpecificHk}
That is for generic choices of the coefficients $h_k$. For special values of $h_k$ the decay of the solutions may be faster. This is what happens at the Toda points $h_k^\rT$, for example.} 

\paragraph{Inverted potential for N odd.}

For $\tilpsi_{+, N}^{\text{odd}}$ the asymptotics is simply
\begin{multline}
     \tilpsi_{+, N}^{\text{odd}}(x, \bh) \simeq
     \re^{\frac{2 \pi x}{\hbar}} \left(\frac{2 \pi \hbar}{{\rm Re}(x)}\right)^{\frac{N}{2}} \br{\frac{{\rm Re}(x)}{\Lambda}}^{N \frac{{\rm Im}(x)}{\hbar}}
     \times
     \\
     u_N(x)
    \sum_{\bs{n}\in \mathcal{W}_{N} \cdot \bs{\gamma}}
    \CQ_{\bn}^0(\bsa, \Lambda, \hbar)
    \, ,
    \qquad
    \Re[x] \to + \infty
    \, ,
\end{multline}
\begin{multline}
    \tilpsi_{+, N}^{\text{odd}}(x, \bh) \simeq
    \re^{\frac{\pi x}{\hbar}}
    \left(\frac{2 \pi \hbar}{|{\rm Re}(x)|}\right)^{\frac{N}{2}}\br{\frac{|{\rm Re}(x)|}{\Lambda}}^{-N \frac{{\rm Im}(x)}{\hbar}}
    \times
    \\
    u_N(x)
    \sum_{\bs{n}\in \mathcal{W}_{N}\cdot \bs{\gamma}}
    \CQ_{\bn}^0(\bsf(\bsa), \Lambda, \hbar)
    \, ,
    \qquad
    \Re[x] \to - \infty
    \, ,
\end{multline}
following from \eqref{eq:Nodd_asymp+} and \eqref{eq:Nodd_asymp-}. To find square-integrable solutions we need once more to impose the quantization condition \eqref{eq:QCodd} so that the asymptotics becomes
\begin{multline}
    \tilpsi_{+, N}^{\text{odd}}(x, \bh) \Big|_{\text{\eqref{eq:QCodd}}} \simeq
    \br{\frac{2 \pi \hbar}{{\rm Re} (x)}}^{\frac{N}{2}} \left\{\br{\frac{{\rm Re}(x)}{\Lambda}}^{N \frac{{\rm Im}(x)}{\hbar}} u_N(x)
    \sum_{\bs{n}\in \mathcal{W}_{N} \cdot \bs{\gamma}}
    \CQ_{\bn}^1(\bsa, \Lambda, \hbar) \right.
    \\
    \left.+\br{\frac{{\rm Re}(x)}{\Lambda}}^{-N \frac{{\rm Im}(x)}{\hbar}}u^{-1}_N(x) 
    \sum_{\bs{n}\in \mathcal{W}_{N} \cdot \bs{\gamma}}
    \CQ_{\bn}^{\frac{N-1}{2}}(\bsf(\bsa), \Lambda, \hbar) \right\}
    \qquad
    \Re[x] \to + \infty
    \, .
\end{multline}
Thus $\tilpsi_{+, N}^{\text{odd}}$ is square-integrable parallel to the real line if and only if the quantization condition \eqref{eq:QCodd} is satisfied and \eqref{eq:toosmall} holds.
Note that this gives square-integrable solutions of the difference equation for a discrete set of complex eigenvalues. These eigenvalues are simply minus the eigenvalues of the non-inverted potential.
On the other hand, it is evident that $\tilpsi_{-, N}^{\text{odd}}$ is square-integrable parallel to the real line if and only if the alternative quantization condition \eqref{eq:QCodd_alt} is satisfied and $\Im[x] < (\hbar/2)(1-1/N)$.

It is important to note that even when $\tilpsi_{\pm, N}^{\text{odd}}$ are entire and square-integrable solutions to the difference equation \eqref{eq:diffeq:invert_pot}, neither of them is generically in the domain $\CD$ in \eqref{eq:domain}, or even in the domain of any symmetric or self-adjoint realization of the difference operator.

\subsection{Kinetic term with sinh(p)}
\label{sec:sinh}

Similarly, one can construct solutions of the difference equation obtained by replacing the kinetic term $\cosh(\mathrm{p})$ with $\sinh(\mathrm{p})$,
\begin{equation}
\label{eq:diffeq:sinh_kin}
    \ri \Lambda^N \pr*{\omega_\pm( x + \ri \hbar, \bh) - \omega_\pm( x - \ri \hbar, \bh)} \pm V_N(x) \omega_\pm(x, \bh)
    = \CE_\pm \omega_\pm(x, \bh)
    \, ,
\end{equation}
\begin{equation}
    \omega_\pm(x) := \exp[\mp \frac{\pi x}{2 \hbar}] \psi(x) \, , \qquad \qquad \CE_\pm := \pm E \, ,
\end{equation}
with the corresponding operator having again the domain $\CD$ in \eqref{eq:domain}.
In this case, the subscript $\pm$ labels solutions of two different equations, with potential $\pm V_N(x)$, respectively. If one requires the operators to be symmetric, the appropriate condition in this case is $\ri\Lambda^N>0$, rather than $\Lambda^N>0$. As before, the analytic properties of $\omega_\pm$ are inherited directly from those of $\psi$.
However, the spectral problems related to \eqref{eq:diffm} and \eqref{eq:diffeq:sinh_kin} are different.

\paragraph{Kinetic term with sinh(p) and confining potential for N even.}
Let us look at the equation \eqref{eq:diffeq:sinh_kin} for the confining potential $+V_N$ and $N$ even.
The asymptotic behaviour for $\Re[x] \to \pm \infty$, with $\Im[x]$ held fixed, is now
\begin{multline}
\label{eq:asymp+_sinh_conf}
    \omega_{+, N}^{\rm even}(x,{\bs h}) \simeq
    \re^{\frac{\pi x}{\hbar}}\left(\frac{2 \pi \hbar}{\abs{\Re[x]}}\right)^{\frac{N}{2}} \br{\frac{\abs{\Re[x]}^N}{\ri \Lambda^N}}^{\frac{{\rm Im}(x)}{\hbar}}
    \times
    \\
    v_N(x)
    \sum_{\bs{n}\in \mathcal{W}_{N}\cdot \bs{\gamma}}
    \CQ_{\bn}^0(\bsa, \Lambda, \hbar)
    \, ,
    \qquad \qquad
    \Re[x] \to + \infty
    \, ,
\end{multline}
\begin{multline}
\label{eq:asymp-_sinh_conf}
    \omega_{+, N}^{\rm even}(x,{\bs h} ) \simeq
    \left(\frac{2 \pi \hbar}{|{\rm Re }(x)|}\right)^{\frac{N}{2}}
    \br{\frac{\abs{\Re[x]}^N}{\ri \Lambda^N}}^{-\frac{{\rm Im}(x)}{\hbar}}
    \times
    \\
    \ri \, v_N(x)
    \sum_{\bs{n}\in \mathcal{W}_{N}\cdot \left(\bs{\gamma}+\bs{e}_N\right)}
    \CQ_{\bn}^{0}(\bsf(\bsa), \Lambda, \hbar)
    \, ,
    \qquad \qquad
    \Re[x] \to - \infty
    \, ,
\end{multline}
where $v_N$ is an $x$-dependent factor of unit modulus, similar to $u_N$ in equation \eqref{eq:cdef},
\begin{equation}
    v_N(x) \deq \re^{\ri \frac{\pi}{4}N}
   \br{\frac{\re^N \ri \Lambda^N}{\abs{{\rm Re} (x)}^N}}^{\ri \frac{\abs{{\rm Re} (x)}}{\hbar}}
    \, .
\end{equation}
Looking for square-integrable solutions leads once again to the quantization condition \eqref{eq:QCeven}, and the asymptotics becomes  
\begin{multline}
    \omega_{+, N}^{\rm even}(x,{\bs h})\Big|_{\text{\eqref{eq:QCeven}}} \simeq
    \re^{-\frac{\pi x}{\hbar}}\left(\frac{2 \pi \hbar}{\abs{\Re[x]}}\right)^{\frac{N}{2}} \br{\frac{\abs{\Re[x]}^N}{\ri \Lambda^N}}^{\frac{{\rm Im}(x)}{\hbar}}
    \times
    \\
    v_N(x)
    \sum_{\bs{n}\in \mathcal{W}_{N}\cdot \bs{\gamma}}
    \CQ_{\bn}^1(\bsa, \Lambda, \hbar)
    \, ,
    \qquad
    \Re[x] \to + \infty
    \, .
\end{multline}
Hence, for a discrete set of eigenvalues satisfying\footnote{However, as said, we now have $\ri \Lambda^N > 0$ rather than $\Lambda^N > 0$, meaning that we will get different $E \in \comps$ satisfying \eqref{eq:QCeven}.} \eqref{eq:QCeven} we have solutions to the difference equation \eqref{eq:diffeq:sinh_kin} which are square-integrable parallel to the real line if
\begin{equation}
   \Im[x] > - \frac{\hbar}{2} \br{1-\frac{1}{N}} \, .
\end{equation}
It is important to note that even when $\omega_{+, N}^{\text{even}}$ is an entire and square-integrable solution to the difference equation \eqref{eq:diffeq:sinh_kin}, it is generically\footnote{That is for generic choices of the parameters $h_k$, see \autoref{footnote:GenericVsSpecificHk}.} not in the original domain $\CD$ of the quantized curve as defined around \eqref{eq:domain}.

\paragraph{Kinetic term with sinh(p) and inverted potential for $N$ even.}
Let us look at the equation \eqref{eq:diffeq:sinh_kin} for the inverted potential potential $-V_N$ and $N$ even.
The asymptotics of $\omega_{-, N}^{\text{even}}$ is simply
\begin{multline}
    \omega_{-, N}^{\rm even}(x,{\bs h}) \simeq
    \re^{\frac{2 \pi x}{\hbar}}
    \left(\frac{2 \pi \hbar}{\abs{\Re[x]}}\right)^{\frac{N}{2}} \br{\frac{\abs{\Re[x]}^N}{\ri \Lambda^N}}^{\frac{{\rm Im}(x)}{\hbar}}
    \times
    \\
    v_N(x)
    \sum_{\bs{n}\in \mathcal{W}_{N}\cdot \bs{\gamma}}
    \CQ_{\bn}^0(\bsa, \Lambda, \hbar)
    \, ,
    \qquad \qquad
    \Re[x] \to + \infty
    \, ,
\end{multline}
\begin{multline}
    \omega_{-, N}^{\rm even}(x,{\bs h} ) \simeq
    \re^{\frac{\pi x}{\hbar}}
    \left(\frac{2 \pi \hbar}{|{\rm Re }(x)|}\right)^{\frac{N}{2}}
    \br{\frac{\abs{\Re[x]}^N}{\ri \Lambda^N}}^{-\frac{{\rm Im}(x)}{\hbar}}
    \times
    \\
    \ri \, v_N(x)
    \sum_{\bs{n}\in \mathcal{W}_{N}\cdot \left(\bs{\gamma}+\bs{e}_N\right)}
    \CQ_{\bn}^{0}(\bsf(\bsa), \Lambda, \hbar)
    \, ,
    \qquad \qquad
    \Re[x] \to - \infty
    \, ,
\end{multline}
following \eqref{eq:asymp+_sinh_conf} and \eqref{eq:asymp-_sinh_conf}. This leads once again to the quantization condition \eqref{eq:QCeven}, and for a discrete set of eigenvalues satisfying \eqref{eq:QCeven} we have solutions to the difference equation \eqref{eq:diffeq:sinh_kin} which are square-integrable parallel to the real line if
\begin{equation}
     \Im[x] < \frac{\hbar}{2} \br{1-\frac{1}{N}} \, .
\end{equation}
However, in this case, we also find a second quantization condition. One should note that $\exp(- 2 \pi x/\hbar) \omega_{-, N}^{\text{even}}(x, \bh)$ is again a solution with now a power law decay near positive real infinity and an exponential growth near negative real infinity, which can be read off from \eqref{eq:asymp+_sinh_conf} and \eqref{eq:asymp-_sinh_conf}. To get square-integrable solutions we need
\begin{equation}
\label{eq:QCeven_alt}
    \sum_{\bs{n}\in \mathcal{W}_{N}\cdot \left(\bs{\gamma}+\bs{e}_N\right)}
    \CQ_{\bn}^{0}(\bsf(\bsa), \Lambda, \hbar)
    = 0
    \, .
\end{equation}
If this quantization condition is satisfied, the asymptotics near negative infinity is
\begin{multline}
     {\exp[- 2 \pi x/\hbar]} \omega_{-,N}^{\rm even}(x,{\bs h} ) \simeq
    \\
    \left(\frac{2 \pi \hbar}{|{\rm Re }(x)|}\right)^{\frac{N}{2}}
    \br{\frac{|{\rm Re}(x)|^N}{\ri \Lambda^N}}^{\frac{{\rm Im}(x)}{\hbar}}
    v_N^{-1}(x)
    \sum_{\bs{n}\in \mathcal{W}_{N}\cdot \bs{\gamma}}
    \CQ_{\bn}^{\frac{N}{2}}(\bsa, \Lambda, \hbar)
    \, ,
    \qquad
    \Re[x] \to - \infty
    \, ,
\end{multline}
so once more we have a discrete set of solutions to the difference equation \eqref{eq:diffeq:sinh_kin} which are square-integrable parallel to the real line if
\be \Im[x] < \frac{\hbar}{2} \br{1-\frac{1}{N}} \, . \ee
Note the power law decay in both directions.

It is important to stress that even when $\omega_{-, N}^{\text{even}}$ is an entire and square-integrable solution to the difference equation \eqref{eq:diffeq:sinh_kin}, it is generically not in the original domain $\CD$ of the  quantized curve as defined around \eqref{eq:domain}.

\paragraph{Kinetic term with sinh(p) and unbounded potential for N odd.}
Let us look at the equation \eqref{eq:diffeq:sinh_kin} for the unbounded potential $+V_N$ and $N$ odd.
The asymptotic for $\Re[x] \to \pm \infty$ for fixed $\Im[x]$ is now
\begin{multline}
    \omega_{+, N}^{\text{odd}}(x,{\bs h}) \simeq
    \re^{\frac{\pi x}{\hbar}} \left(\frac{2 \pi \hbar}{\abs{{\rm Re}(x)}}\right)^{\frac{N}{2}} \br{\frac{\abs{{\rm Re}(x)}^N}{\ri \Lambda^N}}^{\frac{{\rm Im}(x)}{\hbar}}
    \times
    \\
    v_N(x)
    \sum_{\bs{n}\in \mathcal{W}_{N} \cdot \bs{\gamma}}
    \CQ_{\bn}^0(\bsa, \Lambda, \hbar)
    \, ,
    \qquad
    \Re[x] \to + \infty
    \, ,
\end{multline}
\begin{multline}
    \omega_{+, N}^{\text{odd}}(x,{\bs h} ) \simeq
    \re^{-\frac{\pi x}{\hbar}} \left(\frac{2 \pi \hbar}{\abs{{\rm Re}(x)}}\right)^{\frac{N}{2}}\br{\frac{\abs{{\rm Re}(x)}^N}{\ri \Lambda^N}}^{-\frac{{\rm Im}(x)}{\hbar}} 
    \times
    \\
    v_N(x)
    \sum_{\bs{n}\in \mathcal{W}_{N}\cdot \bs{\gamma}}
    \CQ_{\bn}^0(\bsf(\bsa), \Lambda, \hbar)
    \, ,
    \qquad
    \Re[x] \to - \infty
    \, .
\end{multline}
Hence, there is no single quantization condition that gives rise to square-integrable solutions, and the two quantization conditions \eqref{eq:QCodd} and \eqref{eq:QCodd_alt} must be imposed together. However, they do not admit simultaneous solutions for arbitrary values of the coefficients $h_k$, but only for a special subset thereof. It is noteworthy that the Toda points satisfying \eqref{eq:toda-qc} are a part of this subset. This is not surprising, since for odd N the kinetic term in the Baxter equation \eqref{eq:baxter} is given by $\sinh[\rp]$.

\paragraph{Kinetic term with sinh(p) and inverted unbounded potential for N odd.}
Let us look at the equation \eqref{eq:diffeq:sinh_kin} for the inverted unbounded potential $-V_N$ and $N$ odd.
The asymptotics for $\Re[x] \to \pm \infty$ with fixed $\Im[x]$ is now
\begin{multline}
    \omega_{-, N}^{\text{odd}}(x,{\bs h}) \simeq
    \\
    \re^{\frac{2 \pi x}{\hbar}} \left(\frac{2 \pi \hbar}{\abs{{\rm Re}(x)}}\right)^{\frac{N}{2}} \br{\frac{\abs{{\rm Re}(x)}^N}{\ri \Lambda^N}}^{\frac{{\rm Im}(x)}{\hbar}} v_N(x)
    \sum_{\bs{n}\in \mathcal{W}_{N} \cdot \bs{\gamma}}
    \CQ_{\bn}^0(\bsa, \Lambda, \hbar)
    \, ,
    \qquad
    \Re[x] \to + \infty
    \, ,
\end{multline}
\begin{multline}
    \omega_{-, N}^{\text{odd}}(x,{\bs h} ) \simeq
    \\
    \left(\frac{2 \pi \hbar}{\abs{{\rm Re}(x)}}\right)^{\frac{N}{2}}\br{\frac{\abs{{\rm Re}(x)}^N}{\ri \Lambda^N}}^{-\frac{{\rm Im}(x)}{\hbar}} v_N(x)
    \sum_{\bs{n}\in \mathcal{W}_{N}\cdot \bs{\gamma}}
    \CQ_{\bn}^0(\bsf(\bsa), \Lambda, \hbar)
    \, ,
    \qquad
    \Re[x] \to - \infty
    \, .
\end{multline}
The quantization condition giving rise to square integrable solutions is once again \eqref{eq:QCodd}, but where we should now take $\ri \Lambda^N > 0$ rather than $\Lambda^N > 0$. If \eqref{eq:QCodd} is satisfied, one finds
\begin{multline}
    \omega_{-, N}^{\text{odd}}(x,{\bs h}) \Big|_{\text{\eqref{eq:QCodd}}} \simeq
    \\
    \br{\frac{2 \pi \hbar}{ {\abs{\Re[x]}}}}^{\frac{N}{2}} \br{\frac{\abs{{\rm Re}(x)}^N}{\ri \Lambda^N}}^{\frac{{\rm Im}(x)}{\hbar}} v_N(x)
    \sum_{\bs{n}\in \mathcal{W}_{N} \cdot \bs{\gamma}}
    \CQ_{\bn}^1(\bsa, \Lambda, \hbar)
    \, ,
    \qquad
    \Re[x] \to + \infty
    \, .
\end{multline}
So we have solutions to the difference equation \eqref{eq:diffeq:sinh_kin} which are square-integrable parallel to the real line if and only if \eqref{eq:QCodd} holds and ${\abs{\Im[x]}} < (\hbar/2)(1-1/N)$. We have power law-decay both at positive and negative infinity in that case.

However, in this case, we can find a second quantization condition. One can see that $\exp(-2 \pi x/\hbar) \omega_{-, N}^{\text{odd}}(x,{\bs h})$ is again a solution to the difference equation \eqref{eq:diffeq:sinh_kin} with now exponential growth at negative infinity and power-law decay at positive infinity. Hence, to get square-integrability, we should satisfy the alternative quantization condition \eqref{eq:QCodd_alt} and $\abs{\Im[x]} < (\hbar/2)(1-1/N)$.

So we have solutions to the difference equation \eqref{eq:diffeq:sinh_kin} which are square-integrable parallel to the real line if and only if \eqref{eq:QCodd} \emph{or} \eqref{eq:QCodd_alt} holds. It is important to note that even when $\omega_{-, N}^{\text{odd}}(x,{\bs h})$ or $\exp(-2 \pi x/\hbar) \omega_{-, N}^{\text{odd}}(x,{\bs h})$ are entire and square-integrable solutions to the difference equation \eqref{eq:diffeq:sinh_kin}, they are generically not in the original domain $\CD$ of the  quantized curve as defined around \eqref{eq:domain}.

%%%%%%%%%%%%%%%%%%%%%%%%%%%%%%%%%%%%%%%%%%%%%%%%%%%%%%%%%%%%%%%%%%%%%%%%%%%%%%%%%%%%
%%%%%%%%%%%%%%%%%%%%%%%%%%%%%%%%%%%%%%%%%%%%%%%%%%%%%%%%%%%%%%%%%%%%%%%%%%%%%%%%%%%%

\section{Comments on the Toda points}
\label{sec:toda}

In \autoref{sec:spectp} we have discussed the spectral problem \eqref{eq:diffm} from the standpoint of standard quantum mechanics, wherein one first constructs entire off-shell solutions and then imposes square-integrability as a boundary condition.
However, there is another way to view equation \eqref{eq:diffm}, namely from the perspective of integrability. Indeed, the difference equation \eqref{eq:diffm} is closely related to the Baxter equation for the quantum $\mathrm{SU}(N)$ Toda lattice \cite{PasquierGaudin1992,Gutzwiller1981,Gutzwiller1980,Sklyanin1985}, namely
\be \label{eq:baxter}
(\ri\Lambda)^N Q(x+\ri \hbar)+(-\ri\Lambda)^N Q(x-\ri \hbar)-V_N(x)Q(x)=-E Q(x)\, .
\ee
In this setting, one looks for entire solutions to \eqref{eq:baxter} such that:
\be\label{eq:bax}
Q(x)\in L^2(\IR)\quad \text{with} \quad Q(x)\simeq \re^{-\tfrac{N\pi}{2\hbar}|x|}, \qquad x\to \pm\infty\, .
\ee
By defining 
\be\label{eq:todatrasf} Q(x)=\left(\sum_{n_i\in S} d_i \re^{-\frac{2\pi x n_i}{\hbar}}\right)\re^{-\frac{\pi (N+2)}{2 \hbar}x}\psi(x) \, , \quad S\subset
 \IZ, \quad d_i \in \IC \, , \ee
it is easy to show that $Q(x)$ solves \eqref{eq:baxter} for any choice of $S\subset
 \IZ$.  However, the mapping between the Baxter solution \eqref{eq:bax} and our solutions \eqref{eq:eigenfeven}, \eqref{eq:eigenfodd} requires a particular choice of $S$. We will address this question in a separate work \cite{wip}.

Nevertheless, even after taking \eqref{eq:todatrasf} into account, one can see that for generic values of the potential parameters $h_k$ ($k = 2, \ldots, N-1$), the on-shell eigenfunctions \eqref{eq:eigenfeven} and \eqref{eq:eigenfodd} do not reproduce \eqref{eq:bax}.
For odd $N$, this follows directly from the fact that the asymptotics of \eqref{eq:eigenfodd} are outgoing with power-law decay on one side and exponentially decaying on the other, with a rate which is independent of $N$.
Likewise, for $N$ even, this follows from the fact that the decay behaviour of \eqref{eq:eigenfeven} as $x \to \pm \infty $ is the same for both plus and minus infinity, and it is independent of $N$. This is in line with the works of \cite{PasquierGaudin1992,Gutzwiller1981,Gutzwiller1980,Sklyanin1985}, where they found that eigenfunctions satisfying \eqref{eq:bax} exist only for special values of the moduli $h_k$, $k=2,\dots,N$. We refer to these distinguished values as \emph{Toda points} and denote them by $h_k^{\rm T}$, $k=2,\dots,N$. This, in turn, means that for \eqref{eq:bax} to hold, one must impose simultaneously $N-1$ quantization conditions, thereby quantizing all the $h_k$, $k=2,\dots,N$. This was nicely illustrated in \cite{PasquierGaudin1992}. From the gauge theory perspective, it was found in \cite{ns,Kozlowski:2010tv,Mironov:2009dv} that the quantization conditions determining the Toda points are (following the notation of \cite{Grassi:2018bci}):
\begin{equation}
\label{eq:toda-qc}
    \frac{\ri}{\hbar} \frac{\partial F_{\rm NS}}{\partial \bs{a}}
    = \ri 2 \pi \bs{\ell} + \ri \pi \bs{\rho}
    \, ,
\end{equation}
where $\bs{\ell}$ encodes the quantum numbers,
\begin{equation}
    \bs{\ell}= \sum_{k=1}^{N-1} \ell_k \blam_k \, ,
    \qquad \qquad
    \ell_k \in \nnints \, ,
    \qquad \qquad
    \bs{\rho}=\sum_{I=1}^N \br{N-I} \bs{e}_I \, ,
\end{equation}
and $\bs{\rho}$ is a constant vector. It is important to note that \eqref{eq:toda-qc} is a vector equation: it consists of $N-1$ quantization conditions for the $a_I$, the corresponding solutions being denoted by $\bs{a}{\br{\bs{\ell}}}$. After using the generalized Matone relations \eqref{eq:htoam} and \eqref{eq:genmat}, these translate into $N-1$ quantization conditions for the complex moduli $h_k$, $k = 2, \dots, N$. In other words, it constrains the $h_k$ to lie at the Toda points, which we denote by $h_k^{\rm T}({\bs \ell})$.
This stands in sharp contrast with standard quantum mechanics, where square-integrability imposes only a single quantization condition, namely \eqref{eq:QCeven} for $N$ even and \eqref{eq:QCodd} for $N$ odd. This condition quantises the energy $E$, while the other parameters $h_k$, $k = 2, \dots, N-1$, in the potential remain free. However, as shown in \cite{Grassi:2018bci}, if the $h_k$ satisfy the Toda quantization conditions \eqref{eq:toda-qc}, they automatically satisfy the quantum mechanical quantization conditions \eqref{eq:QCeven}, \eqref{eq:QCodd}. Hence, the Toda points correspond to special loci in our quantum mechanical problem, where eigenfunctions have an enhanced decay. As discussed above, this enhancement also comes together with some interesting phenomena, such as degeneracies among bound states.
It is then natural to ask how these phenomena are captured by our explicit eigenfunctions \eqref{eq:eigenfeven} and \eqref{eq:eigenfodd}.

Let us first review what is known about the Toda eigenfunction \eqref{eq:bax} from a gauge-theoretic perspective. It was found in \cite{Kozlowski:2010tv, Sciarappa:2017hds} that, in gauge-theoretic language, the eigenfunctions of the Toda Baxter equation \eqref{eq:baxter} take the form:
\be\label{eq:todeigg}
   Q(x, {\bs h}^{\rT} ({\bs \ell}))=  \re^{-\frac{\pi (N+2)}{  {2} \hbar} x} Z_D(x,{\bs a}({\bs \ell}), \Lambda, \hbar)
   + \xi \,
  \re^{\frac{\pi (N+2)}{  {2} \hbar} x} Z_D(-x,-{\bs a}({\bs \ell}), \Lambda, \hbar),
\ee
where ${\bs a}({\bs \ell})$ are solutions to the $N-1$ quantization conditions \eqref{eq:toda-qc}. The factor $\xi$ in \eqref{eq:todeigg} is an $x$-independent  function that should ensure the cancellation of poles between the two terms in \eqref{eq:todeigg}. It was shown in \cite{Kozlowski:2010tv} that the existence of such {\it{x-independent function}} is guaranteed only if $\bs{a}={\bs a}({\bs \ell})$, where ${\bs a}({\bs \ell})$ are solutions to \eqref{eq:toda-qc}. For example, in the $SU(2)$ case there is only a single quantum number $\ell$, and one finds:
\be
\xi= (-1)^\ell \, .
\ee
Hence, we can see another important difference between \eqref{eq:todeigg} and \eqref{eq:eigenfeven}, \eqref{eq:eigenfodd}:
\begin{itemize}
    \item Our eigenfunctions \eqref{eq:eigenfeven} and \eqref{eq:eigenfodd} are entire in $x$ for generic values of ${\bs a}$, i.e. generic values of $h_k$, $k = 2, \dots, N$. In this case, no quantization condition is required to guarantee pole cancellation. This is made possible by the $x$-dependent factor appearing inside the sums over the Weyl orbit in \eqref{eq:eigenfeven} and \eqref{eq:eigenfodd}.
    \item The eigenfunctions constructed in \cite{Kozlowski:2010tv, Sciarappa:2017hds} are pole-free only if ${\bs a}\equiv {\bs a}({\bs \ell})$, where the ${\bs a}({\bs \ell})$ are fixed by the $N-1$ quantization conditions \eqref{eq:toda-qc}. This is because $\xi$ in \eqref{eq:todeigg} is assumed to be $x$-independent.
\end{itemize}
Given these considerations, let us examine what happens to the eigenfunctions \eqref{eq:eigenfeven} and \eqref{eq:eigenfodd} once we restrict to the Toda points, that is, upon imposing the $N-1$ quantization conditions \eqref{eq:toda-qc}.
Let us first consider the case of even $N>2$.\footnote{For $N=2$ these sums do not vanish, see \cite{Francois:2025wwd}.} We find that the following special combinations vanish identically\footnote{These identities were rigorously proved by Karim Aziz as part of an unpublished bachelor’s project \cite{A26}.}:
 \be \label{eq:problem1} \sum_{\bs{n}\in \mathcal{W}_{N}\cdot \bs{\gamma}} \frac{\exp\br{{\ri}{\left( 2 \pi \bs{\ell}+ \pi \bs{\rho}\right)}\cdot\bs{n}}}{\prod_{\bs{\alpha}\in \Delta_{+}}\br{2\sinh\br{\frac{\pi \bs{a}\cdot\bs{\alpha}}{\hbar}}}^{(\bs{n}\cdot{\bs \alpha})^2}} \prod_{I=1}^{N} \bigl(\re^{\frac{2\pi x}{\hbar}} - \re^{\frac{2\pi {\bs a}\cdot {\bs e}_I}{\hbar}}\bigr)^{\frac{1}{2} - n_I}  =0\,,\ee
  \be \label{eq:problem2}\sum_{\bs{n}\in \mathcal{W}_{N} \cdot \br{\bs{\gamma} + \bs{e}_N}} \frac{\exp\br{{\ri}{\left( 2 \pi \bs{\ell}+ \pi \bs{\rho}\right)}\cdot\bs{n}}}{\prod_{\bs{\alpha}\in \Delta_{+}}\br{2\sinh\br{\frac{\pi \bs{a}\cdot\bs{\alpha}}{\hbar}}}^{(\bs{n}\cdot{\bs \alpha})^2}} \prod_{I=1}^{N} \bigl(\re^{\frac{2\pi x}{\hbar}} - \re^{\frac{2\pi {\bs a}\cdot {\bs e}_I}{\hbar}}\bigr)^{\frac{1}{2} - n_I}  = 0 \,.\ee
Likewise, for $N$ odd we have:
 \be \label{eq:problem3}\sum_{\bs{n}\in \mathcal{W}_{N}\cdot \bs{\gamma}} \frac{\exp\br{{\ri}{\left( 2 \pi \bs{\ell}+ \pi \bs{\rho}+ \ri\frac{\pi}{\hbar} {\bs a}\right)}\cdot\bs{n}}}{\prod_{\bs{\alpha}\in \Delta_{+}}\br{2\sinh\br{\frac{\pi \bs{a}\cdot\bs{\alpha}}{\hbar}}}^{(\bs{n}\cdot{\bs \alpha})^2}} \prod_{I=1}^{N} \bigl(\re^{\frac{2 \pi x}{\hbar}} - \re^{\frac{2 \pi {\bs a}\cdot {\bs e}_I}{\hbar}}\bigr)^{\frac{1}{2} - n_I} = 0 \, .\ee
This means that, strictly speaking, the eigenfunctions \eqref{eq:eigenfeven}, \eqref{eq:eigenfodd} vanish at the Toda points. However, since the two terms in the sums \eqref{eq:eigenfeven} and \eqref{eq:eigenfodd} go to zero at the same rate, an appropriate normalization can be introduced to obtain a non-vanishing result.
A more detailed analysis will be presented in \cite{wip}.

%%%%%%%%%%%%%%%%%%%%%%%%%%%%%%%%%%%%%%%%%%%%%%%%%%%%%%%%%%%%%%%%%%%%%%%%%%%%%%%%%%%%
%%%%%%%%%%%%%%%%%%%%%%%%%%%%%%%%%%%%%%%%%%%%%%%%%%%%%%%%%%%%%%%%%%%%%%%%%%%%%%%%%%%%

\section{Derivation from the TS/ST correspondence}
\label{sec:tsst}

In this section, we derive \eqref{eq:eigenfeven} and \eqref{eq:eigenfodd} from the TS/ST correspondence \cite{ghm,cgm2,Marino:2016rsq,Marino:2017gyg,Francois:2025wwd}.

\subsection{Open TS/ST for the \texorpdfstring{$Y^{N, 0}$}{Y(N, 0)} geometries}

We consider the toric Calabi–Yau threefolds obtained as crepant resolutions of the $Y^{N,0}$ singularities, which geometrically engineer $\mathcal{N} = 1$, SU($N$) SYM in five dimensions or $\mathcal{N} = 2$, SU($N$) SYM in four dimensions \cite{kkv,Klemm:1996bj}. The associated mirror curve \cite{kkv} has the form:
\begin{equation}
\label{eq:5dmirror_kappas}
    \Lambda^N \br{\re^y + \re^{-y}} + \sum_{\ell = 0}^N \kappa_\ell \, \re^{\br{\frac{N}{2} - \ell}x}=0
\end{equation}
where $\Lambda$ and $\kappa_\ell$ are the complex moduli, and we can set without loss of generality $\kappa_0 = 1 = \kappa_N$. It is also useful to introduce the chemical potentials $\mu_\ell$ by
\begin{equation}
    \kappa_\ell = \re^{\mu_\ell}
    \, ,
    \qquad \qquad
    \ell \in \cbr{1, \dots, N-1}
    \, ,
\end{equation}
and the Batyrev coordinates as
\begin{equation}
	z_i= \frac{\kappa_{i-1}\kappa_{i+1}}{\kappa_i^2}, \qquad i \in \{1,\dots N-1\}, \qquad z_N = \frac{\Lambda^{2N}}{\kappa_0\kappa_N}.
\end{equation}
The complex moduli $\kappa_\ell$ are related to the K\"{a}hler parameters $t_k$ through the mirror maps \cite{Hori2003}. Our conventions for the classical mirror maps are as follows:
\begin{equation}
\label{eq:class}
	t_k = \sum_{\ell = 1}^{N-1} C_{k \ell} \mu_\ell + \bigO{z_i}
	\, ,
	\qquad
	k \in \{1, \dots, N-1\}
	\, ,
	\qquad
	t_N = - \log[z_N]
	\, ,
\end{equation}
where $C_{k \ell}$ is the Cartan matrix of SU($N$).

Within the TS/ST correspondence, we consider the quantization of the mirror curve \eqref{eq:5dmirror_kappas}. Upon quantization, this curve is promoted to a difference equation \cite{adkmv,ns,mirmor,acdkv}:
\begin{equation}
\label{eq:diff}
    \Lambda^N \br{\varphi{\br{x + \ri \hbar, \bs{\kappa}}} + \varphi{\br{x - \ri \hbar, \bs{\kappa}}}}+
    \sum_{\ell = 0}^N \kappa_\ell \re^{\br{\frac{N}{2} - \ell}x} \varphi{\br{x, \bs{\kappa}}}
    = 0
    \, ,
\end{equation}
and, simultaneously, the classical K\"ahler parameters $t_k$ are replaced by their quantum counterparts, giving rise to the quantum mirror maps \cite{acdkv,mirmor}, which we denote by
\begin{equation}
\label{eq:quantummir}
    t_k\br{\hbar}, \qquad \qquad k \in \cbr{1, \dots, N-1} \, .
\end{equation}
We refer to \cite{hm, GGH26} for explicit expressions of the quantum mirror maps for the $Y^{N,0}$ geometries. In the limit $\hbar \to 0$, \eqref{eq:quantummir} reduces to the classical mirror map \eqref{eq:class}. For the geometries under consideration, the quantum mirror maps can also be interpreted as the inverses of Wilson loop VEVs of five-dimensional $\mathcal{N}=1$ SYM in the NS limit \cite{Gaiotto:2014ina,Bullimore:2014awa}. Explicit expressions for the relevant Wilson loop VEVs in the case of $Y^{N,0}$ geometries can be found in \cite[app.~D]{GGH26}. In this context, it is often convenient to introduce the 5d Coulomb branch parameters\footnote{We are using the same notation for the 5d and 4d Coulomb branch parameters. The distinction should be clear from the context.}:
\be
    a_I, \qquad I \in \cbr{1, \dots, N} \, ,
    \qquad \sum_{I=1}^N  a_I = 0 \, ,
\ee
which parametrize the VEVs of the complexified scalars in the vector multiplet of 5d SYM, compactified on $\reals^4 \times \mathbb{S}^1$. These are related to the quantum mirror maps by
\be t_k(\hbar)=a_k-a_{k+1}, \qquad k \in \cbr{1, \dots, N-1}, \ee
or equivalently, inverting the relations, by
\be
    a_I = \sum_{\ell = 1}^{I-1} \br{- \frac{\ell}{N}} t_\ell(\hbar) + \sum_{\ell = I}^{N-1} \br{\frac{N - \ell}{N}} t_\ell(\hbar) \, .
\ee
Given the underlying SU($N$) structure, it is convenient to organize these parameters as
\be
{\bs{t}} = \sum_{I = 1}^N a_I \, \bs{e}_I = \sum_{k = 1}^{N-1} t_k(\hbar) \, {\bs \lambda}_k \, ,
\ee
where $\bs{e}_I$ are the weights of the fundamental representation and \( {\bs \lambda}_k \) denote the fundamental weights, see \autoref{app:GT} for details on the conventions.

The TS/ST correspondence for the $Y^{N,0}$ Calabi--Yau geometries has been studied in the closed string sector in \cite{Grassi:2018bci,hm, bgt2, GGH26}. In this work, we also consider the open sector, extending the results of \cite{Marino:2016rsq, Marino:2017gyg, Francois:2025wwd} to this class of geometries. A central object in the correspondence is the full grand potential, which we denote by
\be
\label{eq:totalgrand}
    {\rm J}(x, \bs{t}, \Lambda, \hbar) \;=\; {\rm J}^{\rm closed}(\bs{t}, \Lambda, \hbar)\;+\; {\rm J}^{\rm open}(x, \bs{t}, \Lambda, \hbar) \, .
\ee
The closed topological string grand potential ${\rm J}^{\rm closed}(\bs{t}, \Lambda,\hbar)$ is constructed from a specific combination of the refined topological string partition function in the Gopakumar--Vafa (GV) limit and in the NS limit \cite{Hatsuda:2013oxa,ghm,cgm2}. For the $Y^{N,0}$ geometries, the explicit expression is given in \cite[p.~19]{Grassi:2018bci}. The open topological string grand potential ${\rm J}^{\rm open}(x, \bs{t}, \Lambda,\hbar)$ has an analogous structure, but it involves instead the open string partition functions \cite{Marino:2016rsq,Marino:2017gyg,Francois:2025wwd}. More concretely, for the $Y^{N,0}$ geometries we have:
\be
\label{eq:grandopen}
    {\rm J}^{\rm open}(x, \bs{t}, \Lambda, \hbar)
    = {\rm J}_{\rm p}^{\rm open}(x, \Lambda, \hbar)
    + {\rm J}_{\rm 1\mbox{-}loop}^{\rm open}(x, \bs{t}, \Lambda, \hbar)
    + {\rm J}_{\rm inst}^{\rm open}(x,\bs{t}, \Lambda, \hbar) \, .
\ee
Here, the polynomial part is given by
\begin{equation}
\label{eq:grandopenp}
    {\rm J}_{\rm p}^{\rm open}\br{x, \Lambda, \hbar}
    = \ri \log\br{\Lambda^{N}} \frac{x}{\hbar}
    + \pi \frac{x}{\hbar}
    - \frac{N}{4} \, x \br{1 + \ri \frac{x}{\hbar}} \, .
\end{equation}
This term is obtained by analysing the large-$x$ behaviour of \eqref{eq:diff} and demanding that $\exp{{\rm J}_{\rm p}^{\rm open}}$ provides a formal solution in the large-$x$ asymptotic regime. This procedure fixes the form of \eqref{eq:grandopenp} up to an $\ri \hbar$ periodic factor $\sim \exp{\br{\frac{2 \pi}{\hbar} x}}$, which is then fixed with hindsight by requiring the cancellation of poles between the two terms in \eqref{eq:sigmasum}. The one-loop contribution takes the form  \cite{Francois:2025wwd,Sciarappa:2017hds}
\begin{equation}
    \exp\!{\br{{\rm J}_{\text{1-loop}}^{\rm open}(x, \bs{t}, \Lambda, \hbar)}}
    = \prod_{I=1}^N \Phi_b\br{\frac{- x + \bs{e}_I \cdot \bs{t} + \ri \pi b^2}{2 \pi b}} \, ,
    \qquad \qquad
    \hbar=2\pi b^2,
\end{equation}
where $\Phi_b$ is Faddeev's non-compact quantum dilogarithm, see \autoref{sec:faddev}.
Finally, the instanton contribution ${\rm J}_{\rm inst}^{\rm open}(x,\bs{t}, \Lambda,\hbar)$ is built from a specific combination of the refined open topological string partition function on the $Y^{N,0}$ geometry, taken in the GV and NS limits, see \cite[sec.~4.1]{Marino:2016rsq}, \cite[sec.~2.2]{Marino:2017gyg} and \cite[sec.~3.3]{Francois:2025wwd}.

From the topological string perspective, the open string grand potential \eqref{eq:grandopen} provides a non-perturbative completion of the open topological string free energy in the large-radius frame. This quantity, however, is not background independent and moreover develops poles at finite values of $x$, which is identified with the open string modulus. To remedy this, it was shown in \cite{Marino:2016rsq,Marino:2017gyg,Francois:2025wwd} that one must consider a particular linear combination of the form:
\begin{equation}
\label{eq:sigmasum}
    \varphi{\br{x, \bs{\kappa}}} = \sum_{s \in \cbr{-1, + 1}} \ \sum_{\bs{n} \in Q_{N-1}} \exp{\sbr{{\rm J}_{s}\br{x, {\bs{t}} + \ri 2 \pi \bs{n}, \Lambda, \hbar}}}
    \, ,
\end{equation}
where $Q_{N-1}$ is the root lattice defined in \eqref{eq:defrootlat} and
\( {\rm J}_+(x, {\bs{t}}, \Lambda, \hbar) = {\rm J}(x, {\bs{t}}, \Lambda, \hbar) \),  while ${\rm J}_-$ is obtained from ${\rm J}_+$ by a suitable transformation, as we will see later. Note that in \eqref{eq:sigmasum} we are implicitly using the quantum mirror maps \eqref{eq:quantummir} relating $\bs{\kappa}$ to the quantum K\"{a}hler parameters $\bs t$. In the topological string theory, both summations in \eqref{eq:sigmasum} are essential to achieve background independence in the open and closed string moduli. Indeed, the sum over $\bs{n}$  renders the expression entire in the closed string moduli $\bs{\kappa}$, while the sum over $s$ ensures entireness in the open modulus $x$. From the perspective of spectral theory, equation \eqref{eq:sigmasum} is particularly significant, as it provides an exact entire solution to the quantum mirror curve associated with the underlying Calabi-Yau geometry. Although formal solutions to the quantum mirror curve can be constructed in various ways, \eqref{eq:sigmasum} stands out due to its analyticity in both the open modulus $x$ and the closed moduli ${\bs \kappa}$. Moreover, when evaluated on-shell, it yields genuine square-integrable eigenfunctions, see \cite{Marino:2016rsq,Marino:2017gyg,Francois:2025wwd} for more details.

An important open question in the context of the open TS/ST correspondence is the precise characterization of the second term in \eqref{eq:sigmasum}.
In \cite{Marino:2016rsq,Marino:2017gyg} it was suggested that intuitively this has to do with  moving to a different sheet of the mirror curve. In practice, however, it is challenging to implement this explicitly, that is, to find a simple explicit relation between $\rm J_+$ and $\rm J_-$. For local $\IF_0$, this problem was solved in \cite{Francois:2025wwd}, where, using insights from Painlev\'{e} equations \cite{Francois:2023trm}, it was shown that the two terms are simply related by shifts of the complex moduli and of the variable $x$. At the level of the mirror curve, these shifts are characterized by the fact that they leave the curve itself invariant. Let us now apply the same idea to \eqref{eq:5dmirror_kappas}.
We consider transformations of the form
\begin{equation}
    x \to s\, x - \ri 2 \pi \frac{k_x}{N} \, ,
    \qquad \quad
    y \to \frac{1}{s} \, y - \ri \pi k_y \, ,
    \qquad \quad
    k_x, k_y \in \ints \, ,
    \qquad \quad
    k_x + k_y \in 2 \ints \, ,
\end{equation}
combined with a corresponding transformation of the moduli $\mu_\ell$, which we will introduce below.
Drawing inspiration from the example of local $\IF_0$ \cite{Francois:2025wwd} and the Toda lattice \cite{Kozlowski:2010tv,Sciarappa:2017hds}, it is natural to choose $s = -1$. The classical curve is then invariant if and only if the closed moduli transform as
\begin{equation}\label{eq:shifts}
    \mu_\ell \to \mu_{N-\ell} - \ri 2 \pi \frac{k_x \ell}{N}
    \, .
\end{equation}
Following again the analogy with local $\IF_0$ \cite{Francois:2025wwd},  we set $k_x = 1 = k_y$. This gives
\begin{equation}
\label{eq:tsst}
    \boxed{\varphi(x, {\bs \kappa}) = \varphi_{+}{\br{x,\bs{\kappa}}} + \exp{\br{\frac{\ri}{\hbar} \frac{\pi^2}{N} + \pi \frac{x}{\hbar}}}
    \varphi_{+}{\br{-x-\ri\frac{2\pi}{N}, \, \widetilde{\bs \kappa}}} \, ,}
\end{equation}
\begin{equation}
     \varphi_{+}{\br{x, \bs{\kappa}}} = \sum_{{\bs n}\in Q_{N-1}} \re^{{\rm J}(x,{\bs t}+\ri 2\pi {\bs n}, \Lambda, \hbar)}.
\end{equation}
where $\bs{\kappa}=\{\kappa_1,\cdots,\kappa_{N-1} \}$, $\widetilde{\bs \kappa}=\{\widetilde\kappa_1, \cdots, \widetilde\kappa_{N-1} \} $ with $\widetilde \kappa_\ell=\kappa_{N-\ell}\,\re^{-\ri \frac{2 \pi}{N}\ell}$.
Let us also note that, after applying the quantum mirror map, the transformation \eqref{eq:shifts} is equivalent to
\begin{equation}
    \bs{t} \to \bsf{\br{\bs{t}}} + \ri 2 \pi k_x \bs{e}_N
    \, ,
    \qquad \qquad
    \bsf{\br{\bs{t}}} = \sum_{k=1}^{N-1} t_{N-k} {\bs \lambda}_k=-\sum_{I=1}^N a_{N-I+1} \bs{e}_I
    \, .
\end{equation}
We have carried out preliminary tests of \eqref{eq:tsst} for $N=4$. The next section provides further evidence in favour of \eqref{eq:tsst} for generic $N$ and is the origin of our proposal in \autoref{sec:Eigenfunctions4D}.

\subsection{The four-dimensional limit}

Topological string theory on the $Y^{N,0}$ geometries provides a framework to engineer four-dimensional $\mathcal{N}=2$, SU($N$) SYM \cite{kkv,Klemm:1996bj}. To realize this setup, one performs a specific geometric limit, usually called the four-dimensional limit. In this section, we show that, after carrying out this limit, the eigenfunctions \eqref{eq:tsst} reduce to \eqref{eq:eigenfeven} and \eqref{eq:eigenfodd}, respectively.
The computation of the four-dimensional limit of the eigenfunctions mirrors that for the spectral determinant in \cite[sec.~5]{Grassi:2018bci}, which we follow closely. To take the limit on the mirror curve, we need a different parametrization from the one used in \eqref{eq:5dmirror_kappas}. The reparametrization is implemented by some simple shifts, see also the discussion in \cite[sec.~5]{Grassi:2018bci}.
Combining these shifts with the natural scaling one finds that the four-dimensional limit is
\begin{equation}
\label{eq:Scaling4DLimit}
    x \to R x - \ri \frac{\pi}{N} \indicator{N}{2 \nnints + 1} \, ,
    \quad
    \bs{t} \to R \bs{a} + \ri 2 \pi \bs{\gamma} \, ,
    \quad
    \Lambda^{2 N} \to \br{-1}^N  \br{R \Lambda}^{2N} \, ,
    \quad
    \hbar \to R \hbar \, ,
\end{equation}
and then $R \to 0$ from above, with $\bs \gamma$ defined in \eqref{eq:gammadef}.
One can check that, by taking this limit, the quantum mirror curve \eqref{eq:diff} reduces to the quantum Seiberg-Witten curve of four-dimensional $\CN = 2$ SU($N$) SYM \eqref{eq:diffm}.

For simplicity, we set the four-dimensional reduced Planck constant $\hbar$ to 1. We can reintroduce it by sending $\br{x, \bs{a}, \Lambda} \to \br{x / \hbar, \bs{a} / \hbar, \Lambda / \hbar}$ at the end.
The four-dimensional limit acts similarly on both terms in the right-hand side of \eqref{eq:tsst}. Hence, it is convenient to introduce the variables $s\in \cbr{-1, +1}$ and $k_x = k_y = k \in \cbr{0, 1}$, and look at the $R \to 0$ limit of
\begin{multline}
\label{eq:GenericSaddle5D}
    \hspace{-3mm}
    \sum_{\bs{n} \in Q_{N-1}} \exp\!{\left[
    - \rJ^\text{closed}{\br{R \bs{a} + \ri 2 \pi \bs{\gamma}, \re^{- \ri \frac{\pi}{2} \frac{\indicator{N}{2 \nnints + 1}}{N}} R \Lambda, R}}
    + \ri \frac{\pi^2}{N} \frac{k^2}{R} - s k \pi \br{x - \ri \frac{\pi}{N} \frac{\indicator{N}{2 \nnints + 1}}{R}}
    \right.}
    \\
    + \rJ{\br{s R x - \ri \frac{\pi}{N} \br{2 k + s \indicator{N}{2 \nnints + 1}}, R \bsf_s{\br{\bs{a}}} + \ri 2 \pi \bs{w}, \re^{- \ri \frac{\pi}{2} \frac{\indicator{N}{2 \nnints + 1}}{N}} R \Lambda, R}} \biggr]
    \, ,
\end{multline}
where we introduced the convenient notation
\begin{equation}
\label{eq:Notation4DLimit}
    \bsf_s{\br{\bs{t}}} =
    \begin{cases}
        \bsf{\br{\bs{t}}} & s = -1
        \\
        \bs{t} & s = +1
    \end{cases}
    \, ,
    \qquad \qquad
    \bs{w} = \bs{\gamma} + \br{\frac{1 - s}{2}} \indicator{N}{2 \nnints} \, \bs{e}_N + \bs{n}
    \, .
\end{equation}
The first term on the right-hand side of \eqref{eq:tsst} is then given by $s = 1$ and $k = 0$, while the second term corresponds to $s = -1$ and $k = 1$. Note that we normalized the eigenfunctions by the $x$-independent constant $\exp\br{\rJ^\text{closed}}$, which is convenient in the four-dimensional limit \cite{Grassi:2018bci}. In \eqref{eq:Notation4DLimit}, we have already used the fact that
\begin{equation}
    \bsf{\br{\bs{\gamma}}} + \bs{e}_N + \bs{m}
    = \br{-1}^N \bs{\gamma} + \bs{e}_N + \bs{m}
    = \bs{\gamma} + \indicator{N}{2 \nnints} \, \bs{e}_N + \bs{n} \, ,
\end{equation}
where $\bs{n}, \bs{m} \in Q_{N-1}$ are related by a simple shift of the origin of the root lattice $Q_{N-1}$ when $N$ is odd, and are instead equal when $N$ is even.

\subsubsection{The leading terms and the closed part}

The key observation in the four-dimensional limit is that only finitely many of the $\bs{n} \in Q_{N-1}$ appearing in the sum \eqref{eq:GenericSaddle5D} contribute. As discussed below, which $\bs{n}$ dominate is determined by the polynomial part of the closed grand potential.

Let us first consider the shifts of $\bsf_s{\br{\bs{t}}}$ by $\ri 2\pi\bs{w}$ in \eqref{eq:GenericSaddle5D}. We introduce $\bs{\beta}$ for future convenience,
\begin{equation}
    \bs{\beta} = \bs{\gamma} + \br{\frac{1 - s}{2}} \indicator{N}{2 \nnints} \, \bs{e}_N
    \, ,
    \qquad \qquad
    \bs{w} = \bs{\beta} + \bs{n}
    \, .
\end{equation}
The polynomial part of the closed grand potential then scales as
\begin{multline}
    \rJ_\rp^\text{closed}{\br{R \bsf_s{\br{\bs{a}}} + \ri 2 \pi \bs{w}, \re^{- \ri \frac{\pi}{2} \frac{\indicator{N}{2 \nnints + 1}}{N}} R \Lambda, R}}
    =
    2 \pi N \bs{w}^2 \frac{\log{\br{R}}}{R}
    \\
    + \sbr{\pi \bs{w}^2 \log{\br{\Lambda^{2 N}}}
    + \ri \pi^2 \br{\frac{1}{3} \sum_{\bs{\alpha} \in \Delta_+} \sbr{\bs{\alpha} \cdot \bs{w} - 2 \br{\bs{\alpha} \cdot \bs{w}}^3} + \bs{w}^2 \indicator{N}{2 \nnints + 1}}} \frac{1}{R}
    \\
    - \ri 2 N \br{\bs{w} \cdot \bsf_s{\br{\bs{a}}}} \log{\br{R}}
    + \bigO{R^0}
    \, .
\end{multline}
This is the only part of the total grand potential in \eqref{eq:GenericSaddle5D} with an $\bs{n}$-dependent divergence of order $\log\!{\br{R}}/R$; all other contributions are subleading. Therefore, this part determines which terms in the sum \eqref{eq:GenericSaddle5D} survive.
The $R \to 0$ limit of \eqref{eq:GenericSaddle5D} is then dominated by the minima of the positive definite form
\begin{equation}
    \bs{w}^2 = \left({\bs n}+{\bs \beta}\right)^2
    \, .
\end{equation}
These are precisely the elements of the Weyl orbit of $\bs \beta$ \cite{Grassi:2018bci}, that is we have minima for\footnote{Note that with $\bs{w} = \sum_I w_I \bs{e}_I$ and $\bs{\beta} = \sum_I \beta_I \bs{e}_I$ we have the constraint $\sum_I w_I = \sum_I \beta_I$. Let us now take $\beta_I \in \cbr{\pm 1/2}$ and hence $w_I \in \ints+1/2$ for convenience. One can see that $\bs{w}^2 = \sum_I w_I^2 - \br{\sum_I \beta_I}^2/N$, which gets minimized for $w_I \in \cbr{\pm 1/2}$, and since $\sum_I w_I = \sum_I \beta_I$ we find that the $w_I$'s should be a permutation of the $\beta_I$'s. This means exactly that $\bs{w} \in \mathcal{W}_N \cdot \bs{\beta}$.}
\begin{equation}
    \bs{w} \in \mathcal{W}_N \cdot \bs{\beta} \, .
\end{equation}
For the rest of the closed part, the limit works exactly as outlined in \cite{Grassi:2018bci}.
From the closed 1-loop part we have
\begin{multline}
      \rJ_\text{1-loop}^{\mathrm{closed}}{\br{R \bsf_s{\br{\bs{a}}} + \ri 2 \pi \bs{w}, \re^{- \ri \frac{\pi}{2} \frac{\indicator{N}{2 \nnints + 1}}{N}} R \Lambda, R}} =
     \\
     \ri \frac{\pi^2}{3} \br{\sum_{\bs{\alpha} \in \Delta_+} \bs{\alpha} \cdot \bs{w}} \frac{1}{R} + \ri 2 N \br{\bs{w} \cdot \bsf_s{\br{\bs{a}}}} \log{\br{R}}
     + \bigO{R^0}
     \, ,
\end{multline}
while the instanton part is regular. This means that for the second term in \eqref{eq:tsst}, i.e. for $s=-1$, we get an extra divergent factor which is the exponential of
\begin{multline}
\label{eq:4DLimit:DivergencesClosedSector}
     \left[ \rJ^\text{closed}{\br{R \bsf{\br{\bs{a}}} + \ri 2 \pi \br{\bs{\gamma} + \indicator{N}{2 \nnints} \bs{e}_N}, \re^{- \ri \frac{\pi}{2} \frac{\indicator{N}{2 \nnints + 1}}{N}} R \Lambda, R}} - \right.
    \\
    \left. \rJ^\text{closed}{\br{R \bs{a} + \ri 2 \pi \bs{\gamma}, \re^{- \ri \frac{\pi}{2} \frac{\indicator{N}{2 \nnints + 1}}{N}} R \Lambda, R}} \right]
    =
    - 2 \pi \indicator{N}{2 \nnints} \frac{\log{\br{R}}}{R} - \frac{\pi}{N} \indicator{N}{2 \nnints} \frac{\log{\br{\Lambda^{2 N}}}}{R}
    \\
    + \bigO{R^0} \, .
\end{multline}
The $\bigO{R^0}$ contribution coming from the closed sector can then be obtained directly from \cite[p.~34]{Grassi:2018bci}, and the standard four-dimensional limit becomes
\begin{multline}
    \lim_{R \to 0} \eqref{eq:GenericSaddle5D}
    \propto \sum_{\bs{w}\in \CW_{N}\cdot \bs{\beta}} \frac{\exp{\br{\br{\ri {\partial_{\bs{a}} F_{\rm NS}} - \pi \bs{a} \indicator{N}{2 \nnints + 1}}\cdot\bs{w}}}}{\prod_{\bs{\alpha}\in \Delta_{+}}\br{2\sinh\br{\pi \bs{a}\cdot\bs{\alpha}}}^{(\bs{w}\cdot{\bs \alpha})^2}} \cdot
    \lim_{R \to 0} \re^{\br{\frac{1 - s}{2}} \eqref{eq:4DLimit:DivergencesClosedSector} + \ri \frac{\pi^2}{N} \frac{k^2}{R}}
    \\
    \cdot\lim_{R \to 0} \re^{- s k \pi \br{x - \ri \frac{\pi}{N} \frac{\indicator{N}{2 \nnints + 1}}{R}} +
    \rJ^\text{open}{\bigl(s R x - \ri \frac{\pi}{N} \br{2 k + s \indicator{N}{2 \nnints + 1}}, R \bsf_s{\br{\bs{a}}} + \ri 2 \pi \bs{w}, \re^{- \ri \frac{\pi}{2} \frac{\indicator{N}{2 \nnints + 1}}{N}} R \Lambda, R}\bigr)}
    \, .
\end{multline}
It should be noted that the proportionality constant we are neglecting here is finite and independent of $x$, $s$ and $k$. Hence, it is an overall constant of the eigenfunctions and the same for both saddles. One can compare with \cite[p.~34]{Grassi:2018bci}.

\subsubsection{The open part}

One finds for the open polynomial part:
\begin{multline}
\label{eq:4DLimit:DivergencesOpenSector}
    \hspace{-3mm}
    \ri \frac{\pi^2}{N} \frac{k^2}{R} - s k \pi \br{x - \ri \frac{\pi}{N} \frac{\indicator{N}{2 \nnints + 1}}{R}}
     +
    \rJ_{\mathrm{p}}^\text{open}{\br{s R x - \ri \frac{\pi}{N} \br{2 k + s \indicator{N}{2 \nnints + 1}}, \re^{- \ri \frac{\pi}{2} \frac{\indicator{N}{2 \nnints + 1}}{N}} R \Lambda, R}}
    \\
    \!
    =
    \begin{cases}
        2 \pi k \frac{\log\!{\br{R}}}{R} + 2 \pi \frac{k}{N} \frac{\log\!{\br{\Lambda^N}}}{R} + \ri s N x \log\!{\br{R}} + \ri \frac{\pi}{2} k + \ri s x \log\!{\br{\Lambda^N}} + s \pi \br{1 - 2 k} x &
        \\
        \pi \frac{\log\!{\br{R}}}{R} + \frac{\frac{\pi}{N} \log\!{\br{\Lambda^N}} - \ri \frac{5}{4} \frac{\pi^2}{N}}{R} + \ri s N x \log\!{\br{R}} + \ri \frac{\pi}{4} + \ri s x \log\!{\br{\Lambda^N}} + s \pi \br{1 - k} x  &
    \end{cases}
\end{multline}
up to $\bigO{R}$ corrections, and where the first case is for $N$ even and the second for $N$ odd. We will come back to the divergent terms near the end of the section.

For the open 1-loop part, we need the $b \to 0$ limit of
\begin{multline}
    \Phi_b\br{- b \br{s x - \bs{e}_I \cdot \bsf_s\br{\bs{a}}} + \ri \frac{b}{2} + \ri \br{\bs{e}_I \cdot \bs{w} + \frac{k}{N} + s \frac{\indicator{N}{2 \nnints + 1}}{2 N}} b^{-1}}
    \\
    = \Phi_b\br{- \ri \br{\frac{1}{2} - w_I} b^{-1} + \ri b + \ri \frac{b^{-1}}{2} - \ri b z_I}
    \, ,
\end{multline}
where $b = \sqrt{R / 2 \pi}$ and we used that
\begin{equation}
    \bs{e}_I \cdot \bs{w} - \frac{1}{2} + \frac{k}{N} + s \frac{\indicator{N}{2 \nnints + 1}}{2 N}
    = -\br{\frac{1}{2} - w_I}\,,
    \qquad \quad
    z_I = \frac{1}{2} - \ri \br{s x - \bs{e}_I \cdot \bsf_s\br{\bs{a}}}
    \, ,
\end{equation}
with $\bs{w} = \sum_I w_I \bs{e}_I$. It should be noted that $1/2 - w_I$ is either 0 or 1.
Using the quasi-periodicity of the quantum dilogarithm \eqref{eq:NCQDiLogQuasiPer} under shifts of $\ri b$ and $-\ri b^{-1}$ gives
\begin{multline}
    \Phi_b\br{- \ri \br{\frac{1}{2} - w_I} b^{-1} + \ri b + \ri \frac{b^{-1}}{2} - \ri b z_I}
    \\
    = \frac{\br{1 + \br{\frac{1}{2} - w_I} \exp\br{- \ri 2 \pi z_I}}}{1 - \exp\br{- \ri 2 \pi b^2 \br{z_I - \frac{1}{2}}}} \Phi_b\br{\ri \frac{b^{-1}}{2} - \ri b z_I} \, .
\end{multline}
Expanding around $b \to 0$ and using \cite[eq.~(4.29)]{Francois:2025wwd}, we find
\begin{multline}
    \frac{\br{1 + \br{\frac{1}{2} - w_I} \exp\br{- \ri 2 \pi z_I}}}{1 - \exp\br{- \ri 2 \pi b^2 \br{z_I - \frac{1}{2}}}} \Phi_b\br{\ri \frac{b^{-1}}{2} - \ri b z_I} = \exp\br{- \ri \frac{\pi}{12} b^{-2} + \br{z_I-1} \log\br{2 \pi b^2}}
    \\
    \br{-\frac{\ri}{\sqrt{2 \pi}}} \br{1 + \br{\frac{1}{2} - w_I} \exp\br{- \ri 2 \pi z_I}} \exp\br{\ri \frac{\pi}{2} z_I} \Gamma\br{z_I - \frac{1}{2}} \br{1 + \bigO{b^2}}
\end{multline}
where we also used the integral representations of the quantum dilogarithm \eqref{eq:NonCompQuantDiLog:IntRepsBLimit} and of the log gamma function \eqref{eq:LogGamma}. The leading contribution from the 1-loop part in the four-dimensional limit is then
\begin{multline}
    \lim_{R \to 0} \exp\!{\sbr{\rJ_\text{1-loop}^\text{open}{\br{s R x - \ri \frac{\pi}{N} \br{2 k + s \indicator{N}{2 \nnints + 1}}, R \bsf_s{\br{\bs{a}}} + \ri 2 \pi \bs{w}, R}}}}
    \\
    = \br{\frac{\re^{-\ri \frac{\pi}{4}}}{\sqrt{2 \pi}}}^N \exp\br{- \ri \frac{\pi^2}{6} \frac{N}{R} - N \br{\frac{1}{2} + \ri s x} \log\br{R}} \exp\br{s \frac{\pi}{2} N x}
    \\
    \prod_{I = 1}^N
    \cbr{1 - \br{\frac{1}{2} - w_I} \exp\br{- 2 \pi \sbr{s x - \bs{e}_I \cdot \bsf_s\br{\bs{a}}}}}
    \,
    \Gamma\br{- \ri  \sbr{s x - \bs{e}_I \cdot \bsf_s\br{\bs{a}}}}
\end{multline}
Regarding the instanton part, the contributions from the 5d GV free energy vanish in the 4d limit, while the contributions from the 5d NS free energy lead directly to their 4d counterparts \cite{Iqbal:2003zz,Iqbal:2007ii}. This is exactly as noted in \cite{Grassi:2018bci} for the closed sector and in \cite{Francois:2025wwd} for the open sector in the $N = 2$ case. That is
\begin{multline}
\label{eq:4dinstanton}
    \lim_{R \to 0} \exp\!{\sbr{\rJ_\text{inst}^\text{open}{\br{s R x - \ri \frac{\pi}{N} \br{2 k + s \indicator{N}{2 \nnints + 1}}, R \bsf_s{\br{\bs{a}}} + \ri 2 \pi \bs{w}, \re^{- \ri \frac{\pi}{2} \frac{\indicator{N}{2 \nnints + 1}}{N}} R \Lambda, R}}}}
    \\
    = Z_{\rm D}^{\text{inst}}\br{s x, \bs{f_s}{\br{\bs{a}}}, \Lambda, 1}
    \, ,
\end{multline}
and there are no divergences coming from the instanton part.

Let us briefly come back to the divergent factors.
For $N$ odd, we have no divergences coming from the closed part of the grand potential, and only an overall, $x$-independent divergence coming from the open part. This can be dealt with by introducing an appropriate, overall normalization.
For $N$ even on the other hand, the polynomial part of the closed grand potential gives rise to a divergent constant for the second saddle \eqref{eq:4DLimit:DivergencesClosedSector} that is not present for the first saddle. However, this gets exactly cancelled by the divergences from the polynomial part of the open grand potential \eqref{eq:4DLimit:DivergencesOpenSector}, leaving only an $x$-independent, overall divergent factor. This can again be dealt with by an appropriate overall normalization.

\subsubsection{The result of the four-dimensional limit}

Hence, up to an $x$-independent, divergent overall factor, we obtain for both saddles
\begin{equation}
\label{eq:4DLimit_Result}
    \re^{\ri \frac{\pi}{2} k \indicator{N}{2 \nnints}}
    \re^{- s \pi \br{1 + \indicator{N}{2 \nnints}} k x} Z_{D}\br{s x, \bsf_s{\br{\bs{a}}}, \Lambda, 1}
    \sum_{\bs{w} \in \mathcal{W}_N \cdot \bs{\beta}} \CP_{\bs{w}}\br{s x, \bsf_s{\br{\bs{a}}}, \Lambda, 1}
\end{equation}
where we have $s = 1$ and $k = k_x = k_y = 0$ for the first saddle, while $s = -1$ and $k = k_x = k_y = 1$ for the second saddle. We used the expressions\footnote{Note that $\CP_{\bs{w}}$ in \eqref{eq:SpecialFactorFrom5D} is indeed equal to \eqref{eq:Hndef}, since $1/2-w_I$ is either $0$ or $1$.}
\begin{equation}
    Z_{D}\br{x, \bs{a}, \Lambda, 1} =
    \re^{\ri \log\br{\Lambda^N} x + \pi \br{\frac{N}{2} + 1} x}
    \br{\prod_{I = 1}^N \Gamma{\br{\ri \sbr{\bs{e}_I \cdot \bs{a} - x}}}}
    Z_{\rm D}^{\text{inst}}{\br{ x, \bs{a}, \Lambda, 1}}
    \, ,
\end{equation}
\begin{equation}
\label{eq:SpecialFactorFrom5D}
    \CP_{\bs{w}}\br{x, \bs{a}, \Lambda, 1} = \frac{\re^{ \ri \partial_{\bs{a}} F_\mathrm{NS} \cdot \bs{w} - \pi \bs{a} \cdot \bs{w} \indicator{N}{2 \nnints + 1}}}{\prod_{\bs{\alpha} \in \Delta_+} \br{2 \sinh\br{\pi \bs{a} \cdot \bs{\alpha}}}^{\br{\bs{w} \cdot \bs{\alpha}}^2}}
    \prod_{I = 1}^N  \br{1 - \br{\frac{1}{2} - w_I} \re^{2 \pi \br{\bs{e}_I \cdot \bs{a} - x}}}
    \, ,
\end{equation}
\begin{equation}
\label{eq:beta_gamma}
    \bs{\beta} = \bs{\gamma} + \br{\frac{1 - s}{2}} \indicator{N}{2 \nnints} \, \bs{e}_N
    \, .
\end{equation}
The number of non-trivial factors in \eqref{eq:SpecialFactorFrom5D} is $N/2$ for $s = 1$ and $(N-2)/2$ for $s = -1$ when $N$ is even, and $(N-1)/2$ when $N$ is odd. This can be verified directly from the expression of $\bs{\beta}$ in \eqref{eq:beta_gamma}. Let us end with the comment that
\begin{equation}
    \br{\frac{1}{2} - w_I} = \frac{1}{2} - \bs{e}_I \cdot \bs{w} - \br{\frac{k}{N} + s \frac{\indicator{N}{2 \nnints + 1}}{2 N}}
    = \frac{1}{2} - \bs{e}_I \cdot \bs{w} - \br{\frac{1 - s \indicator{N}{2 \nnints}}{2 N}}
\end{equation}
where the latter two expressions are coordinate invariant and independent of the choice $w_I \in \cbr{\pm 1/2}$, and they are always 0 or 1.

%%%%%%%%%%%%%%%%%%%%%%%%%%%%%%%%%%%%%%%%%%%%%%%%%%%%%%%%%%%%%%%%%%%%%%%%%%%%%%%%%%%%
%%%%%%%%%%%%%%%%%%%%%%%%%%%%%%%%%%%%%%%%%%%%%%%%%%%%%%%%%%%%%%%%%%%%%%%%%%%%%%%%%%%%

\section{Conclusion and outlook}

In this work, we investigated the open TS/ST correspondence for the family of $Y^{N,0}$ toric Calabi–Yau geometries. Following the general framework of \cite{Marino:2016rsq,Marino:2017gyg,Francois:2025wwd}, we constructed a background-independent quantity, given in equation \eqref{eq:tsst}, which plays the role of the non-perturbative open topological string partition function. This object is entire in both the open and closed string moduli and provides an exact analytic solution to the difference equation obtained by quantizing the mirror curve of the $Y^{N,0}$ geometry. The resulting wavefunction becomes square-integrable only at discrete energy values, where it reproduces the physical eigenfunctions of the associated operator. We then analysed the four-dimensional limit of this construction. In this limit, the quantum mirror curve \eqref{eq:diff} reduces to the deformed quantum-mechanical Hamiltonian \eqref{eq:hamil}, which can be regarded as a deformation of the standard Schr\"odinger operator with an arbitrary polynomial potential. Through the open TS/ST correspondence, we derived explicit analytic expressions for the corresponding eigenfunctions, given in equations \eqref{eq:eigenfeven} and \eqref{eq:eigenfodd}. These functions are entire in $x$ for arbitrary values of the moduli $h_k$ and become $L^2$-normalizable only for a discrete set of values of the energy, in perfect agreement with the quantization conditions previously derived in \cite{Grassi:2018bci}.
Furthermore, starting from the explicit eigenfunctions \eqref{eq:eigenfeven} and \eqref{eq:eigenfodd}, we show how to construct analytic solutions for related spectral problems, including the case of the inverted potential and the deformation in which the kinetic term \(2\cosh(p)\) is replaced by a \(\sinh(p)\)-type operator.

Our analysis opens up several directions for future investigation, some of which we briefly outline below.

\begin{itemize}

    \item[-] Within the open TS/ST correspondence, the structure of the non-perturbative open partition function still poses some interesting challenges. As shown in \cite{Marino:2016rsq,Marino:2017gyg,Francois:2025wwd} and summarized in equation \eqref{eq:tsst}, this function can be expressed as a sum of two contributions, or saddles. Although in some cases we have a technical prescription to construct the second saddle from the first, its geometrical and physical meaning remains elusive and calls for a deeper understanding.

    \item[-] It would be interesting to investigate how our solutions \eqref{eq:tsst}, \eqref{eq:eigenfeven} and \eqref{eq:eigenfodd} emerge from a resurgent analysis of the open-string wavefunction and an exact WKB analysis; see for instance \cite{DM68a, DM68b, Kashani-Poor:2016edc, Grassi:2022zuk, Alim:2022oll, DelMonte:2024dcr, Hao:2025azt}.

    \item[-] The deformed Hamiltonian \eqref{eq:hamil} exhibits several novel spectral features absent in standard Schrödinger operators. Due to its difference-equation nature, the usual oscillation theorem \cite[p.~66, thm.~3.5]{BS91} does not apply. Moreover, the system can display ground-state degeneracies and, remarkably, bound states with real energies even in potentials unbounded from below. These phenomena reveal a rich and unconventional spectral structure whose physical and mathematical interpretation deserves further study. 

    \item[-] It is important to undertake a rigorous study of the spectral properties of the operators considered in this work. The case with a $\cosh(p)$ kinetic term and a confining potential was analysed in \cite{LST19}, where suitable self-adjoint realizations with purely discrete spectra were established.\footnote{{See also \cite{ZILS25} for extensions to certain non-self-adjoint operators with complex potentials.}} However, many of the spectral problems discussed here fall outside this framework, including operators with potentials unbounded from below (see \autoref{sec:spectp} and \autoref{sec:inverted}), as well as those with a $\sinh(p)$ kinetic term (see \autoref{sec:sinh}). While similar questions have been addressed for standard Schr\"odinger operators with unbounded potentials (see \cite{Caliceti, Caliceti1983}), an analogous understanding for the finite-difference operators studied here is still missing, and questions regarding the natural choice of operator domains and the characterization of the spectra remain open. The explicit eigenfunctions obtained in this work may provide a possible starting point.

    \item [-] It was pointed out in \cite{bgt, bgt2} that there exists another four-dimensional limit, the so-called dual 4d limit, which can be implemented on the mirror curve \eqref{eq:5dmirror_kappas} and leads to a family of integral operators. The corresponding spectral problems make contact with the 4D gauge theory in the self-dual phase of the $\Omega$-background, rather than the NS-phase, as is the case for the spectral problems studied in this paper. For \(N = 2\), this limit on the eigenfunctions was analysed in \cite{Francois:2023trm, Francois:2025wwd}, where a new functional relation between the modified Mathieu and the McCoy-Tracy-Wu operators was found. It would be interesting to extend this analysis to \(N > 2\) in light of our new results on the eigenfunction.

    \item [-] Finally, it would be interesting to generalize our construction of the eigenfunctions of the deformed Hamiltonian \eqref{eq:hamil} to other four-dimensional quantum mirror curves, including those associated with SW theories with matter or with different gauge groups. Such extensions could reveal new classes of solvable quantum systems and provide further insights into the open TS/ST correspondence.

\end{itemize}
We hope to report on some of these problems in the future.

\begin{comment}
\paragraph{Rights and permissions.}

This article is licensed under a Creative Commons Attribution 4.0 International Licence. You are free to use, share, copy, redistribute, reproduce, adapt, remix, transform, and build upon the material in any medium or format for any purpose, under the condition that you give appropriate credit to the original authors and the source, provide a link to the licence, and indicate if changes were made.
%You may do so in any reasonable manner, but not in any way that suggests the licensor endorses you or your use.
%You may not apply legal terms or technological measures that legally restrict others from doing anything the licence permits. The licensor cannot revoke the freedoms specified in the licence as long as you follow the licence terms. You do not have to comply with the licence for elements of the material in the public domain or where your use is permitted by an applicable exception or limitation.
%The licence may not give you all the permissions necessary for your intended use. For example, other rights such as publicity, privacy, or moral rights may limit how you use the material.
To view a copy of this licence, visit \url{https://creativecommons.org/licenses/by/4.0/}.

\doclicenseThis
\end{comment}

\clearpage

%%%%%%%%%%%%%%%%%%%%%%%%%%%%%%%%%%%%%%%%%%%%%%%%%%%%%%%%%%%%%%%%%%%%%%%%%%%%%%%%%%%%
%%%%%%%%%%%%%%%%%%%%%%%%%%%%%%%%%%%%%%%%%%%%%%%%%%%%%%%%%%%%%%%%%%%%%%%%%%%%%%%%%%%%

\appendix

\section{Conventions for the \texorpdfstring{SU($N$)}{SU(N)} root system}
\label{app:GT}

Let us introduce some objects from the Lie algebra $\mathfrak{su}\br{N}$:
\begin{equation}\label{eq:latticedef}
    \begin{cases}
        \cbr{\bs{b}_I}_{I \in \cbr{1, \cdots, N}}  & \text{standard euclidean orthonormal basis;}
        \\
        \cbr{\bs{e}_I}_{I \in \cbr{1, \cdots, N}}  & \text{weights of the fundamental representation;}
        \\
        \cbr{\bs{\lambda}_k}_{k \in \cbr{1, \cdots, N-1}} & \text{fundamental weights, basis of the weight lattice;}
        \\
        \cbr{\bs{\alpha}_\ell}_{\ell \in \cbr{1, \cdots, N-1}}  & \text{simple roots, dual to the fundamental weights;}
    \end{cases}
\end{equation}
These quantities are related to each other by:
\begin{subequations}
\begin{align}
&\bs{e}_I = \bs{b}_I - \frac{1}{N} \sum_{J = 1}^N \bs{b}_J
    \, ,
    \qquad
    \bs{\alpha}_k = \bs{e}_k - \bs{e}_{k+1} = \bs{b}_k - \bs{b}_{k+1}
    \, ,
    \qquad
    \bs{\lambda}_k = \sum_{I = 1}^k \bs{e}_I
    \, ,\\
&\bs{b}_I \cdot \bs{b}_J = \delta_{I, J}
     \, ,
     \qquad \qquad
     \bs{e}_I \cdot \bs{e}_J = \delta_{I, J} - \frac{1}{N}
     \, ,\\
&\bs{\alpha}_k \cdot \bs{\alpha}_l = C_{k,\ell} = - \delta_{k,\ell-1} + 2 \delta_{k,\ell} - \delta_{k,\ell+1}
     \, ,
     \qquad \qquad
     \bs{\lambda}_k \cdot \bs{\alpha}_l = \delta_{k, \ell}
     \, ,
\end{align}
\end{subequations}
where $C_{k, \ell}$ is the SU($N$) Cartan matrix. The set of positive roots is given by:
\be \label{eq:positroot}\Delta^+=\left\{\bs{\alpha}_{k, \ell} = \bs{e}_k - \bs{e}_\ell\, \Big| \,  1 \leqslant k < \ell \leqslant N \right\},\ee
while the root lattice is denoted by:
\be \label{eq:defrootlat}Q_{N-1} = \cbr{\sum_{\ell=1}^{N-1} m_\ell \bs{\alpha}_\ell \,\bigg|\, m_\ell \in \mathbb{Z}} = \cbr{\sum_{I=1}^{N} m_I \bs{e}_I \, \bigg| \, m_I \in \mathbb{Z} \land \sum_{I=1}^{N} m_I = 0} \, .\ee
In our construction of the eigenfunctions, we also need
\begin{equation}
    \bs{\gamma}
    = \frac{1}{2} \sum_{I=1}^N (-1)^{I-1} \bs{e}_I
    = \sum_{k = 1}^{N-1}(-1)^{k-1} \bs{\lambda}_k
    = \frac{1}{2} \sum_{\ell = 1}^{N-1} \br{\indicator{\ell}{2 \nnints + 1} - \frac{\ell}{N} \indicator{N}{2 \nnints + 1}} \pmb{\alpha}_\ell
    \, .
\end{equation}

\section{Faddeev's non-compact quantum dilogarithm}
\label{sec:faddev}

The defining representation of the quantum dilogarithm is often taken to be \cite[eq.~(42)]{EllegaardAndersen:2011vps}
\begin{equation}
    {\Phi_b\br{z}} =
    \exp\br{\frac{1}{4} \int_{\reals + \ri 0} \frac{\re^{- \ri 2 z u}}{{\sinh\br{b u}} {\sinh\br{b^{-1}u}}} \frac{\rd u}{u}}
    \, ,
    \qquad \quad
    2 \abs{\imaginary{z}} < \abs{\real{b + b^{-1}}}
    \, .
\end{equation}
It can be analytically continued to a meromorphic function of $z$ on the whole complex plane with poles and roots at \cite[eq.~(45)]{EllegaardAndersen:2011vps}
\begin{equation}
    z =
    \begin{cases}
        + \ri \sbr{\br{k + \frac{1}{2}} b + \br{\ell + \frac{1}{2}} b^{-1}} & \text{poles}
        \\
        - \ri \sbr{\br{k + \frac{1}{2}} b + \br{\ell + \frac{1}{2}} b^{-1}} & \text{roots}
    \end{cases}
    \, ,
    \qquad \qquad
    k , \ell \in \mathbb{N} \, ,
\end{equation}
and with an essential singularity at complex infinity \cite[p.~34]{EllegaardAndersen:2011vps}. One can determine the order of the poles and roots when $b^2 \in \prationals$, based on \cite[eq.~(21)]{Garoufalidis:2014ifa}, see \cite[eq.~(A.10)]{Francois:2025wwd}.
The parameter $b$ is in general such that $b^2 \in \mathbb{C} \setminus \mathbb{R}_{\leqslant 0}$, but we are mostly interested in $b^2 > 0$, since this corresponds to $\hbar, g_s > 0$. The quantum dilogarithm satisfies the following quasi-periodicity relations in $z$ \cite[eq.~(48)]{EllegaardAndersen:2011vps} \cite[eq.~(77)]{Garoufalidis:2014ifa},
\begin{equation}
\label{eq:NCQDiLogQuasiPer}
    \Phi_b(z + s \ri b^{\pm}) = \left( 1 + \re^{s \ri \pi b^{\pm 2}} \re^{2 \pi b^{\pm} z} \right)^{-s} \Phi_b(z)
    \qquad \qquad
    s \in \cbr{-1, +1}
    \, .
\end{equation}

When taking the 4d limits the following representation of Faddeev's quantum dilogarithm comes in useful \cite[eqs.~(3.2)-(3.8)]{Hatsuda:2015owa}\footnote{There appears to be a constant term missing in \cite[eqs.~(3.2)-(3.8)]{Hatsuda:2015owa}.}
\begin{equation}
\label{eq:NonCompQuantDiLog:IntRepsBLimit}
    \log\Phi_b\left( z \right) = - \frac{\ri}{2 \pi b^2} \mathrm{Li_2}\left( - \re^{2 \pi b z} \right) - \ri \int_0^{+\infty} \frac{\rd u}{1 + \re^{2 \pi u}} \log \left( \frac{1 + \re^{2 \pi b z - 2 \pi b^2 u}}{1 + \re^{2 \pi b z + 2 \pi b^2 u}} \right)
    \, ,
\end{equation}
under the condition that $2 \abs{\imaginary{z}} < b^{-1}$ when $b > 0$.\footnote{One can note from the behaviour under shifts of $z$ by $\ri b^{-1}$ that the representation above has a limited domain of validity: the quantum dilogarithm is quasi-periodic under such shifts while the same shifts act trivially on the right-hand side above. Some numerical checks seem to suggest that the representation above is only valid for $ 2 \abs{\imaginary{z}} < b^{-1}$ when $\mathrm{Re}(z) \geqslant 0$.}
In taking the standard or dual 4d limit on the integral representation above, one may use the following integral representation of the log gamma function \cite[p.~8]{adamchik2003contributions},
\begin{equation}
\label{eq:LogGamma}
    \log{\Gamma{\br{z + \frac{1}{2}}}} = \frac{\log \br{2 \pi}}{2} - z + z \log \br{z} - 2 \int_{0}^{+ \infty} \frac{ \rd u}{1 + \re^{2 \pi u}} \arctan\!{\br{\frac{u}{z}}}
    \, ,
    \qquad
    \real{z} > 0
    \, ,
\end{equation}
which can be obtained from Binet's second formula for the log gamma function.

\section{Special functions from gauge theory}
\label{sec:sf}

The Nekrasov–Shatashvili (NS) functions capture physical information about supersymmetric gauge theories in the NS limit of the gravitational $\Omega$-background, with the case relevant here being four-dimensional $\mathcal N=2$, SU($N$) SYM. It is convenient to organize the Coulomb branch parameters $a_I$ into the vector\footnote{The $a_I$ parameters here are denoted by $\alpha_I$ in \cite{Grassi:2018bci}. } \be
    {\bs a}=\sum_{I=1}^N a_I {\bs e}_I,
    \qquad \sum_{I=1}^N a_I=0 \, ,
\ee
where ${\bs e}_I$ are the weights of the fundamental representation of SU($N$) (see \autoref{app:GT} for our conventions).

Note on convention: when dealing with Nekrasov functions, the term “instanton partition function” refers to the instantons of the supersymmetric gauge theory that appear in localization computations \cite{Moore:1997dj,Lossev:1997bz,Nekrasov:2002qd,Flume:2002az,Bruzzo:2002xf}. These should not be confused with quantum-mechanical instantons.

\subsection{The NS free energy}
Let ${\bs Y}=(Y_1,\dots,Y_N)$ denote an $N$-tuple of Young diagrams, each one being equivalently specified by an ordered partition $Y=\br{y_1, y_2,\dots}$, $y_1\geqslant y_2 \geqslant\dots \geqslant0$. The total size of the $N$-tuple is given by
\begin{equation}
    \ell({\bs Y})=\sum_{I=1}^N |Y_I|, \qquad \left| Y \right| = \sum_{i \geqslant 1}y_i.
\end{equation}
For a box $s=(i,j)\in Y$, we define the arm and leg lengths as
\begin{equation}
A_Y(s)=y_i-j,
\qquad
L_Y(s)=y^T_j-i,
\end{equation}
where $Y^T=\br{y_1^T,y_2^T,\dots}$ denotes the transposed partition. Let $Y,W$ be two Young diagrams, we introduce the building block
\begin{equation}
    N_{Y,W}(z;\epsilon_1,\epsilon_2)
    = \prod_{s\in Y}\!\left(z - \epsilon_1 L_W(s)+\epsilon_2\big(A_Y(s)+1\big)\right)
    \prod_{t\in W}\!\left(z + \epsilon_1\big(L_Y(t)+1\big)-\epsilon_2 A_W(t)\right).
\end{equation}
The 4d instanton Nekrasov partition function for this gauge theory is given by \cite{Moore:1997dj,Lossev:1997bz,Nekrasov:2002qd,Flume:2002az,Bruzzo:2002xf}
\begin{equation}
\label{eq:inst}
    Z^{\rm inst}({\bs a}, \Lambda, \epsilon_1,\epsilon_2)
    = \sum_{{\bs Y}} \big( (-1)^N \Lambda^{2N} \big)^{\ell({\bs Y})}\, \mathcal{Z}_{{\bs Y}} ({\bs a}, \epsilon_1,\epsilon_2) \, ,
\end{equation}
with
\be
\mathcal{Z}_{\bs Y}({\bs a}, \epsilon_1,\epsilon_2)
= \prod_{I, J=1}^N \frac{1}{N_{Y_I,Y_J}(a_I-a_J; \epsilon_1, \epsilon_2)} \, .
\ee
The instanton part of the NS free energy is then defined as
\be
\label{fns4d}
F_{\rm NS}^{\rm inst}({\bs a}, \Lambda, \hbar)
= \ri \hbar \, \lim_{\epsilon_2 \rightarrow 0} \,
   \epsilon_2 \log Z^{\rm inst}({\bs a}, \Lambda, \ri \hbar, \epsilon_2).
\ee
The first few orders in the $\Lambda$ expansion of \eqref{fns4d} for N = 2, 3 and 4 are provided in the accompanying file \texttt{ancillary\char`_ file.nb}. Note that the functions \eqref{eq:inst} and \eqref{fns4d} are expressed as series in $\Lambda$. These series are, however, convergent; see, e.g. \cite{ilt,Arnaudo:2022ivo,Desiraju:2024fmo}.

\subsection{The NS partition function in presence of a surface defect}

In the presence of a type-II chiral defect, the partition function was computed in \cite{Gaiotto:2014ina,Bullimore:2014awa} and is nicely summarized in \cite[app.~A]{Sciarappa:2017hds}. By evaluating the residues in these expressions explicitly, the partition function can be written as
\be
Z^{(c),\mathrm{inst}}_{2d/4d}(x,{\bs a}, \Lambda, \epsilon_1,\epsilon_2)
= \sum_{{\bs Y}} \big( (-1)^N \Lambda^{2N} \big)^{\ell({\bs Y})}\,
   \mathcal{Z}_{{\bs Y}} ({\bs a}, \epsilon_1,\epsilon_2)\,
   \mathcal{Z}_{{\bs Y}}^{(c)} (x,{\bs a},\epsilon_1,\epsilon_2) \, ,
\ee
where the defect factor is
\begin{equation}
    \mathcal{Z}_{{\bs Y}}^{(c)} (x,{\bs a},\epsilon_1,\epsilon_2)
    = \prod_{I=1}^N \prod_{(i,j)\in Y_I}
    \frac{  a_I-x + i\,\epsilon_1 + j\,\epsilon_2}{  a_I-x+ i\,\epsilon_1 + (j-1)\epsilon_2}.
\end{equation}
Here $x$ denotes the two-dimensional twisted mass parameter. We then define the full instanton defect partition function in the NS limit as
\be
\label{eq:Zdinst}
Z_{\rm D}^{\rm inst} (x,{\bs a}, \Lambda, \hbar)
= \lim_{\epsilon_2\to 0}
  \frac{Z^{(c),\mathrm{inst}}_{2d/4d}(x,{\bs a}, \Lambda, \ri \hbar,\epsilon_2)}
       {Z^{\rm inst}({\bs a}, \Lambda, \ri \hbar,\epsilon_2)} \, .
\ee
The first few orders in the $\Lambda$ expansion of \eqref{eq:Zdinst} for $N = 2$, $3$ and $4$ are provided in the accompanying file \texttt{ancillary\char`_ file.nb}.

\subsection{The Wilson loops}\label{sec:wl}

Another ingredient we need is the four-dimensional limit of Wilson-loop expectation values \cite{Gaiotto:2014ina,Bullimore:2014awa}; here we closely follow \cite{Grassi:2018bci}. As before, let ${\boldsymbol Y}=(Y_1,\dots,Y_N)$ denote an $N$-tuple of Young diagrams. We define the equivariant Chern character as
\begin{equation}
\label{cher-char}
     {\rm Ch}_{\boldsymbol{Y}}({\bs a}, \epsilon_1,\epsilon_2) = \CW- (1-\re^{R\epsilon_1})(1-\re^{R \epsilon_2}) \CV_{\boldsymbol{Y}} \, ,
\end{equation}
\begin{equation}
     \CW = \sum_{I=1}^N \re^{ R a_I} \, ,
    \qquad \qquad
    \CV_{\boldsymbol{Y}} =\sum_{I=1}^N \re^{R a_I} \sum_{(k, l)\in Y_I} \re^{(k-1) R \epsilon_1 + (l-1) R\epsilon_2} \, .
\end{equation}
In the $R\to 0$ limit, the Chern character behaves as
\be
{\rm Ch}_{\boldsymbol{Y}}({\bs a}, \epsilon_1,\epsilon_2)=N+\sum_{k \geqslant 2}
   \mathcal{C}^{(k)}_{\boldsymbol{Y}}({\bs a}, \epsilon_1,\epsilon_2)\,R^k,
\ee
where, for example:
\begin{equation}
\begin{aligned}
\mathcal{C}^{(2)}_{\bs{Y}}(\bs{a},\epsilon_1,\epsilon_2)&=\frac{1}{2}\sum_{I=1}^N a_I^2-\epsilon_1\epsilon_2\ell\br{\bs{Y}},\\
\mathcal{C}^{(3)}_{\bs{Y}}(\bs{a},\epsilon_1,\epsilon_2)&=\frac{1}{6}\sum_{I=1}^N a_I^3-\epsilon_1\epsilon_2\left(\frac{\epsilon_1+\epsilon_2}{2}\ell\br{\bs{Y}}+\epsilon_1\sum_{I=1}^{N}c_2\br{Y_I^T}\right. \\
& \left. \quad +\epsilon_2\sum_{I=1}^N c_2\br{Y_I}+\sum_{I=1}^N a_I\left|Y_I \right| \right),
\end{aligned}
\end{equation}
while
\begin{equation}
    c_2\br{Y}=\frac{1}{2}\sum_{i\geqslant1}y_i\br{y_i-1}.
\end{equation}
We define the “4d Wilson loops'' as
\be
{\rm W}^{(k)}({\bs a}, \Lambda, \epsilon_1,\epsilon_2)
= \frac{1}{Z^{\rm inst}} \sum_{\boldsymbol{Y}}
((-1)^N \Lambda^{2 N})^{\ell (\boldsymbol{Y})}\,
  \mathcal{C}^{(k)}_{\boldsymbol{Y}}({\bs a}, \epsilon_1,\epsilon_2)\,
  Z_{\boldsymbol{Y}}({\bs a}, \epsilon_1,\epsilon_2) \, .
\ee
Here we are interested in their NS limit, which we denote by
\be \label{eq:wsNs}{\rm W}^{(k)}_{\rm NS}({\bs a}, \Lambda, \hbar)=\lim_{\epsilon_2\to 0} {\rm W}^{(k)}({\bs a}, \Lambda, \ri\hbar, \epsilon_2),\quad k=1, \dots N-1.\ee
These observables are directly related to the complex moduli parameters $h_k$ appearing in the mirror curve, and they generalise the Matone relation to higher rank gauge theories. More precisely, we have
\be \label{eq:genmat}\boxed{h_j=\sum_{\bs k} \frac{(-1)^{|{\bs k}|-\ell({\bs k})}}{z_{\bs k}} {\rm W}_{\rm NS}^{\bs {k}}({\bs a}, \Lambda, \hbar), \qquad j \in \cbr{2, \cdots, N}} \ee
where $\bs{k}=(0, k_2,\cdots)$ satisfies
\be\ell(\bs{k})= \sum_{m \geqslant 2 } m k_{m}= j \, , \ee
and we used
\be \ba
&{\rm W}_{\rm NS}^{\bs {k}}({\bs a}, \Lambda, \hbar)=\prod_{j\geqslant 1}(j!{\rm W}_{\rm NS}^{(j)}({\bs a}, \Lambda, \hbar))^{k_j} \, ,\quad z_{\bs k}= \prod_{j\geqslant 1} k_j! j^{k_j}\, , \quad
 |\bs{k}|=\sum_m k_m\, .
 \ea\ee
Hence, for instance:\footnote{We omit the dependence on $(\bs{a}, \Lambda, \hbar)$ in $W_{\rm NS}^{(2)}$ for brevity.}
\be\ba
h_2 &= -\,{\rm W}_{\rm NS}^{(2)}, \\
h_3 &= 2 {\rm W}_{\rm NS}^{(3)}, \\
h_4 &= \tfrac{1}{2} \bigl({\rm W}_{\rm NS}^{(2)}\bigr)^{2} - 6 {\rm W}_{\rm NS}^{(4)}, \\
h_5 &= -2\, {\rm W}_{\rm NS}^{(2)} {\rm W}_{\rm NS}^{(3)} + 24 {\rm W}_{\rm NS}^{(5)}, \\
h_6 &= -\tfrac{1}{6} \bigl({\rm W}_{\rm NS}^{(2)}\bigr)^{3} + 6 {\rm W}_{\rm NS}^{(2)} {\rm W}_{\rm NS}^{(4)} + 2 \bigl({\rm W}_{\rm NS}^{(3)}\bigr)^{2} - 120 {\rm W}_{\rm NS}^{(6)}.
\ea\ee
The first few orders in the $\Lambda$ expansion of $h_j$ for $N = 2$, $3$ and $4$ are provided in the accompanying file \texttt{ancillary\char`_ file.nb}. One can further check that
\begin{equation}
    {\rm W}^{(2)}_{\rm NS} = -\frac{1}{2 N}\Lambda \partial_\Lambda F_{\rm NS} \, ,
\end{equation}
giving rise to the usual quantum Matone relation \cite{matone,Flume:2004rp}.

\section{Figures and numerical evidence}
\label{sec:figs}

In this appendix, we present additional numerical evidence supporting our results. We focus on the cubic  and quartic potentials, displaying, for various choices of parameters, the ground, first, and second excited states for the cubic case, the first and second excited states for the quartic case.

\begin{figure}[ht]
    \centering
    \includegraphics[width=0.91\textwidth]{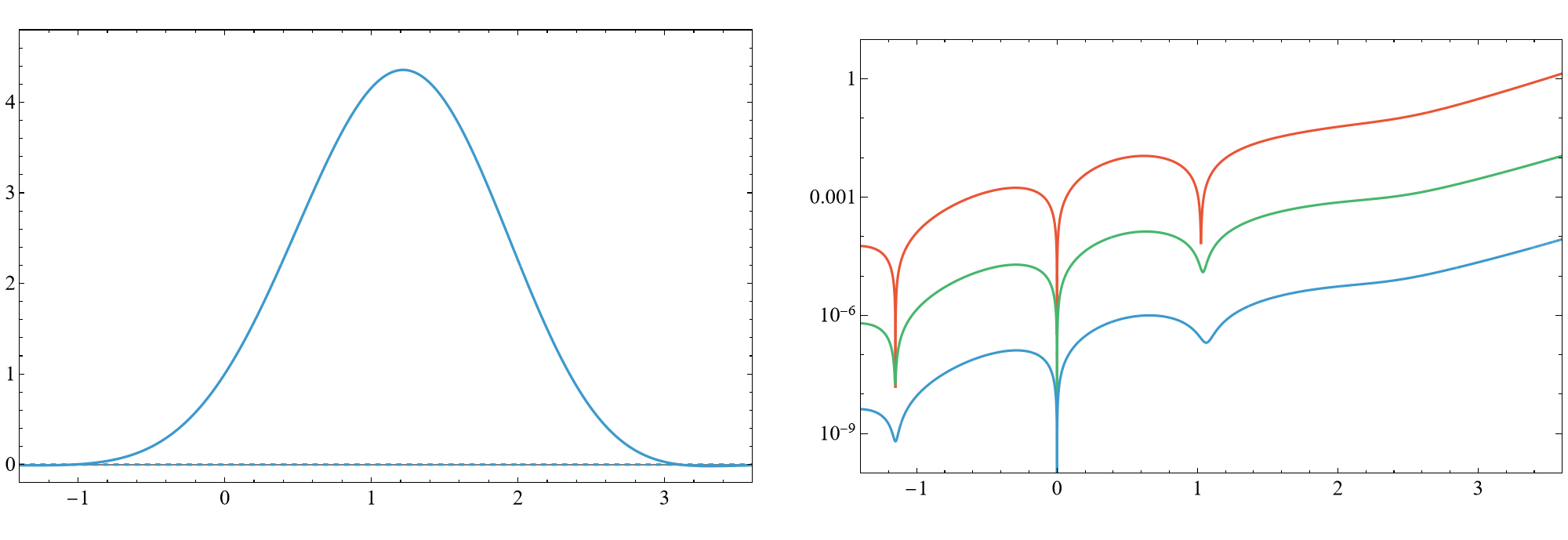}
    \caption{Left: ground state of the $V_3(x)$ potential with $h_2=-5$, $\hbar=1$ and $\Lambda=(3/5)^{1/2}$; the corresponding complex energy is $E_0 \approx -1.855383 + \ri\,5.697572\cdot10^{-9} $. Dashed lines denote the imaginary part of the eigenfunction, while solid lines denote the real part. Right: difference between the numerical eigenfunction and the analytic expression from \eqref{eq:eigenfodd}. The coloured curves show the effect of including an increasing number of terms in the $\Lambda$-expansion of the eigenfunction: red (0 terms), green (1 term), blue (2 terms).}
    \label{fig:su3_ground}
\end{figure}

\begin{figure}[ht]
    \centering
    \includegraphics[width=0.91\textwidth]{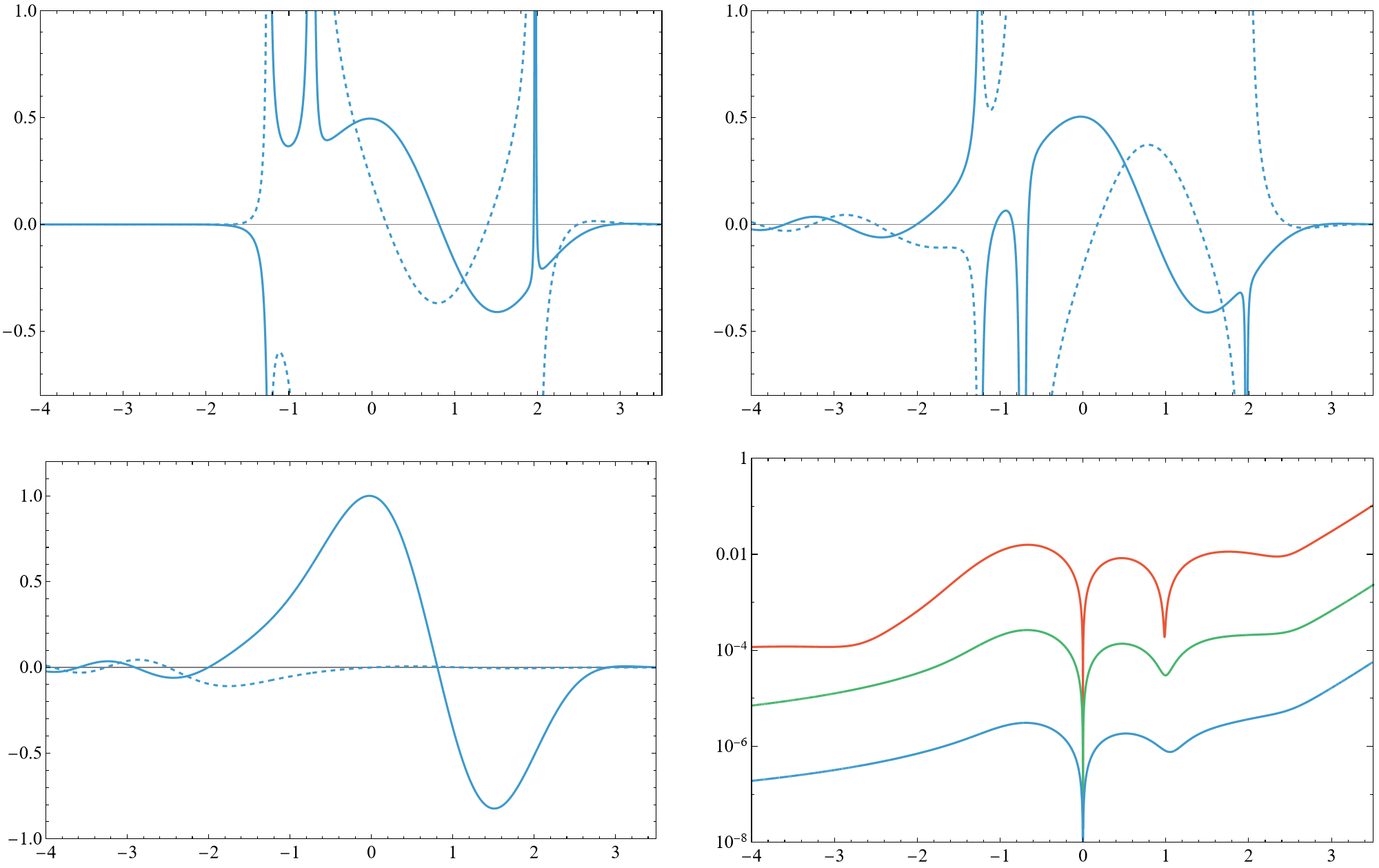}
    \caption{The first excited state of the $V_3(x)$ potential with $h_2 = -3$, $\hbar = 1$, and $\Lambda = \tfrac{2}{3}$; the corresponding complex energy is $E_1\approx 1.758084 + \ri\,0.010614 $. Dashed lines denote the imaginary part of the eigenfunction, while solid lines denote the real part. Top panels: first (left) and second (right) saddle contributions in the analytic expression \eqref{eq:eigenfodd}, evaluated on-shell. The two functions develop poles individually, which cancel in the sum. Bottom left: full on-shell eigenfunction from \eqref{eq:eigenfodd}. Bottom right: difference between the numerical eigenfunction and the analytic result from \eqref{eq:eigenfodd}. The coloured curves show the effect of including an increasing number of terms in the $\Lambda$-expansion of the eigenfunction: red (0 terms), green (1 term), blue (2 terms).}
    \label{fig:su3_1st}
\end{figure}

\begin{figure}[ht]
    \centering
    \includegraphics[width=0.91\textwidth]{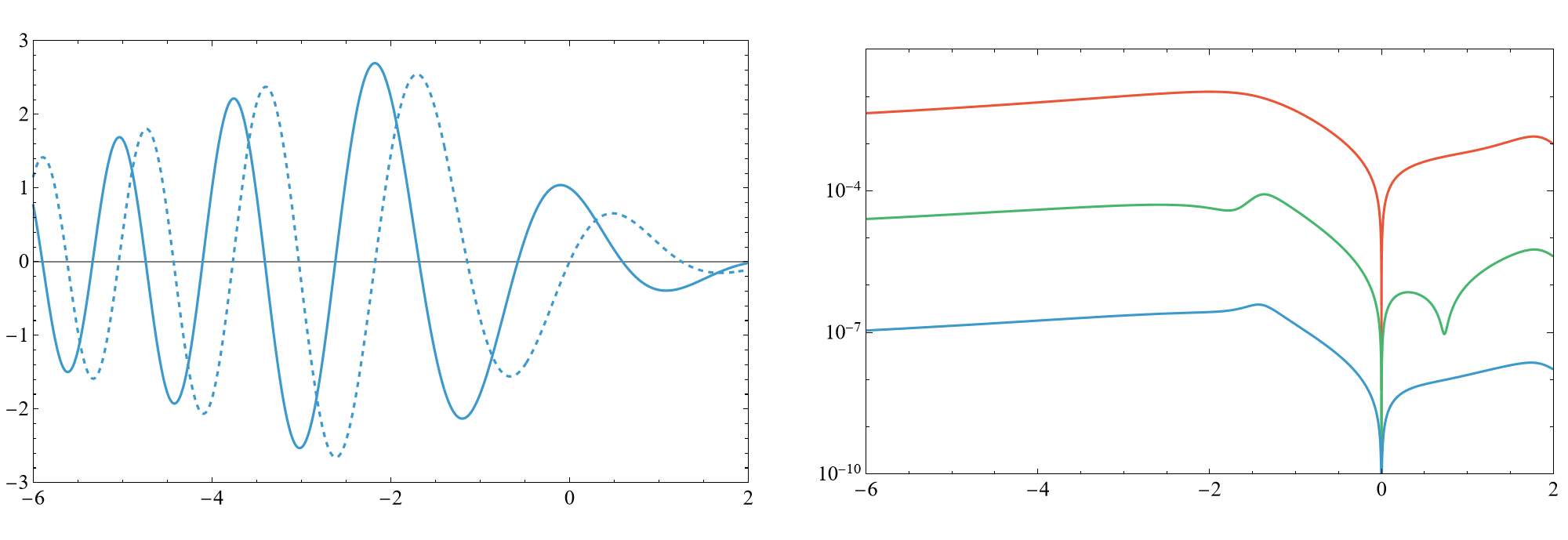}
    \caption{Left: second excited state of the $V_3(x)$ potential with $h_2 = 2$, $\hbar = 1$ and $\Lambda = \tfrac{9}{10}$; the corresponding complex energy is $E_2\approx 7.842147 + \ri\,8.389762 $. Dashed lines denote the imaginary part of the eigenfunction, while solid lines denote the real part. Right: difference between the numerical eigenfunction and the analytic expression from \eqref{eq:eigenfodd}. The coloured curves show the effect of including an increasing number of terms in the $\Lambda$-expansion of the eigenfunction: red (0 terms), green (1 term), blue (2 terms).}
    \label{fig:su3_2nd}
\end{figure}

\begin{figure}[ht]
    \centering
    \includegraphics[width=0.91\textwidth]{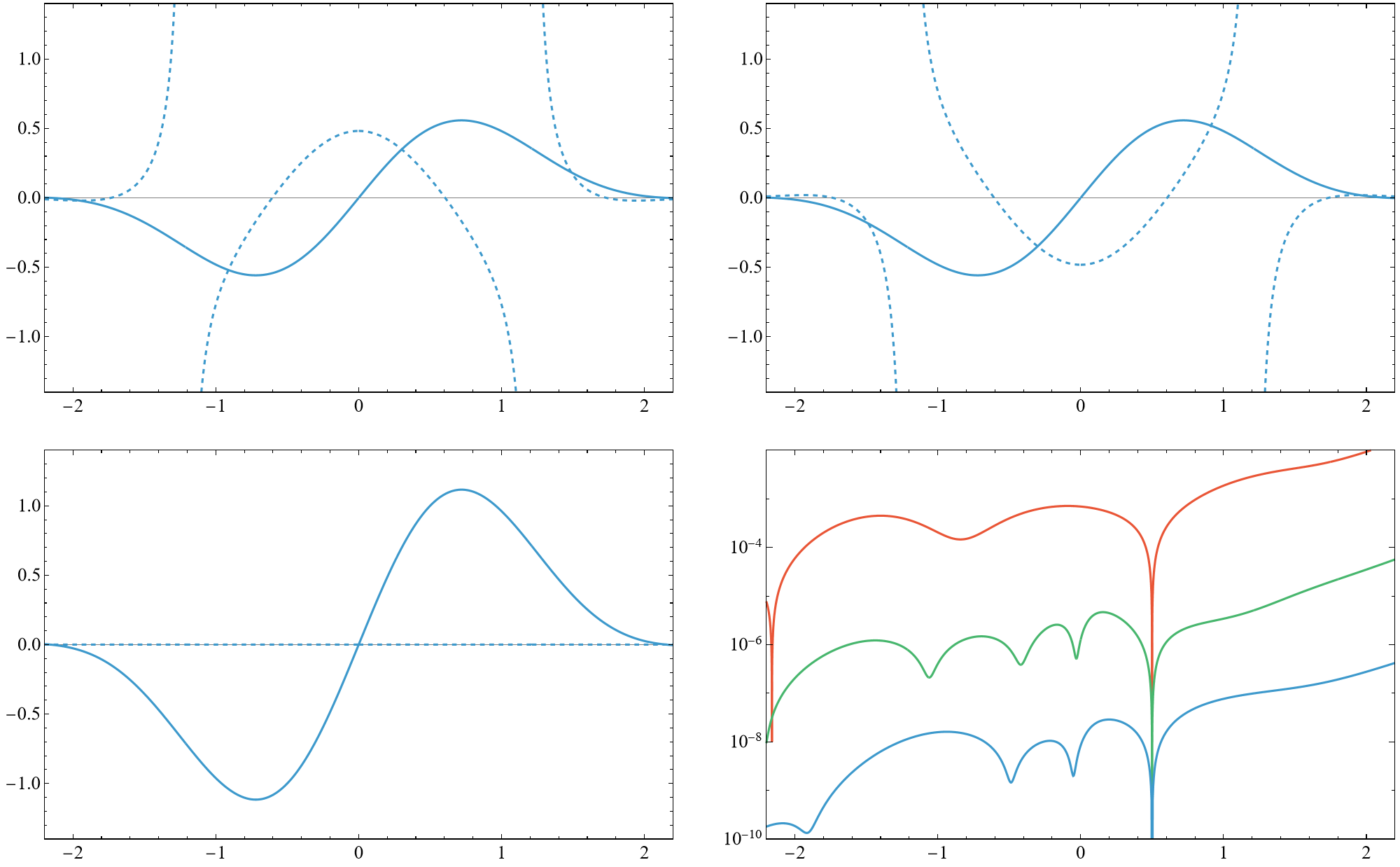}
    \caption{First excited state of the symmetric $V_4(x)$ potential with $h_2 = 3.5$, $h_3 = 0$, $\hbar = 1$ and $\Lambda = \tfrac{7}{8}$; the corresponding energy is $E_1\approx 7.324917$. Dashed lines denote the imaginary part of the eigenfunction, while solid lines denote the real part. Top panels: first (left) and second (right) saddle contributions in the analytic expression \eqref{eq:eigenfeven}, evaluated on-shell. The two functions develop poles individually, which cancel in the sum. Bottom left: full on-shell eigenfunction from \eqref{eq:eigenfeven}. Bottom right: difference between the numerical eigenfunction and the analytic result from \eqref{eq:eigenfeven}. The coloured curves show the effect of including an increasing number of terms in the $\Lambda$-expansion of the eigenfunction: red (0 terms), green (1 term), blue (2 terms).}
    \label{fig:su4_1st}
\end{figure}

\begin{figure}[ht]
    \centering
    \includegraphics[width=0.91\textwidth]{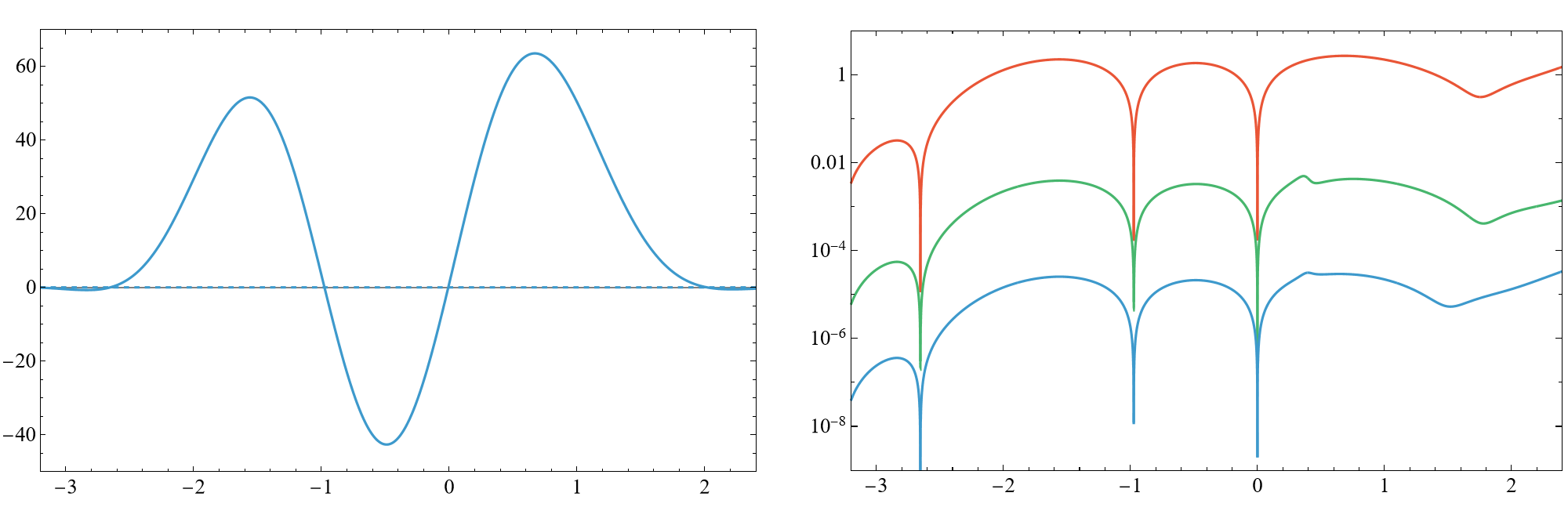}
    \caption{Left: second excited state of the $V_4(x)$ potential with $h_2 = 1.2$, $h_3 = -4.7$, $\hbar = 1$ and $\Lambda = \tfrac{4}{5}$; the corresponding energy is $E_2\approx 8.094909$. Dashed lines represent the imaginary part of the eigenfunction, while solid lines represent the real part. Right: difference between the numerical eigenfunction and the analytic expression from \eqref{eq:eigenfeven}. The coloured curves show the effect of including an increasing number of terms in the $\Lambda$-expansion of the eigenfunction: red (0 terms), green (1 term), blue (2 terms).}
    \label{fig:su4_2nd}
\end{figure}

In \autoref{fig:su3_ground} we show  the ground-state wavefunction for the cubic potential for $h_2=-5$. Due to the deep local minimum of the  potential, the wavefunction closely resembles that of a bound state. In \autoref{fig:su3_1st} we show the first excited state for $h_2=-3$, together with the two singular saddles, making the cancellation of poles in their sum evident.
Finally, for the cubic case, \autoref{fig:su3_2nd} shows the wavefunction corresponding to the second excited state for the cubic potential with $h_2=2$. In this case, the potential has no real stationary points, yet resonant eigenfunctions still exist. Moving on to the quartic case, in \autoref{fig:su4_1st} we show the wavefunction of the first excited state for the symmetric ($h_3=0$) single well potential with $h_2=3.5$, together with the two singular saddle contributions. The on-shell solution is antisymmetric and normalized to $1$ at $x = 1/2$. Finally, \autoref{fig:su4_2nd} displays the wavefunction corresponding to the second excited state of the quartic potential with $h_2=1.2$ and $h_3=-4.7$. The corresponding potential is an asymmetric quartic single well with a unique minimum.

\clearpage

\clearpage

\bibliographystyle{JHEP}

% To create a "References" entry in the contents and/or the pdf bookmarks.
\clearpage
\phantomsection % Creates an anchor for the hyperlink
\addcontentsline{toc}{section}{References} % Adds an entry to both the TOC & pdf bookmarks
%\pdfbookmark[1]{References}{References} % Adds an entry to the pdf bookmarks only

\bibliography{biblio.bib}

\providecommand{\href}[2]{#2}\begingroup\raggedright\begin{thebibliography}{10}

\bibitem{Marino:2015nla}
M.~Mari\~{n}o, \emph{{Spectral Theory and Mirror Symmetry}},  in
  \emph{Proceedings of Symposia in Pure Mathematics}, vol.~98, pp.~259--294,
  American Mathematical Society, 2018,
  \href{https://doi.org/https://doi.org/10.1090/pspum/098}{DOI}
  [\href{https://arxiv.org/abs/1506.07757}{{\ttfamily 1506.07757}}].

\bibitem{Marino:2024tbx}
M.~Mari{\~n}o, \emph{Les {{Houches}} lectures on non-perturbative topological
  strings},
  \href{https://doi.org/10.21468/SciPostPhysLectNotes.112}{\emph{SciPost Phys.
  Lect. Notes} (2026) 112} [\href{https://arxiv.org/abs/2411.16211}{{\ttfamily
  2411.16211}}].

\bibitem{TURBINER1987181}
A.~Turbiner and A.~Ushveridze, \emph{Spectral singularities and quasi-exactly
  solvable quantal problem},
  \href{https://doi.org/https://doi.org/10.1016/0375-9601(87)90456-7}{\emph{Physics
  Letters A} {\bfseries 126} (1987) 181}.

\bibitem{adkmv}
M.~Aganagic, R.~Dijkgraaf, A.~Klemm, M.~Marino and C.~Vafa, \emph{{Topological
  strings and integrable hierarchies}},
  \href{https://doi.org/10.1007/s00220-005-1448-9}{\emph{Commun. Math. Phys.}
  {\bfseries 261} (2006) 451}
  [\href{https://arxiv.org/abs/hep-th/0312085}{{\ttfamily hep-th/0312085}}].

\bibitem{acdkv}
M.~Aganagic, M.C.~Cheng, R.~Dijkgraaf, D.~Krefl and C.~Vafa, \emph{{Quantum
  Geometry of Refined Topological Strings}},
  \href{https://doi.org/10.1007/JHEP11(2012)019}{\emph{JHEP} {\bfseries 1211}
  (2012) 019} [\href{https://arxiv.org/abs/1105.0630}{{\ttfamily 1105.0630}}].

\bibitem{mirmor}
A.~Mironov and A.~Morozov, \emph{{Nekrasov Functions and Exact Bohr-Sommerfeld
  Integrals}}, \href{https://doi.org/10.1007/JHEP04(2010)040}{\emph{JHEP}
  {\bfseries 1004} (2010) 040}
  [\href{https://arxiv.org/abs/0910.5670}{{\ttfamily 0910.5670}}].

\bibitem{Ito:2021boh}
K.~Ito, T.~Kondo, K.~Kuroda and H.~Shu, \emph{{WKB periods for higher order ODE
  and TBA equations}},
  \href{https://doi.org/10.1007/JHEP10(2021)167}{\emph{JHEP} {\bfseries 10}
  (2021) 167} [\href{https://arxiv.org/abs/2104.13680}{{\ttfamily
  2104.13680}}].

\bibitem{Yan:2020kkb}
F.~Yan, \emph{{Exact WKB and the quantum Seiberg-Witten curve for 4d N = 2 pure
  SU(3) Yang-Mills. Abelianization}},
  \href{https://doi.org/10.1007/JHEP03(2022)164}{\emph{JHEP} {\bfseries 03}
  (2022) 164} [\href{https://arxiv.org/abs/2012.15658}{{\ttfamily
  2012.15658}}].

\bibitem{ns}
N.A.~Nekrasov and S.L.~Shatashvili, \emph{{Q}uantization of {I}ntegrable
  {S}ystems and {F}our {D}imensional {G}auge {T}heories},  in \emph{XVIth
  International Congress On Mathematical Physics}, pp.~265--289, World
  Scientific, 2010, \href{https://doi.org/10.1142/9789814304634_0015}{DOI}
  [\href{https://arxiv.org/abs/0908.4052}{{\ttfamily 0908.4052}}].

\bibitem{ghm}
A.~Grassi, Y.~Hatsuda and M.~Marino, \emph{{Topological Strings from Quantum
  Mechanics}}, \href{https://doi.org/10.1007/s00023-016-0479-4}{\emph{Ann.
  Henri Poincar{\'e}} {\bfseries 17} (2016) 3177}
  [\href{https://arxiv.org/abs/1410.3382}{{\ttfamily 1410.3382}}].

\bibitem{gmn}
D.~Gaiotto, G.W.~Moore and A.~Neitzke, \emph{{Wall-crossing, Hitchin systems,
  and the WKB approximation}},
  \href{https://doi.org/10.1016/j.aim.2012.09.027}{\emph{Adv. Math.} {\bfseries
  234} (2013) 239} [\href{https://arxiv.org/abs/0907.3987}{{\ttfamily
  0907.3987}}].

\bibitem{Gaiotto:2014bza}
D.~Gaiotto, \emph{{Opers and TBA}},
  \href{https://doi.org/10.48550/arXiv.1403.6137}{\emph{arXiv preprint} (2014)
  } [\href{https://arxiv.org/abs/1403.6137}{{\ttfamily 1403.6137}}].

\bibitem{Grassi:2018spf}
A.~Grassi and J.~Gu, \emph{{Argyres-Douglas theories, Painlev{\'e} II and
  quantum mechanics}},
  \href{https://doi.org/10.1007/JHEP02(2019)060}{\emph{JHEP} {\bfseries 02}
  (2019) 060} [\href{https://arxiv.org/abs/1803.02320}{{\ttfamily
  1803.02320}}].

\bibitem{Ito:2017ypt}
K.~Ito and H.~Shu, \emph{{ODE/IM correspondence and the Argyres-Douglas
  theory}}, \href{https://doi.org/10.1007/JHEP08(2017)071}{\emph{JHEP}
  {\bfseries 08} (2017) 071}
  [\href{https://arxiv.org/abs/1707.03596}{{\ttfamily 1707.03596}}].

\bibitem{Ito:2019twh}
K.~Ito, S.~Koizumi and T.~Okubo, \emph{{Quantum Seiberg-Witten curve and
  Universality in Argyres-Douglas theories}},
  \href{https://doi.org/10.1016/j.physletb.2019.03.024}{\emph{Phys. Lett. B}
  {\bfseries 792} (2019) 29}
  [\href{https://arxiv.org/abs/1903.00168}{{\ttfamily 1903.00168}}].

\bibitem{Hollands:2021itj}
L.~Hollands, P.~R\"uter and R.J.~Szabo, \emph{{A geometric recipe for twisted
  superpotentials}}, \href{https://doi.org/10.1007/JHEP12(2021)164}{\emph{JHEP}
  {\bfseries 12} (2021) 164}
  [\href{https://arxiv.org/abs/2109.14699}{{\ttfamily 2109.14699}}].

\bibitem{Ito:2018eon}
K.~Ito, M.~Mari{\~n}o and H.~Shu, \emph{{TBA equations and resurgent Quantum
  Mechanics}}, \href{https://doi.org/10.1007/JHEP01(2019)228}{\emph{JHEP}
  {\bfseries 01} (2019) 228}
  [\href{https://arxiv.org/abs/1811.04812}{{\ttfamily 1811.04812}}].

\bibitem{Fucito:2023txg}
F.~Fucito, A.~Grassi, J.F.~Morales and R.~Savelli, \emph{{Partition functions
  of non-Lagrangian theories from the holomorphic anomaly}},
  \href{https://doi.org/10.1007/JHEP07(2023)195}{\emph{JHEP} {\bfseries 07}
  (2023) 195} [\href{https://arxiv.org/abs/2306.05141}{{\ttfamily
  2306.05141}}].

\bibitem{Ito:2024wxw}
K.~Ito and J.~Yang, \emph{{TBA equations and quantum periods for D-type
  Argyres-Douglas theories}},
  \href{https://doi.org/10.1007/JHEP01(2025)047}{\emph{JHEP} {\bfseries 01}
  (2025) 047} [\href{https://arxiv.org/abs/2408.01124}{{\ttfamily
  2408.01124}}].

\bibitem{Bonelli:2025owb}
G.~Bonelli, A.~Shchechkin and A.~Tanzini, \emph{Refined {{Painlev\'e}}/{{Gauge
  Theory Correspondence}} and {{Quantum Tau Functions}}},
  \href{https://doi.org/10.1007/s00023-025-01621-8}{\emph{Ann. Henri
  Poincar{\'e}} (2025) } [\href{https://arxiv.org/abs/2502.01499}{{\ttfamily
  2502.01499}}].

\bibitem{Grassi:2018bci}
A.~Grassi and M.~Mari\~no, \emph{{A Solvable Deformation of Quantum
  Mechanics}}, \href{https://doi.org/10.3842/SIGMA.2019.025}{\emph{SIGMA}
  {\bfseries 15} (2019) 025}
  [\href{https://arxiv.org/abs/1806.01407}{{\ttfamily 1806.01407}}].

\bibitem{ggm}
A.~Grassi, J.~Gu and M.~Mari\~no, \emph{{Non-perturbative approaches to the
  quantum Seiberg-Witten curve}},
  \href{https://doi.org/10.1007/JHEP07(2020)106}{\emph{JHEP} {\bfseries 07}
  (2020) 106} [\href{https://arxiv.org/abs/1908.07065}{{\ttfamily
  1908.07065}}].

\bibitem{Chakrabarti:2023czz}
S.~Chakrabarti and M.~Raman, \emph{{Exploring T-Duality for Self-Dual Fields}},
  \href{https://doi.org/10.1002/prop.202400023}{\emph{Fortsch. Phys.}
  {\bfseries 72} (2024) 2400023}
  [\href{https://arxiv.org/abs/2311.09153}{{\ttfamily 2311.09153}}].

\bibitem{Marino:2016rsq}
M.~Marino and S.~Zakany, \emph{{Exact eigenfunctions and the open topological
  string}}, \href{https://doi.org/10.1088/1751-8121/aa791e}{\emph{J. Phys.}
  {\bfseries A50} (2017) 325401}
  [\href{https://arxiv.org/abs/1606.05297}{{\ttfamily 1606.05297}}].

\bibitem{Marino:2017gyg}
M.~Marino and S.~Zakany, \emph{{Wavefunctions, integrability, and open
  strings}}, \href{https://doi.org/10.1007/JHEP05(2019)014}{\emph{JHEP}
  {\bfseries 05} (2019) 014}
  [\href{https://arxiv.org/abs/1706.07402}{{\ttfamily 1706.07402}}].

\bibitem{Francois:2025wwd}
M.~Fran{\c c}ois and A.~Grassi, \emph{On the {{Open TS}}/{{ST
  Correspondence}}},
  \href{https://doi.org/10.1007/s00220-026-05608-2}{\emph{Commun. Math. Phys.}
  {\bfseries 407} (2026) 146}
  [\href{https://arxiv.org/abs/2503.21762}{{\ttfamily 2503.21762}}].

\bibitem{BS91}
F.A.~Berezin and M.A.~Shubin, \emph{The {{Schr\"odinger Equation}}}, no.~66 in
  Mathematics and Its {{Applications}}, Springer, Dordrecht (May, 1991),
  \href{https://doi.org/10.1007/978-94-011-3154-4}{10.1007/978-94-011-3154-4}.

\bibitem{Gutzwiller1980}
M.C.~Gutzwiller, \emph{The quantum mechanical {T}oda lattice},
  \href{https://doi.org/https://doi.org/10.1016/0003-4916(80)90214-6}{\emph{Ann.
  Phys.} {\bfseries 124} (1980) 347}.

\bibitem{Gutzwiller1981}
M.C.~Gutzwiller, \emph{The quantum mechanical {T}oda lattice, {II}},
  \href{https://doi.org/https://doi.org/10.1016/0003-4916(81)90253-0}{\emph{Ann.
  Phys.} {\bfseries 133} (1981) 304}.

\bibitem{Sklyanin1985}
E.K.~Sklyanin, \emph{The quantum {T}oda chain},  in \emph{Nonlinear Equations
  in Classical and Quantum Field Theory}, N.~Sanchez, ed., vol.~226 of
  \emph{Lecture Notes in Physics}, (Berlin, Heidelberg), pp.~196--233, Springer
  (1985), \href{https://doi.org/10.1007/3-540-15213-X_80}{DOI}.

\bibitem{PasquierGaudin1992}
V.~Pasquier and M.~Gaudin, \emph{The periodic {T}oda chain and a matrix
  generalization of the {Be}ssel function recursion relations},
  \href{https://doi.org/10.1088/0305-4470/25/20/007}{\emph{J. Phys. A: Math.
  Gen.} {\bfseries 25} (1992) 5243}.

\bibitem{Kozlowski:2010tv}
K.K.~Kozlowski and J.~Teschner, \emph{{TBA} for the {T}oda chain},  in
  \emph{{N}ew {T}rends in {Q}uantum {I}ntegrable {S}ystems}, pp.~195--219,
  World Scientific, 2010,
  \href{https://doi.org/10.1142/9789814324373_0011}{DOI}
  [\href{https://arxiv.org/abs/1006.2906}{{\ttfamily 1006.2906}}].

\bibitem{LST19}
A.~Laptev, L.~Schimmer and L.A.~Takhtajan, \emph{Weyl asymptotics for perturbed
  functional difference operators},
  \href{https://doi.org/10.1063/1.5093401}{\emph{Journal of Mathematical
  Physics} {\bfseries 60} (2019) 103505}.

\bibitem{LST16}
A.~Laptev, L.~Schimmer and L.A.~Takhtajan, \emph{{Weyl type asymptotics and
  bounds for the eigenvalues of functional-difference operators for mirror
  curves.}}, \href{https://doi.org/10.1007/s00039-016-0357-8}{\emph{Geom.
  Funct. Anal.} {\bfseries 26} (2016) 288}
  [\href{https://arxiv.org/abs/1510.00045}{{\ttfamily 1510.00045}}].

\bibitem{RS75}
M.~Reed and B.~Simon, \emph{Fourier {{Analysis}}, {{Self-Adjointness}}}, no.~2
  in Methods of {{Modern Mathematical Physics}}, Academic Press, San Diego, US
  (Sept., 1975).

\bibitem{S12}
K.~Schm{\"u}dgen, \emph{Unbounded {{Self-adjoint Operators}} on {{Hilbert
  Space}}}, vol.~265 of \emph{Graduate {{Texts}} in {{Mathematics}}}, Springer,
  Dordrecht (2012),
  \href{https://doi.org/10.1007/978-94-007-4753-1}{10.1007/978-94-007-4753-1}.

\bibitem{Caliceti}
E.~Caliceti, S.~Graffi and M.~Maioli, \emph{Perturbation theory of odd
  anharmonic oscillators},
  \href{https://doi.org/10.1007/BF01962591}{\emph{Commun. Math. Phys.}
  {\bfseries 75} (1980) 51}.

\bibitem{complexdila}
R.~Yaris, J.~Bendler, R.A.~Lovett, C.M.~Bender and P.A.~Fedders,
  \emph{Resonance calculations for arbitrary potentials},
  \href{https://doi.org/10.1103/PhysRevA.18.1816}{\emph{Phys. Rev. A}
  {\bfseries 18} (1978) 1816}.

\bibitem{Caliceti1983}
E.~Caliceti and M.~Maioli, \emph{Odd anharmonic oscillators and shape
  resonances}, {\emph{Annales de l'I.H.P. Physique théorique} {\bfseries 38}
  (1983) 175}.

\bibitem{Maioli}
M.~Maioli, \emph{Exponential perturbations of the harmonic oscillator},
  \href{https://doi.org/10.1063/1.525141}{\emph{Journal of Mathematical
  Physics} {\bfseries 22} (1981) 1952}.

\bibitem{Mironov:2009dv}
A.~Mironov and A.~Morozov, \emph{{Nekrasov Functions from Exact BS Periods: The
  Case of SU(N)}},
  \href{https://doi.org/10.1088/1751-8113/43/19/195401}{\emph{J. Phys. A}
  {\bfseries 43} (2010) 195401}
  [\href{https://arxiv.org/abs/0911.2396}{{\ttfamily 0911.2396}}].

\bibitem{Gaiotto:2014ina}
D.~Gaiotto and H.-C.~Kim, \emph{{Surface defects and instanton partition
  functions}}, \href{https://doi.org/10.1007/JHEP10(2016)012}{\emph{JHEP}
  {\bfseries 10} (2016) 012} [\href{https://arxiv.org/abs/1412.2781}{{\ttfamily
  1412.2781}}].

\bibitem{Bullimore:2014awa}
M.~Bullimore, H.-C.~Kim and P.~Koroteev, \emph{{Defects and Quantum
  Seiberg-Witten Geometry}},
  \href{https://doi.org/10.1007/JHEP05(2015)095}{\emph{JHEP} {\bfseries 05}
  (2015) 095} [\href{https://arxiv.org/abs/1412.6081}{{\ttfamily 1412.6081}}].

\bibitem{matone}
M.~Matone, \emph{{Instantons and recursion relations in $\mathcal{N}=2$ SUSY
  gauge theory}},
  \href{https://doi.org/10.1016/0370-2693(95)00920-G}{\emph{Phys. Lett.}
  {\bfseries B357} (1995) 342}
  [\href{https://arxiv.org/abs/hep-th/9506102}{{\ttfamily hep-th/9506102}}].

\bibitem{Flume:2004rp}
R.~Flume, F.~Fucito, J.F.~Morales and R.~Poghossian, \emph{{Matone's relation
  in the presence of gravitational couplings}},
  \href{https://doi.org/10.1088/1126-6708/2004/04/008}{\emph{JHEP} {\bfseries
  04} (2004) 008} [\href{https://arxiv.org/abs/hep-th/0403057}{{\ttfamily
  hep-th/0403057}}].

\bibitem{Alday:2010vg}
L.F.~Alday and Y.~Tachikawa, \emph{{Affine SL(2) conformal blocks from 4d gauge
  theories}}, \href{https://doi.org/10.1007/s11005-010-0422-4}{\emph{Lett.
  Math. Phys.} {\bfseries 94} (2010) 87}
  [\href{https://arxiv.org/abs/1005.4469}{{\ttfamily 1005.4469}}].

\bibitem{Kanno:2011fw}
H.~Kanno and Y.~Tachikawa, \emph{{Instanton counting with a surface operator
  and the chain-saw quiver}},
  \href{https://doi.org/10.1007/JHEP06(2011)119}{\emph{JHEP} {\bfseries 06}
  (2011) 119} [\href{https://arxiv.org/abs/1105.0357}{{\ttfamily 1105.0357}}].

\bibitem{Sciarappa:2017hds}
A.~Sciarappa, \emph{{Exact relativistic Toda chain eigenfunctions from
  Separation of Variables and gauge theory}},
  \href{https://doi.org/10.1007/JHEP10(2017)116}{\emph{JHEP} {\bfseries 10}
  (2017) 116} [\href{https://arxiv.org/abs/1706.05142}{{\ttfamily
  1706.05142}}].

\bibitem{Jeong:2021rll}
S.~Jeong, N.~Lee and N.~Nekrasov, \emph{{Intersecting defects in gauge theory,
  quantum spin chains, and Knizhnik-Zamolodchikov equations}},
  \href{https://doi.org/10.1007/JHEP10(2021)120}{\emph{JHEP} {\bfseries 10}
  (2021) 120} [\href{https://arxiv.org/abs/2103.17186}{{\ttfamily
  2103.17186}}].

\bibitem{Gu:2026xgp}
J.~Gu and M.~Marino, \emph{{Thou shalt not tunnel: Complex instantons and
  tunneling suppression in deformed quantum mechanics}},
  \href{https://doi.org/10.48550/arXiv.2602.20576}{\emph{arXiv preprint} (2026)
  } [\href{https://arxiv.org/abs/2602.20576}{{\ttfamily 2602.20576}}].

\bibitem{wip}
M.~Fran{\c{c}}ois, A.~Grassi and T.~Pedroni, ``Work in progress.''.

\bibitem{A26}
K.~Aziz, ``{Group-Theoretic Identities and the Quantum Toda Lattice}.'' 2026.

\bibitem{cgm2}
S.~Codesido, A.~Grassi and M.~Marino, \emph{{Spectral Theory and Mirror Curves
  of Higher Genus}},
  \href{https://doi.org/10.1007/s00023-016-0525-2}{\emph{Ann. Henri
  Poincar{\'e}} {\bfseries 18} (2017) 559}
  [\href{https://arxiv.org/abs/1507.02096}{{\ttfamily 1507.02096}}].

\bibitem{kkv}
S.H.~Katz, A.~Klemm and C.~Vafa, \emph{{Geometric engineering of quantum field
  theories}},
  \href{https://doi.org/10.1016/S0550-3213(97)00282-4}{\emph{Nucl.Phys.}
  {\bfseries B497} (1997) 173}
  [\href{https://arxiv.org/abs/hep-th/9609239}{{\ttfamily hep-th/9609239}}].

\bibitem{Klemm:1996bj}
A.~Klemm, W.~Lerche, P.~Mayr, C.~Vafa and N.P.~Warner, \emph{{Selfdual strings
  and N=2 supersymmetric field theory}},
  \href{https://doi.org/10.1016/0550-3213(96)00353-7}{\emph{Nucl. Phys. B}
  {\bfseries 477} (1996) 746}
  [\href{https://arxiv.org/abs/hep-th/9604034}{{\ttfamily hep-th/9604034}}].

\bibitem{Hori2003}
K.~Hori, S.~Katz, A.~Klemm, R.~Pandharipande, R.~Thomas, C.~Vafa et~al.,
  \emph{Mirror Symmetry}, vol.~1 of \emph{Clay Mathematics Monographs},
  American Mathematical Society, Providence, USA (2003).

\bibitem{hm}
Y.~Hatsuda and M.~Marino, \emph{{Exact quantization conditions for the
  relativistic Toda lattice}},
  \href{https://doi.org/10.1007/JHEP05(2016)133}{\emph{JHEP} {\bfseries 05}
  (2016) 133} [\href{https://arxiv.org/abs/1511.02860}{{\ttfamily
  1511.02860}}].

\bibitem{GGH26}
P.~Gavrylenko, A.~Grassi and Q.~Hao, \emph{Connecting topological strings and
  spectral theory via non-autonomous {{Toda}} equations},
  \href{https://doi.org/10.4310/ATMP.260521223306}{\emph{Adv. Theor. Math.
  Phys.} {\bfseries 30} (2026) 335}
  [\href{https://arxiv.org/abs/2304.11027}{{\ttfamily 2304.11027}}].

\bibitem{bgt2}
G.~Bonelli, A.~Grassi and A.~Tanzini, \emph{{New results in $\mathcal{N}=2$
  theories from non-perturbative string}},
  \href{https://doi.org/10.1007/s00023-017-0643-5}{\emph{Ann. Henri
  Poincar{\'e}} {\bfseries 19} (2018) 743}
  [\href{https://arxiv.org/abs/1704.01517}{{\ttfamily 1704.01517}}].

\bibitem{Hatsuda:2013oxa}
Y.~Hatsuda, M.~Marino, S.~Moriyama and K.~Okuyama, \emph{{Non-perturbative
  effects and the refined topological string}},
  \href{https://doi.org/10.1007/JHEP09(2014)168}{\emph{JHEP} {\bfseries 09}
  (2014) 168} [\href{https://arxiv.org/abs/1306.1734}{{\ttfamily 1306.1734}}].

\bibitem{Francois:2023trm}
M.~Fran{\c c}ois and A.~Grassi, \emph{Painlev\'e {{Kernels}} and {{Surface
  Defects}} at {{Strong Coupling}}},
  \href{https://doi.org/10.1007/s00023-024-01469-4}{\emph{Ann. Henri
  Poincar{\'e}} {\bfseries 26} (2025) 2117}
  [\href{https://arxiv.org/abs/2310.09262}{{\ttfamily 2310.09262}}].

\bibitem{Iqbal:2003zz}
A.~Iqbal and A.-K.~Kashani-Poor, \emph{{SU(N) geometries and topological string
  amplitudes}}, \href{https://doi.org/10.4310/ATMP.2006.v10.n1.a1}{\emph{Adv.
  Theor. Math. Phys.} {\bfseries 10} (2006) 1}
  [\href{https://arxiv.org/abs/hep-th/0306032}{{\ttfamily hep-th/0306032}}].

\bibitem{Iqbal:2007ii}
A.~Iqbal, C.~Kozcaz and C.~Vafa, \emph{{The Refined topological vertex}},
  \href{https://doi.org/10.1088/1126-6708/2009/10/069}{\emph{JHEP} {\bfseries
  10} (2009) 069} [\href{https://arxiv.org/abs/hep-th/0701156}{{\ttfamily
  hep-th/0701156}}].

\bibitem{DM68a}
R.B.~Dingle and G.J.~Morgan, \emph{{WKB} methods for difference equations {I}},
  \href{https://doi.org/10.1007/BF00382348}{\emph{Appl. Sci. Res.} {\bfseries
  18} (1968) 221}.

\bibitem{DM68b}
R.B.~Dingle and G.J.~Morgan, \emph{{WKB} methods for difference equations
  {II}}, \href{https://doi.org/10.1007/BF00382349}{\emph{Appl. Sci. Res.}
  {\bfseries 18} (1968) 238}.

\bibitem{Kashani-Poor:2016edc}
A.-K.~Kashani-Poor, \emph{{Quantization condition from exact WKB for difference
  equations}}, \href{https://doi.org/10.1007/JHEP06(2016)180}{\emph{JHEP}
  {\bfseries 06} (2016) 180}
  [\href{https://arxiv.org/abs/1604.01690}{{\ttfamily 1604.01690}}].

\bibitem{Grassi:2022zuk}
A.~Grassi, Q.~Hao and A.~Neitzke, \emph{{Exponential Networks, WKB and
  Topological String}},
  \href{https://doi.org/10.3842/SIGMA.2023.064}{\emph{SIGMA} {\bfseries 19}
  (2023) 064} [\href{https://arxiv.org/abs/2201.11594}{{\ttfamily
  2201.11594}}].

\bibitem{Alim:2022oll}
M.~Alim, L.~Hollands and I.~Tulli, \emph{{Quantum Curves, Resurgence and Exact
  WKB}}, \href{https://doi.org/10.3842/SIGMA.2023.009}{\emph{SIGMA} {\bfseries
  19} (2023) 009} [\href{https://arxiv.org/abs/2203.08249}{{\ttfamily
  2203.08249}}].

\bibitem{DelMonte:2024dcr}
F.~Del~Monte and P.~Longhi, \emph{{Monodromies of Second Order $q$-difference
  Equations from the WKB Approximation}},
  \href{https://doi.org/10.48550/arXiv.2406.00175}{\emph{arXiv preprint} (2024)
  } [\href{https://arxiv.org/abs/2406.00175}{{\ttfamily 2406.00175}}].

\bibitem{Hao:2025azt}
Q.~Hao, \emph{{Exact WKB of solutions by Borel summation and open TBA}},
  \href{https://doi.org/https://doi.org/10.48550/arXiv.2507.06922}{\emph{arXiv
  preprint} (2025) } [\href{https://arxiv.org/abs/2507.06922}{{\ttfamily
  2507.06922}}].

\bibitem{ZILS25}
A.~Zernova, A.~Ilyin, A.~Laptev and L.~Schimmer, \emph{Eigenvalues of
  {{Non-Selfadjoint Functional Difference Operators}}},
  \href{https://doi.org/10.1134/S1234567825030048}{\emph{Funct. Anal. Its
  Appl.} {\bfseries 59} (2025) 258}
  [\href{https://arxiv.org/abs/2504.06858}{{\ttfamily 2504.06858}}].

\bibitem{bgt}
G.~Bonelli, A.~Grassi and A.~Tanzini, \emph{{Seiberg\textendash{}Witten theory
  as a Fermi gas}},
  \href{https://doi.org/10.1007/s11005-016-0893-z}{\emph{Lett. Math. Phys.}
  {\bfseries 107} (2017) 1} [\href{https://arxiv.org/abs/1603.01174}{{\ttfamily
  1603.01174}}].

\bibitem{EllegaardAndersen:2011vps}
J.~Ellegaard~Andersen and R.~Kashaev, \emph{{A TQFT from Quantum Teichm\"uller
  Theory}}, \href{https://doi.org/10.1007/s00220-014-2073-2}{\emph{Commun.
  Math. Phys.} {\bfseries 330} (2014) 887}
  [\href{https://arxiv.org/abs/1109.6295}{{\ttfamily 1109.6295}}].

\bibitem{Garoufalidis:2014ifa}
S.~Garoufalidis and R.~Kashaev, \emph{{Evaluation of state integrals at
  rational points}},
  \href{https://doi.org/10.4310/CNTP.2015.v9.n3.a3}{\emph{Commun. Num. Theor.
  Phys.} {\bfseries 09} (2015) 549}
  [\href{https://arxiv.org/abs/1411.6062}{{\ttfamily 1411.6062}}].

\bibitem{Hatsuda:2015owa}
Y.~Hatsuda and K.~Okuyama, \emph{{Resummations and Non-Perturbative
  Corrections}}, \href{https://doi.org/10.1007/JHEP09(2015)051}{\emph{JHEP}
  {\bfseries 09} (2015) 051}
  [\href{https://arxiv.org/abs/1505.07460}{{\ttfamily 1505.07460}}].

\bibitem{adamchik2003contributions}
V.S.~Adamchik, \emph{Symbolic and numeric computations of the {B}arnes
  function},
  \href{https://doi.org/https://doi.org/10.1016/S0010-4655(03)00498-3}{\emph{Computer
  Physics Communications} {\bfseries 157} (2004) 181}
  [\href{https://arxiv.org/abs/math/0308086}{{\ttfamily math/0308086}}].

\bibitem{Moore:1997dj}
G.W.~Moore, N.~Nekrasov and S.~Shatashvili, \emph{{Integrating over Higgs
  branches}}, \href{https://doi.org/10.1007/PL00005525}{\emph{Commun. Math.
  Phys.} {\bfseries 209} (2000) 97}
  [\href{https://arxiv.org/abs/hep-th/9712241}{{\ttfamily hep-th/9712241}}].

\bibitem{Lossev:1997bz}
A.~Losev, N.~Nekrasov and S.L.~Shatashvili, \emph{Testing {S}eiberg-{W}itten
  {S}olution},  in \emph{Strings, Branes and Dualities}, L.~Baulieu,
  P.~Di~Francesco, M.~Douglas, V.~Kazakov, M.~Picco and P.~Windey, eds.,
  vol.~520 of \emph{NATO ASI Series}, (Dordrecht), pp.~359--372, Springer
  (1999), \href{https://doi.org/10.1007/978-94-011-4730-9_13}{DOI}
  [\href{https://arxiv.org/abs/hep-th/9801061}{{\ttfamily hep-th/9801061}}].

\bibitem{Nekrasov:2002qd}
N.A.~Nekrasov, \emph{{Seiberg-Witten prepotential from instanton counting}},
  \href{https://doi.org/10.4310/ATMP.2003.v7.n5.a4}{\emph{Adv. Theor. Math.
  Phys.} {\bfseries 7} (2004) 831}
  [\href{https://arxiv.org/abs/hep-th/0206161}{{\ttfamily hep-th/0206161}}].

\bibitem{Flume:2002az}
R.~Flume and R.~Poghossian, \emph{{An Algorithm for the microscopic evaluation
  of the coefficients of the Seiberg-Witten prepotential}},
  \href{https://doi.org/10.1142/S0217751X03013685}{\emph{Int. J. Mod. Phys.}
  {\bfseries A18} (2003) 2541}
  [\href{https://arxiv.org/abs/hep-th/0208176}{{\ttfamily hep-th/0208176}}].

\bibitem{Bruzzo:2002xf}
U.~Bruzzo, F.~Fucito, J.F.~Morales and A.~Tanzini, \emph{{Multiinstanton
  calculus and equivariant cohomology}},
  \href{https://doi.org/10.1088/1126-6708/2003/05/054}{\emph{JHEP} {\bfseries
  0305} (2003) 054} [\href{https://arxiv.org/abs/hep-th/0211108}{{\ttfamily
  hep-th/0211108}}].

\bibitem{ilt}
A.~Its, O.~Lisovyy and Y.~Tykhyy, \emph{Connection problem for the
  sine-gordon/painlevé iii tau function and irregular conformal blocks},
  \href{https://doi.org/10.1093/imrn/rnu209}{\emph{Int. Math. Res. Notices}
  {\bfseries 2015} (2014) 8903}
  [\href{https://arxiv.org/abs/1403.1235}{{\ttfamily 1403.1235}}].

\bibitem{Arnaudo:2022ivo}
P.~Arnaudo, G.~Bonelli and A.~Tanzini, \emph{{On the convergence of Nekrasov
  functions}}, \href{https://doi.org/10.1007/s00023-023-01349-3}{\emph{Ann.
  Henri Poincar{\'e}} {\bfseries 25} (2024) 2389–2425}
  [\href{https://arxiv.org/abs/2212.06741}{{\ttfamily 2212.06741}}].

\bibitem{Desiraju:2024fmo}
H.~Desiraju, P.~Ghosal and A.~Prokhorov, \emph{{Proof of Zamolodchikov
  conjecture for semi-classical conformal blocks on the torus}},
  \href{https://doi.org/https://doi.org/10.48550/arXiv.2407.05839}{\emph{arXiv
  preprint} (2024) } [\href{https://arxiv.org/abs/2407.05839}{{\ttfamily
  2407.05839}}].

\end{thebibliography}\endgroup

\end{document}